\documentclass{elsart}

\usepackage{graphicx}

\usepackage{latexsym,amsmath,amssymb}

\newcommand{\pt}{\partial}
\newcommand{\nn}{\nonumber}
\newcommand{\be}{\begin{equation}}
\newcommand{\ee}{\end{equation}}
\newcommand{\bea}{\begin{eqnarray}}
\newcommand{\eea}{\end{eqnarray}}
\newcommand{\eg}{{\it e.g.}~}
\newcommand{\ie}{{\it i.e.}~}
\newcommand{\tr}{\textrm{tr}}

\def\a{\alpha}
\def\b{\beta}

\renewcommand{\d}{\mathrm{d}}
\def\e{\epsilon}

\def\f{\phi}

\def\k{\kappa}

\def\o{\omega}

\def\s{\sigma}

\def\t{\tau}

\def\cN{{\cal N}}
\def\cP{{\cal P}}

\begin{document}

\begin{frontmatter}



\title{Ekpyrotic and Cyclic Cosmology}


\author{Jean-Luc Lehners}

\address{Princeton Center for Theoretical Science, Jadwin Hall, Princeton University, Princeton NJ 08544, USA}

\ead{jlehners@princeton.edu}

\begin{abstract}

Ekpyrotic and cyclic cosmologies provide theories of the very
early and of the very late universe. In these models, the big bang
is described as a collision of branes - and thus the big bang is
not the beginning of time. Before the big bang, there is an
ekpyrotic phase with equation of state $w=\frac{\cP}{\rho}\gg 1$
(where $\cP$ is the average pressure and $\rho$ the average energy
density) during which the universe slowly contracts. This phase
resolves the standard cosmological puzzles and generates a nearly
scale-invariant spectrum of cosmological perturbations containing
a significant non-gaussian component. At the same time it produces
small-amplitude gravitational waves with a blue spectrum. The dark
energy dominating the present-day cosmological evolution is
reinterpreted as a small attractive force between our brane and a
parallel one. This force eventually induces a new ekpyrotic phase
and a new brane collision, leading to the idea of a cyclic
universe. This review discusses the detailed properties of these
models, their embedding in M-theory and their viability, with an
emphasis on open issues and observational signatures.

\end{abstract}

\begin{keyword}
cosmology \sep branes \sep big bang \sep dark energy
\PACS 98.80.Es \sep 98.80.Cq \sep 03.70.+k
\end{keyword}
\end{frontmatter}

\tableofcontents{}
\section{Introduction}
\label{section intro}

The evolution of our universe is very well understood from the
time of big bang nucleosynthesis until the present. However, if we
go beyond these time frontiers, almost nothing is known with a
great amount of certainty. We know that the early universe must
have been in a very special state (very homogeneous and flat, but
with tiny fluctuations in curvature), but we do not know why. The
theory of inflation \cite{Guth:1980zm,Linde:1981mu,Albrecht:1982wi} provides a possible explanation by means of a
period of rapid expansion preceding nucleosynthesis, but it has
not yet been possible to test inflation
conclusively, which is why it is important to keep an open mind. Also,
we know that dark energy has come to dominate the energy density
of the universe \cite{Riess:1998cb,Perlmutter:1998np} and that it will determine the expansion of the
universe in the near future. If dark energy turns out to be a
cosmological constant, which is the theoretically simplest way of
modeling it, then the universe will continue to expand at an
accelerating pace and become increasingly dilute and cold, making
life very difficult, if not impossible. However, it should be borne in mind that we have no fully convincing explanation of dark energy at present.

This review will be concerned with a set of ideas, strongly
inspired by string theory, that suggests alternative solutions to
the early universe puzzles mentioned above, and considers an alternative fate
for the future of our universe. What makes these ideas
exciting is that, on the one hand, they provide a theoretical
playground for applying string theory to cosmology, and, on the
other hand, certain observational signatures predicted from these
models are in a range that will be tested by near-future satellite
experiments. This provides a certain timeliness to a subject
concerned with eras far removed from everyday experience.

\begin{figure}[t]
\begin{center}
\includegraphics{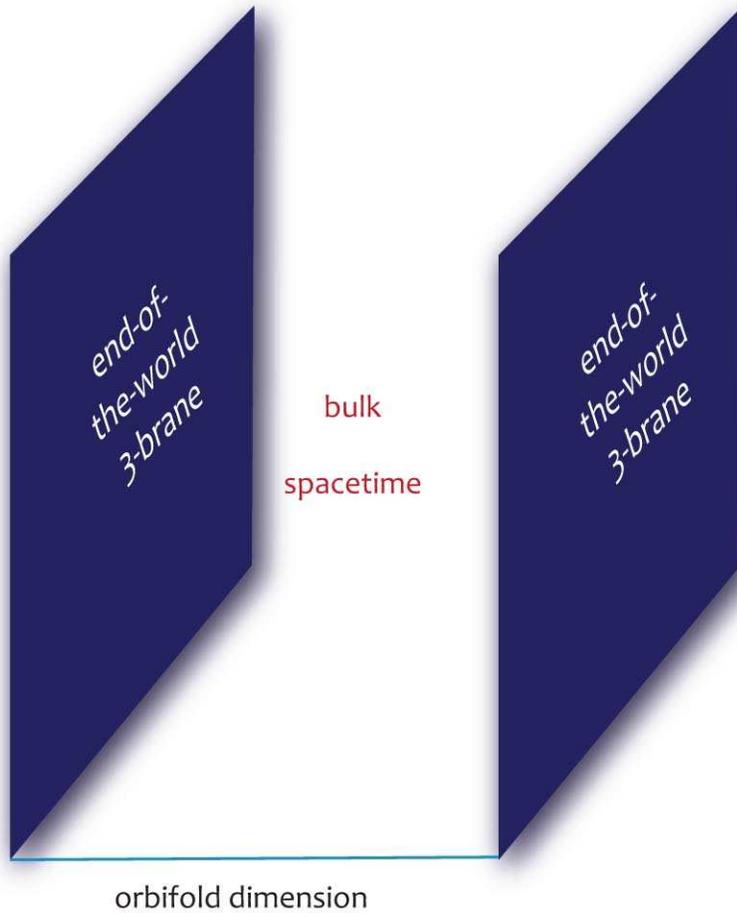}
\caption{\label{figure braneworld} {\small The braneworld picture
of our universe. Think of a sandwich: the filling is the
5-dimensinonal bulk spacetime, which is bounded by the two pieces
of bread a.k.a. the 4-dimensional boundary branes. There is no
space ``outside'' of the sandwich, but the branes can be infinite
in all directions perpendicular to the line segment. In the M-theory embedding, there are 6 additional internal dimensions at each 5-dimensional spacetime point.}}
\end{center}
\end{figure}

Ekpyrotic\footnote{The name ekpyrosis can be translated as all-engulfing cosmic fire. In Stoic philosophy, it represents the contractive phase of eternally-recurring destruction and re-creation \cite{Wiki}.} and cyclic cosmology are based on the braneworld picture
of the universe, in which spacetime is effectively 5-dimensional,
but with one dimension not extending indefinitely, but being a
line segment\footnote{This setting will be discussed in detail in
section \ref{section fundamental}, along with its motivation from
and embedding in heterotic M-theory.}, see Fig. \ref{figure
braneworld}. The endpoints of this line segment (orbifold) are two
$(3+1)$-dimensional boundary branes.
All matter and forces, except for gravity, are localized on the branes, while
gravity can propagate in the whole spacetime. Our universe, as we
see it, is identified with one of the boundary branes and, as long
as the branes are far apart, can interact with the other brane
only via gravity. The ekpyrotic model assumes that there is an
attractive force between the two branes, which causes the branes
to approach each other. This ekpyrotic phase has the rather
non-intuitive property that it flattens the branes to a very high
degree. Eventually the two branes collide and move through each
other (since there is no space ``outside'' of the boundary branes,
it makes no difference whether we say that the branes bounce off
each other or move through each other). It is this collision that,
from the point of view of someone living on one of the branes,
looks like the big bang. The collision is slightly inelastic and
produces matter and radiation on the branes, where the standard
cosmological evolution now takes place. However, due to quantum
fluctuations, the branes are slightly rippled and do not collide
everywhere at exactly the same time. In some places, the branes
collide slightly earlier, which means that the universe has a
little bit more time to expand and cool. In other places, the
collision takes place slightly later, and those regions remain a
little hotter. This provides a heuristic picture of the way
temperature fluctuations are naturally produced within the model.
Shortly after the brane collision, the distance between the
boundary branes gets almost stabilized, but the branes start
attracting each other again very slightly. This very slight
attraction acts as quintessence, and is identified with the
dark energy observed in the universe. After a long time and as the
branes become closer again, they start attracting each other more
strongly so that we get a new ekpyrotic phase and eventually a new
brane collision with the creation of new matter. In this way, a
cyclic model of the universe emerges.

The preceding description of course only provided a very rough
outline of the main ideas. All of this will be made precise in the
following. However, at this stage it is already clear that
conceptually the cyclic universe differs substantially from the
standard big bang picture. For one, the big bang is seen as a
physical event and not a mysterious moment of creation. As such,
it does not represent the beginning of time (note that if quantum
gravity is unitary, this point of view seems unavoidable). Thus
there was plenty of time before the big bang for the universe to
be in causal contact over large regions, and in this way the
horizon problem is automatically solved. The long timescales
existing before the big bang also imply that other cosmological
puzzles don't have to be solved in the extremely short time
interval between the big bang and nucleosynthesis. What's more,
properties of the very early universe (such as cosmological
perturbations) and properties of the late universe (such as dark
energy) are described by the same ingredients, namely the motion
of branes.

In the description above, the higher-dimensional
interpretation was emphasized. This higher-dimensional viewpoint
is useful for understanding the origin of the ideas, and provides
a concrete realization of them. However, it is important to
realize that many of the processes mentioned above can also be
discussed purely in a 4-dimensional effective theory. Some
processes, such as the ekpyrotic period of slow contraction before
the big bang, are quite general and could have a different origin
than the brane motion just described. Therefore, the four
subsequent sections of this review, which describe the
fundamentals of ekpyrotic and cyclic cosmology, are phrased mostly
in terms of a 4-dimensional effective theory. Section \ref{section
ekpyrotic} describes the ekpyrotic phase and the way it resolves
the standard cosmological puzzles. Section \ref{section crunch} describes the approach to the big crunch in more detail. The cosmological fluctuations
generated by the model are discussed in section \ref{section
perturbations}. This includes a treatment of scalar perturbations
with their second-order non-gaussian corrections as well as tensor
perturbations. The following section is concerned with the details
of the cyclic universe, such as its relationship to the second law
of thermodynamics and its global structure. Section \ref{section
fundamental} explains the origin of the 4-dimensional effective
theory used in the preceding sections and its relation to string
and M-theory. Finally, section \ref{section outlook} provides a brief conclusion and an outlook to possible future developments.

It should be noted from the outset that ekpyrotic and cyclic
cosmology are theories that are still in full development. Certain
aspects of the theories have gained or lost in importance over
time. Here are two examples: in the original ekpyrotic proposal,
the brane collision was between a boundary and a bulk brane,
the second boundary brane being a spectator. It was later realized
that the model could be simplified by considering a collision of
the boundary branes. It was also thought that the dark energy
phase was crucial for smoothing and isotropizing the
universe. Again, it was later realized that the ekpyrotic phase
can achieve this on its own. In this review the subject will be
discussed only according to its present understanding. Moreover,
every effort will be made to highlight assumptions that are in
need of a precise calculation, or that are due for example to
limitations in our understanding of string theory. This should
hopefully provide the reader with opportunities to contribute to
the subject.


\section{The Ekpyrotic Phase}
\label{section ekpyrotic}

The ekpyrotic phase is the central insight that shows how a
contracting phase preceding the big bang can solve the standard
cosmological puzzles \cite{Khoury:2001wf,Erickson:2006wc}. This is a surprising statement since one
would naively think that a contracting, gravitating system
would lead to large curvatures near a singularity. And we know
from the near-flatness of our current universe that shortly
after the big bang the universe must have been extremely flat.
This is usually referred to as the flatness problem. Let us
briefly quantify this problem; at the same time this will serve
to set up notation. Consider a Friedmann-Robertson-Walker (FRW)
metric\footnote{I will mostly use natural units $\hbar=c=1$ and
$8\pi G = M_{Pl}^{-2}=1.$} \be
\d s^2=-\d t^2+a(t)^2\left(\frac{\d r^2}{1-\kappa r^2}+r^2
\d \Omega_2^2 \right) \label{FRWmetric}\ee where $a(t)$ denotes
the scale factor of the universe and $\kappa=-1,0,1$ for an
open, flat or closed universe respectively. If we consider the
universe to be filled with a perfect fluid (which is a good
approximation for many types of matter), with energy-momentum
tensor \be T^{\mu}_{\nu} = diag[-\rho,\cP,\cP,\cP],\ee where $\rho$ denotes the energy density and $\cP$ the pressure, then the
Einstein equations reduce to the so-called Friedmann equations
\bea H^2 &=& \frac{1}{3} \rho - \frac{\kappa}{a^2}
\label{Friedmann1} \\ \frac{\ddot{a}}{a} &=&
-\frac{1}{6}(\rho+3\cP), \label{Friedmann2} \eea where the Hubble
parameter is defined by $H\equiv \frac{\dot{a}}{a}$ and a dot
denotes a derivative w.r.t. time $t.$ The Bianchi identity on
the Einstein tensor $\nabla^{\nu} G_{\mu \nu}=0$ requires for
consistency that the energy-momentum tensor be covariantly
conserved, \ie $\nabla^{\nu} T_{\mu \nu}=0,$ which leads to the
continuity equation \be \dot{\rho} + 3
\frac{\dot{a}}{a}(\rho+\cP)=0. \label{continuity}\ee For a fluid
with a constant equation of state $w\equiv \frac{\cP}{\rho},$ the
continuity equation can be integrated to give \be \rho \propto
a^{-3(1+w)}. \label{fluidscaling}\ee We will use this scaling
relation on many occasions. Now define the quantity
$\Omega(t)\equiv\frac{\rho(t)}{\rho_{crit}(t)}$ where
$\rho_{crit}=3H^2$ denotes the critical density. Then the first
Friedmann equation (\ref{Friedmann1}) can be rewritten as \be
\Omega-1=\frac{\kappa}{(aH)^2} \label{flatnessscaling}\ee and it expresses how close
the universe is to flatness. At the present time, observations
show that \cite{Komatsu:2008hk} \be |\Omega-1|_0 \lesssim 10^{-2}. \ee
Extrapolating back in time, this means that at the Planck time
\be
\frac{|\Omega-1|_{Pl}}{|\Omega-1|_0}=\frac{(aH)_0^2}{(aH)_{Pl}^2}.
\ee If we assume a radiation-dominated universe ($w=1/3$), which is a
good approximation for the present calculation, then the
continuity and Friedmann equations lead to $\rho \propto
a^{-4}$ and $a\propto t^{1/2},$ so that \be |\Omega-1|_{Pl}
\lesssim \frac{t_{Pl}}{t_0}10^{-2}\sim 10^{-62}.
\label{flatness-constraint}\ee Even though extrapolating all
the way back to the Planck time is probably exaggerated, this
simple estimate shows that the universe must have been
extremely flat at early times. Clearly, this peculiar
observation asks for an explanation.

In order to understand how the ekpyrotic phase can resolve this
problem, it is useful to briefly review the solution proposed
in inflationary theory. Inflation postulates a period of rapid
expansion immediately following the big bang, during which the
equation of state $w=\frac{\cP}{\rho}$ of the universe is close
to $-1$ \cite{Guth:1980zm,Linde:1981mu,Albrecht:1982wi}. From the acceleration equation \be
\frac{\ddot{a}}{a}=-\frac{1}{6}(\rho+3\cP)=-\frac{\rho}{6}(1+3w),\ee
we can see that such an equation of state indeed leads to
accelerated expansion. One way to model a matter component with
the required equation of state is to have a scalar field $\phi$
with canonical kinetic energy and with a very flat potential
$V(\phi),$ \ie we add the following terms to the Lagrangian \be
\sqrt{-g}\left( - \frac{1}{2}g^{\mu \nu}\pt_{\mu}\phi
\pt_{\nu}\phi -V(\phi)\right). \ee Then a quick calculation
shows that (in the absence of spatial gradients) the equation
of state is given by \be w_{\phi}=
\frac{\cP_{\phi}}{\rho_{\phi}}=\frac{\frac{1}{2}\dot{\phi}^2-V(\phi)}{\frac{1}{2}\dot{\phi}^2+V(\phi)},
\label{scalareqofstate} \ee which is close to $-1$ if the
potential is sufficiently flat so that the fields are rolling
very slowly, \ie if $\frac{1}{2}\dot{\phi}^2 \ll V(\phi).$ In
the presence of different matter types, represented here by
their energy densities $\rho,$ the Friedmann equation
(\ref{Friedmann1}) generalizes to \be H^2 = \frac{1}{3}
\left(\frac{-3\kappa}{a^2}+ \frac{\rho_m}{a^3} +
\frac{\rho_r}{a^4}+ \frac{\sigma^2}{a^6} + \ldots +
\frac{\rho_{\phi}}{a^{3(1+w_{\phi})}} \right) \label{Friedmann}
\ee The subscript $m$ refers to non-relativistic matter and
includes dark matter, $r$ refers to radiation and $\sigma$
denotes the energy density of anisotropies in the curvature of
the universe (the associated scaling with $a$ will be calculated in the following
section). In an expanding universe, as the scale factor $a$
grows, matter components with a slower fall-off of their energy
density come to dominate. Eventually, the inflaton, whose
energy density is roughly constant, dominates the cosmic
evolution and determines the (roughly constant) Hubble parameter while causing the
scale factor to grow exponentially, \be a \propto e^{Ht}. \ee
We can define the relative energy density in the curvature as
$\Omega_{\kappa} \equiv -\kappa/(a^2H^2)$ and in the
anisotropies as $\Omega_{\sigma} \equiv \sigma^2/(3a^6H^2).$
During inflation, these relative densities fall off quickly,
and the universe is rendered exponentially flat; according to
(\ref{flatness-constraint}) the flatness puzzle is then
resolved as long as the scale factor grows by at least 60
e-folds.

\begin{figure}[t]
\begin{center}
\includegraphics{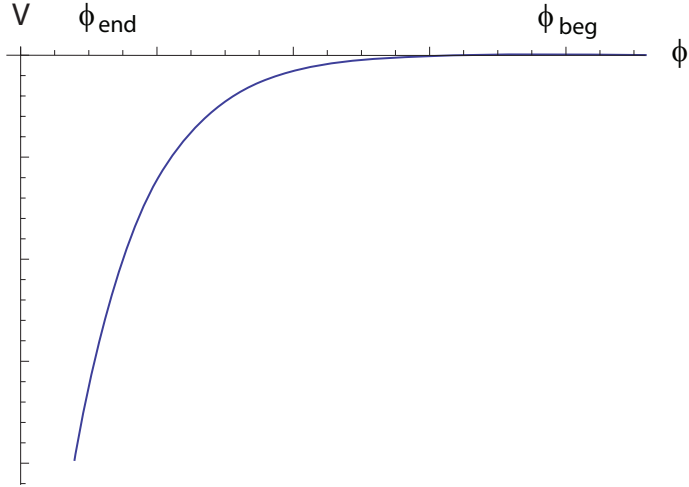}
\caption{\label{figure ekpotential} {\small The potential during
ekpyrosis is negative and steeply falling; it can be modeled by the exponential form $V(\phi)=-V_0 e^{-c\phi}.$}}
\end{center}
\end{figure}

Now we will show how the same problem can be solved by having a
contracting phase before the standard expanding phase of the
universe. The Friedmann equation relates the Hubble parameter
to the total energy density in the universe, which is the sum
of kinetic and potential energy. Now suppose that, instead of a flat potential,
the scalar $\phi$ has a very steep, negative potential, as
shown in Fig. \ref{figure ekpotential}. As a concrete example,
one can model the potential with a negative exponential \be
V(\phi)=-V_0 e^{-c \phi}, \ee where $V_0$ and $c$ are constants.
Let us give one brief motivation for a negative potential: foreshadowing a cyclic picture, we can ask if it is possible for the universe to revert from an
expanding to a contracting phase in a non-singular way. Then at some point
the Hubble parameter must go through zero, which can be achieved by having a negative potential in order to cancel the positive
kinetic energy of matter. From (\ref{scalareqofstate}) this
automatically implies an equation of state $w>1,$ and will be
seen to have very powerful consequences. In a contracting
universe, the argument presented in the previous paragraph is
reversed, and one would initially expect the anisotropy term
(proportional to $a^{-6}$) to come to dominate the cosmic
evolution. However, if there is a matter component with $w>1,$ then (\ref{fluidscaling}) implies that this component will scale with an even larger negative power of $a,$ and hence will come to dominate the cosmic evolution in a contracting universe in the same way as the inflaton comes to dominate in an expanding universe.

In fact
it is straightforward to generalize the treatment to having
many scalars $\phi_i$ with potentials $V_i(\phi_i).$ Then, in a
flat FRW background, the equations of motion become \be
\ddot{\phi}_i + 3H\dot{\phi}_i + V_{i,\phi_i} = 0 \label{Eomscalar}\ee
 and \be H^2 =\frac{1}{3} \left[\frac{1}{2}  \sum_i
 \dot\phi_i^{~2}+ \sum_i V_i(\phi_i)
 \right], \label{EomFriedmann}\ee
where $V_{i,\phi_i} = (\partial V_i/\partial \phi_i)$ with no
summation implied. A useful relation is \be \dot{H} = -
\frac{1}{2}  \sum_i \dot\phi_i^{~2}. \label{eq-Hdot}\ee If all
the fields have negative exponential potentials
$V_i(\phi_i)=-V_i\, e^{-c_i \phi_i}$ and if $c_i \gg 1$ for all
$i,$ then the Einstein-scalar equations admit the scaling
solution \be a = (-t)^{p}, \qquad \phi_i = {2\over c_i} \ln
(-\sqrt{c_i^2V_i/2} t), \qquad p=\sum_i {2 \over c_i^2}.
\label{ekpyrosis-scaling} \ee Thus, we have a very slowly
contracting universe with (constant) equation of state \be
w=\frac{2}{3p}-1 \gg 1. \label{ekpyrosis-eq-of-state}\ee We are using a coordinate system in which the big crunch occurs at $t=0;$ in other words, the time coordinate is negative during the ekpyrotic phase. The
steeply falling scalar fields act as a very stiff fluid, and,
in fact, one can take the condition $w\gg1$ (or equivalently
$p\ll \frac{1}{3}$) to be the defining feature of ekpyrosis.
The matter content does not necessarily have to be composed of
scalar fields, but it is easy to model the ekpyrotic phase that
way, and scalar fields commonly appear in effective theories
arising from higher dimensions. The main consequence is that
the extra term in the Friedmann equation (\ref{Friedmann}) with
$w \gg 1$ comes to dominate the cosmic evolution, and once
more, the fractional energy densities $\Omega_{\kappa} \propto
a^{-2}H^{-2}$ and $\Omega_{\sigma} \propto a^{-6}H^{-2}$ quickly decay. Thus,
neglecting quantum effects, the universe is left exponentially
flat and isotropic as it approaches the big crunch. As we will
see in section \ref{section perturbations}, the inclusion of
quantum effects superposes small fluctuations on this classical
background.

What are the requirements on the ekpyrotic phase for the
flatness problem to be solved? First, we must ensure that $w>1,$ which means that the potentials have to be steep enough. In the single-field case for example,
from (\ref{ekpyrosis-eq-of-state}) we see that we need
$c_1>\sqrt{6}.$ Also, from the scaling solution
(\ref{ekpyrosis-scaling}), we see that the scale factor $a$
remains almost constant, while the Hubble parameter $H\propto
t^{-1}$ during ekpyrosis. In order to solve the flatness
problem, $aH$ has to grow by at least 60 e-folds in magnitude,
see (\ref{flatnessscaling}) and (\ref{flatness-constraint}), which means that \be |t_{beg}|
\geqslant e^{60} |t_{end}|, \label{flatness-constraint-time}
\ee where the subscripts $beg$ and $end$ refer to the beginning
and the end of the ekpyrotic phase respectively. As will be
discussed in section \ref{section perturbations}, we need
$t_{end} \approx -10^3 M_{Pl}^{-1}$ in order to obtain the
observed amplitude of cosmological perturbations, so that \be
|t_{beg}| \geqslant 10^{30} M_{Pl}^{-1} = 10^{-13} s. \ee This
is the minimum time the ekpyrotic phase has to last in order to
solve the flatness problem, under the assumption that a flat
universe undergoing a crunch/bang transition can reemerge as a
flat universe afterwards. We will discuss this assumption
repeatedly in  what follows.

In the cyclic picture of the
universe, the potential $V$ interpolates between the GUT scale
and the dark energy scale. From the scaling solution
(\ref{ekpyrosis-scaling}), we have that $V \propto t^{-2},$ and
this leads to \be |t_{beg}| =
\sqrt{\frac{V_{end}}{V_{beg}}}|t_{end}| \approx
\sqrt{10^{112}}10^3 M_{Pl}^{-1} = 10^{16} s. \ee In this case,
the ekpyrotic phase lasts on the order of a billion years, and
it easily resolves the flatness puzzle.

From the higher-dimensional point of view, one of the scalar
fields $\phi_i$ is the radion, which determines the distance
between the branes (the higher-dimensional picture will be
discussed in detail in section \ref{section fundamental}). The
potential then represents a conjectured attractive force
between the end-of-the-world branes. By diluting
inhomogeneities in the brane curvature, the ekpyrotic phase
causes the branes to become very flat and parallel over large
patches. Thus the homogeneity puzzle, namely the question of
how different parts of the sky can have approximately the same
temperature despite the fact that they could not have been in
causal contact since the big bang, is also solved by the
ekpyrotic phase.

Moreover, as will be discussed in sections \ref{section crunch}
and \ref{section fundamental}, the brane scale factors remain
finite even at the collision. Semi-classical calculations
confirm the intuition that the collision should be slightly
inelastic, so that matter is produced at a finite temperature
\cite{Turok:2004gb}. And as long as this temperature is below
the GUT scale (which might depend on the collision rapidity
being low enough), topological defects will not form, and thus
the monopole problem is avoided. In this way all the standard
cosmological puzzles can be resolved by the ekpyrotic phase.


\section{Towards the Crunch}
\label{section crunch}

We have just seen how the simple assumption of a contracting
phase with equation of state $w>1,$ preceding the usual
expanding phase of the universe, manages to solve the flatness
and homogeneity puzzles of standard big bang cosmology. But now
we will see that it does much more than that: it provides a way
of avoiding Belinsky-Khalatnikov-Lifshitz (BKL) chaotic
mixmaster behavior \cite{Belinsky:1970ew}, it prepares the
universe in the most favorable state as it approaches the big
crunch and, on top of that, it generates a nearly
scale-invariant spectrum of scalar perturbations that act as
the seeds for the large-scale structure of the universe.

\subsection{Avoiding Chaos}
\label{subsection chaos}

Any cosmological theory involving a contracting phase must
address the issue of chaos: from the studies of BKL
\cite{Belinsky:1970ew}, it is known that if all matter
components have an equation of state $w<1,$ then a contracting
universe is unstable w.r.t small perturbations. The metric
becomes highly anisotropic, and of Kasner form (see below).
Typically all spatial dimensions except for one shrink away,
but repeatedly the metric jumps from one Kasner form to another
one, thus leading to BKL oscillations or ``chaotic mixmaster''
behavior \cite{Misner:1974qy}. In this case all predictability
in going towards a big crunch seems to be lost.

In order to better understand the mechanism of the ekpyrotic
phase, and to address both the issues of chaos and cosmological
perturbations, it is useful to look in more detail at the
evolution of anisotropies. In synchronous gauge, we can write
the metric as \be \d s^2=-\d t^2 + g_{ij}(x^{\mu}) \d x^i \d x^j. \ee
In this gauge, we can choose a time coordinate such that the
big crunch occurs everywhere at $t=0.$ From the work of BKL
\cite{Belinsky:1970ew} we know that in a contracting universe
spatial gradients quickly become irrelevant compared to time
gradients, so that we can consider the simplified metric \be
\d s^2=-\d t^2+a(t)^2 \sum_i e^{2\beta_i(t)}\d x^i \d x^i,\ee which is
of Kasner form and where we require $\sum_i \beta_i =0.$ In
other words, the dynamics becomes ultralocal and the Einstein
equations reduce to the Friedmann equations \bea && 3H^2 =
\frac{1}{2}\sum_i \dot{\beta}_i^2 + \cdots \\ && \ddot{\beta}_i
+ 3H \dot{\beta}_i = 0. \eea Then the growing-mode solution to
the last equation is $\dot{\beta}_i \propto a^{-3},$ which
means that the first Friedmann equation can be rewritten as \be
3H^2 = \frac{\sigma^2}{a^6} + \cdots \ee for a constant
$\sigma$ parameterizing the anisotropic energy density at unit
scale factor. Hence, as stated in the previous section,
anisotropies blueshift proportionally to $a^{-6}$ and would
quickly dominate the cosmological evolution if it wasn't for
the ekpyrotic scalar whose energy density grows faster, namely
as $a^{-2/p}.$ From the scaling solution
(\ref{ekpyrosis-scaling}) it is easy to see that the
time-dependent part of the Kasner exponents $\beta_i$ scales as
$(-t)^{1-3p}$ so that these exponents approach constant values
during the ekpyrotic phase where $p<\frac{1}{3}.$ This gives
rise to the so-called cosmic no-hair theorem for ekpyrosis
\cite{Erickson:2003zm}, according to which a contracting
universe with anisotropy and inhomogeneous curvature converges
to a homogeneous, flat and isotropic universe if it contains
energy with equation of state $w>1.$ From (\ref{Friedmann}) and
(\ref{flatness-constraint-time}) we can see that the relative
importance of the anisotropy term in the Friedmann equation
drops by a factor of at least \be \frac{(a^6 H^2)_{beg}}{(a^6
H^2)_{end}}\geqslant e^{120} \ee during the ekpyrotic phase
(note that the relative importance of the curvature term
$\propto a^{-2}H^{-2}$ drops off by a similar amount). Thus, at
the end of the ekpyrotic phase, the size of pre-ekpyrotic
anisotropies and spatial curvatures has become exponentially
small!

The ekpyrotic phase does not last right up to the big crunch.
Indeed, on general grounds one expects the potential to turn
off and become irrelevant as the big crunch is neared, see
section \ref{section fundamental}. Then the universe enters a
kinetic energy dominated phase, with equation of state $w=1.$
The equations of motion reduce to \be
3H^2=\frac{1}{2}\dot{\phi}^2=-\dot{H} \qquad
\ddot{\phi}+3H\dot{\phi}=0. \ee  They immediately lead to $a
\propto e^{\phi/\sqrt{6}}$ and are solved by \be a=a_0
(-t)^{1/3} \qquad \phi=\sqrt{\frac{2}{3}} \ln(-t)+\phi_0,
\label{kinetic solution}\ee for some integration constants $a_0, \, \f_0.$ As will be derived in section
\ref{section perturbations}, the ekpyrotic phase lasts until
about $10^3$ Planck times before the big crunch; the subsequent
kinetic phase is relevant until the quantum gravity regime is
reached around $t \sim -1 M_{Pl}^{-1},$ and also again in the
first moments after the big bang. Now of course one has to
answer the question of whether this kinetic phase undoes what
the ekpyrotic phase has achieved. From the solution above, one
can see immediately that $\dot{\phi}^2 \propto a^{-6},$ which
means that the scalar field energy density scales in the same
way as the anisotropies. Thus the relative importance of the
anisotropic term remains constant during the kinetic phase. In
a similar way \cite{Erickson:2003zm}, one can show that the
curvature terms are harmless during this $w=1$ phase. In
the presence of p-forms, depending on their coupling to the
scalar $\phi,$ dangerous modes can grow, but only as a power of
time $t$ and for not longer than about 7 e-folds. In the
relevant case of heterotic M-theory for example
\cite{Lukas:1998tt}, one such coupling is critical
\cite{Erickson:2003zm} implying that the approach to the big
crunch could be chaotic if there was sufficient energy density
in the p-form modes.

Now it is clear how the ekpyrotic phase can avoid the
uncertainties of a chaotic crunch: even when dangerous modes
are present their energy density gets diluted to such an extent
during ekpyrosis that they don't have time to come to dominate
before less than a Planck time before the big crunch\footnote{What's more, some dangerous modes are further
suppressed by the topology of the internal manifold, see
\cite{Wesley:2005bd}.}\cite{Erickson:2003zm}. Thus chaos is avoided by being delayed
until we have entered the quantum gravity regime, where the
equations of general relativity break down. At that point, one might
heuristically expect the spread of the wavefunction of the universe to prevent
chaotic behavior from resurfacing. This last speculation will
be discussed further in section \ref{section fundamental}.

\subsection{The Milne Universe}
\label{subsection Milne}

After the ekpyrotic phase, the universe enters a kinetic energy
dominated phase as it approaches the big crunch. Thus, from a
4-dimensional point of view, the universe seems to be headed
towards the worst kind of singularity. However, in approaching
the crunch at $t=0,$ the universe really starts looking
5-dimensional, and the higher-dimensional description reveals
that the singularity is actually much milder. We will give a
simplified description here; a more complete treatment is
postponed to section \ref{section fundamental}.

\begin{figure}[t]
\begin{center}
\includegraphics[width=0.3\textwidth]{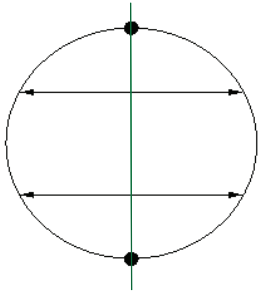}
\caption{\label{figure orbifold} {\small A line segment represented as the orbifold $S^1/\mathbb{Z}_2.$ The thick dots represent the orbifold fixed points.}}
\end{center}
\end{figure}

Consider gravity in 5 dimensions, but assume that one spatial
dimension is a line segment, as suggested by Ho\v{r}ava-Witten
theory (see section \ref{section fundamental}). This line
segment can be described as a circle modded out by a
$\mathbb{Z}_2$ reflection symmetry across an axis, see Fig.
\ref{figure orbifold}.
Then we can write the metric as \cite{Khoury:2001bz} \be \d s_5^2 =
e^{-\sqrt{2/3}\phi} \d s_4^2 + e^{2\sqrt{2/3}\phi} \d y^2,
\label{5dmetric}\ee where $-y_0<y<y_0$ denotes the orbifold
coordinate and $\phi$ is the radion field parameterizing the
size of the orbifold. The $\phi$-dependent prefactor in front
of the 4-dimensional metric $\d s_4^2$ ensures that after
dimensional reduction we are left with the canonically
normalized Lagrangian \be {\cal L} = \sqrt{-g} [\frac{1}{2}
R - \frac{1}{2}(\partial \phi)^2]. \ee We know that the
kinetic phase described in equation (\ref{kinetic solution}) is
a solution to this theory. This means that we can now plug the solution
(\ref{kinetic solution}) into the original 5-dimensional metric, to find (up to trivial
rescalings) \bea \d s_5^2 &=& (-t)^{-2/3} [-\d t^2 + (-t)^{2/3}
\d x_3^2] + (-t)^{4/3} \d y^2 \\ &=& -\d T^2 + T^2 \d y^2 +\d x_3^2, \eea
after changing coordinates to $T \propto (-t)^{2/3}.$ This
spacetime goes by the name of ``compactified Milne mod
$\mathbb{Z}_2$'' $\times \mathbb{R}_3$ \cite{Tolley:2003nx}. It describes the two
orbifold planes (\ie the $(3+1)$-dimensional boundaries of the
line segment) approaching each other, colliding, and receding
away from each other again, as shown in Fig. \ref{figure
Milne}.
\begin{figure}[t]
\begin{center}
\includegraphics{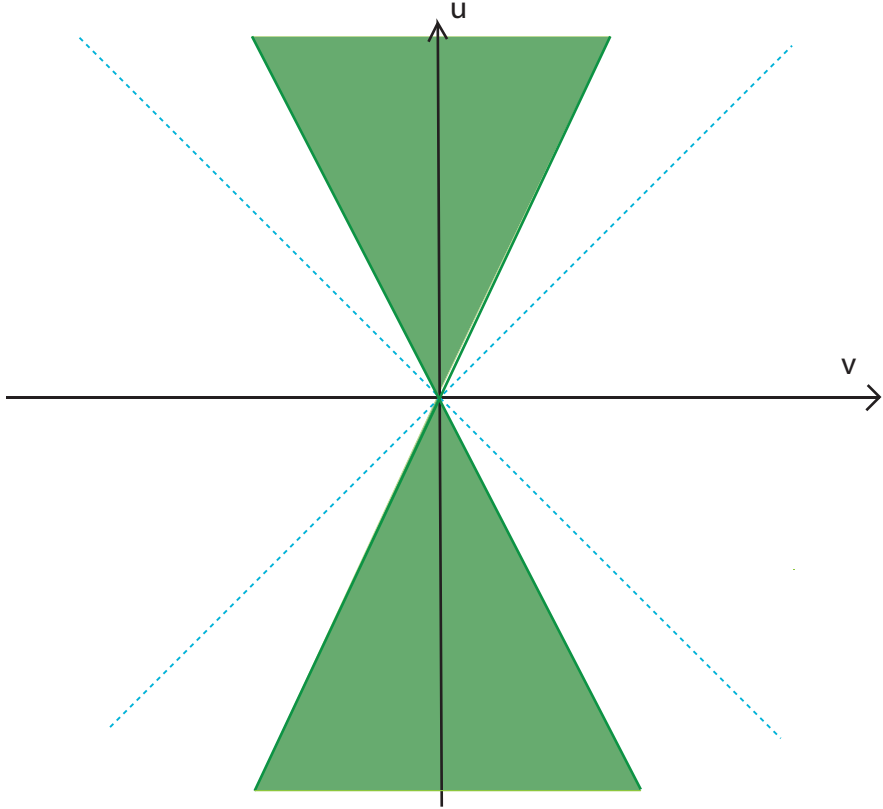}
\caption{\label{figure Milne} {\small The compactified Milne mod $\mathbb{Z}_2$ space describes the collision of two boundary branes, embedded in Minkowski space.}}
\end{center}
\end{figure}
Under the further change of coordinates $u=T \cosh y, \, v=T
\sinh y,$ the metric becomes \be \d s_5^2 = -\d u^2 + \d v^2 +\d x_3^2, \label{metric Minkowski}
\ee which shows that the embedding spacetime is simply
Minkowski space. Thus, except at the collision itself, this
spacetime is flat and therefore automatically a solution of any
theory of gravity without a cosmological constant. This also
means that higher-derivative corrections are small for a
spacetime that is close to Milne. In this sense the
compactified Milne space is the simplest model spacetime for a
brane collision. Note that, from the 5-dimensional point of
view, the big crunch singularity is just the momentary
shrinking away of the orbifold dimension, while the other
dimensions remain at finite size. This can best be seen by
examining equation (\ref{5dmetric}) together with the solution
(\ref{kinetic solution}): even though the 4-dimensional scale
factor $a \rightarrow 0$ at the collision, the brane scale
factors are really given by $e^{-\phi/\sqrt{6}} a,$ which is a
constant. Thus the density of matter (which is stuck on the
branes) and the temperature remain finite at the collision.

The general idea here is that the ekpyrotic phase drives the
universe to being exponentially flat and isotropic in a
4-dimensional sense, which means that the subsequent kinetic
phase and big crunch/big bang transition can be well
approximated in 5 dimensions by the compactified Milne
spacetime. Of course one can only trust the spacetime
description up to about a Planck time on either side of the
collision. For a proper treatment, we would need a complete
theory of quantum gravity. But in a sense, the ekpyrotic phase
prepares the universe in the best possible state to enter the
collision phase.

There have been a number of studies in string/M-theory of the
Milne spacetime near the collision: Turok {\it et al.}
\cite{Turok:2004gb} have shown how the equations for winding
membranes remain well behaved near the collision. And in
\cite{Niz:2006ef} it was argued that from the string theory
perspective strings evolve as sets of weakly coupled ``string
bits'' smoothly across the singularity. However it is not clear
to what extent these treatments encompass all the relevant
degrees of freedom near the collision. Furthermore, it was
argued in \cite{Turok:2004gb} that near the collision the
theory should be described in terms of an expansion in the
string tension $1/\alpha'$ rather than in $\alpha';$ such an
expansion is not well understood at present. It is also known
that perturbative string theory breaks down in these and
similar settings as soon as interactions are taken into account
\cite{Liu:2002kb,Berkooz:2002je,Craps:2003ai}. But again it is
not clear that perturbative string theory is really the right
approach. The most promising development so far is the
treatment of a toy model crunch/bang transition in the
framework of the AdS/CFT correspondence by Craps {\it et al.}
\cite{Turok:2007ry,Craps:2007ch}. However, in this toy model, the bulk spacetime is 5-dimensional and of Kasner type close to the crunch. But most importantly, the radius of curvature of the AdS spacetime is smaller than the string length, in sharp contrast to ekpyrotic models, in which we rather expect the curvature to be small in approaching the crunch. Hence the results of this work cannot directly be applied to ekpyrotic models. Nevertheless, one might hope that the techniques used in \cite{Craps:2007ch} can be extended in the future to include the treatment of more realistic spacetimes.

Of course this is a topic of crucial importance for ekpyrotic
theory. In order to discuss cosmological perturbation theory,
it is vital to know how to match the universe going into the
big crunch to the one coming out of the big bang. In light of
the ambiguities discussed above, two broadly different
attitudes can be adopted (see
\cite{Battefeld:2004mn,Brandenberger:2001bs,Creminelli:2004jg,Durrer:2002jn,Hwang:2001zt,Lyth:2001pf,Martin:2004pm,Martin:2001ue,Tolley:2002cv,Tolley:2003nx,Tsujikawa:2001ad}
for many discussions of these ``matching conditions'' in the
literature):

In analogy with the treatment of reheating in inflation,
Creminelli {\it et al.} \cite{Creminelli:2004jg} have advocated
the model-independent assumption that the spacetime metric can
simply go through the collision essentially unchanged. This
assumption is based on the following argument: the ekpyrotic
phase renders the universe isotropic up to exponentially small
terms. Also, as we will see in the next section, quantum
effects lead to small anisotropies in the metric. Thus, in
synchronous gauge ($g_{0\mu} = \eta_{0\mu}$), the metric reads \be \d s^2=-\d t^2+a(t)^2
e^{2\zeta(x^{\mu)}} e^{2 h_{ij}(x^{\mu})} \d x^i \d x^j, \ee where
we are allowing for scalar curvature perturbations $\zeta$ and
tensor perturbations $h_{ij}.$ Then locally the small,
long-wavelength perturbations we're interested in can be gauged
away by a local re-scaling of the spatial coordinates. In other
words, locally these fluctuations are just integration
constants. Thus the universe undergoes the same history at
every point in space up to exponentially small terms. If we
then assume that the dynamics is not sensitive to these
exponentially small terms (on super-horizon scales) then the
metric will reemerge from the crunch/bang transition with the
same integration constants, \ie with the same perturbations.
Note that this argument, if correct, applies not only to linear
perturbations but to the full non-linear metric.

The second possibility is that the dynamics of the crunch/bang
transition is important: for example it was shown in
\cite{Tolley:2002cv,Tolley:2003nx} that for free fields in a
compactified Milne universe, one can simply analytically
continue the evolution of the fields around the singularity at
$t=0.$ Such a prescription corresponds to matching
perturbations on surfaces of constant energy density (see also
\cite{Khoury:2001zk,Turok:2004yx}) and it leads to a mixing of different
perturbation modes, such as the dominant and subdominant components of the Newtonian potential and the
curvature perturbation. As the energy density perturbation is
exponentially small, this represents a concrete example of a
situation in which the history of the universe does depend
significantly on the exponentially small perturbation terms. It
should be noted however that this treatment breaks down once
interactions are taken into account, as noted above. It has
also been suggested by McFadden {\it et al.}
\cite{McFadden:2005mq} that a mixing of modes can occur when
the effective 4-dimensional description breaks down near the
collision. It should be clear that in this case as well, the
evolution of the metric and its perturbations depend
sensitively on the detailed dynamics of the bounce.

We will adopt the ``nothing happens'' assumption of
Creminelli {\it et al.} in this review (\ie we will assume that
the curvature perturbation is non-linearly conserved through
the collision), but with the proviso that this assumption is
subject to revision once the crunch/bang transition is
understood in more detail.

\subsection{New Ekpyrotic Models}
\label{subsection newekpyrotic}

The idea behind the ``new ekpyrotic'' models
\cite{Buchbinder:2007ad,Buchbinder:2007tw,Creminelli:2007aq} is
to build cosmological models in which the ekpyrotic contracting
phase is joined with the standard expanding phase by a smooth,
non-singular bounce that can be described entirely within a
4-dimensional effective field theory. Thus, the energy never
reaches the Planck scale, and cosmological perturbations can be
evolved unambiguously through the bounce phase. Of course, in
light of the singularity theorems of Hawking and Penrose, it is
clear that in order to achieve a smooth bounce, one of the
assumptions going into these theorems has to give; in the
models considered so far this has been the null energy
condition (NEC), which is violated by means of a ghost
condensate \cite{ArkaniHamed:2003uy}.\footnote{We should note
though that it is not clear yet if such a ghost condensate can
be realized in string theory \cite{Adams:2006sv} or in a UV complete theory in general \cite{Kallosh:2007ad}.} In the presence of a fluid with energy density $\rho$ and pressure ${\cal P},$ the Friedmann equations can be combined to yield \be
\dot{H}=-\frac{1}{2}(\rho + {\cal P}). \label{Hdot}\ee The null energy condition requires that $\rho + {\cal P}\geqslant 0,$ or, in covariant form, $T_{\mu \nu} n^{\mu} n^{\nu} \geqslant 0$ for any non-spacelike vector $n^{\mu}.$ Thus, in order to have $\dot{H}>0,$ which is a necessary condition to revert from a contracting phase to an expanding one, the NEC must be violated.

We will briefly review the main ideas of ghost condensation, in order to see how the NEC can be violated: we
start by considering an effective Lagrangian of the form \be
{\cal L} = \sqrt{-g}M^4 P(X), \ee where \be
X=-\frac{1}{2m^4}(\partial \phi)^2 \ee and $P(X)$ is a function
that remains to be specified. $M$ and $m$ are two mass scales
that must be determined by the underlying microscopic theory,
and, as we will see, that have to satisfy certain consistency
conditions. Note that the theory admits the constant shift
symmetry $\phi \mapsto \phi + constant.$ In a cosmological background,
the resulting equation of motion reads \be \frac{\d}{\d t}(a^3
P_{,X} \dot{\phi})=0, \ee which is automatically solved at an
extremum $P_{,X}=0.$ We will assume that the extremum lies at
$X_0=\frac{1}{2}$ say, which in turn can be solved by giving
the scalar $\phi$ the time-dependent expectation value \be
\phi=-m^2 t.\ee With this expression for $\phi,$ we are
justified {\it a posteriori} to have omitted higher-derivative
terms like $\Box \phi$ in the Lagrangian, since these would
vanish in any case.

The ghost condensate by itself does not violate the null energy
condition yet; in fact it has the same equation of state as a
cosmological constant. To see this, note that the
energy-momentum tensor is given by \be T_{\mu \nu} = M^4 P
g_{\mu \nu} + \frac{M^4}{m^4} P_{,X} \pt_{\mu} \phi \pt_{\nu} \phi. \ee Thus
the energy density is given by $\rho=M^4(2P_{,X}X-P),$ while
the pressure is ${\cal P}=M^4 P.$ Hence, at the extremum, we
have ${\cal P}=-\rho$ or $w=-1.$ However, the fluctuations
around the extremum can violate the NEC. We define the
fluctuations $\pi$ via \be \phi = -m^2 t + \pi. \ee Their
dynamics can be expressed by the effective Lagrangian \be {\cal
L} \propto 2X_0 P_{,XX}(X_0)\dot{\pi}^2, \ee which shows
that the kinetic term has the correct sign if the extremum is a
minimum, as we will assume (in other words, there are no ghosts around the ghost
condensate). If we then expand the expression for the energy density, $\rho=M^4(2P_{,X}X-P),$ to linear order in $\pi,$ we obtain $\rho \approx -K M^4 \dot{\pi}/m^2.$ We can provoke such a fluctuation by adding a potential $V,$ implying
that the energy density and the pressure are now approximately given by \be
\rho \approx -\frac{K M^2}{m^2}\dot{\pi}+V \qquad {\cal
P} \approx -V, \ee where $K\equiv P_{,XX}(X_0)>0$ and $P(X_0)=0.$ From (\ref{Hdot}) we can
see that this immediately implies \be \dot{H} \approx \frac{KM^4}{2m^2}\dot{\pi},
\ee and hence, since $\dot{\pi}$ can take either sign, we now have the possibility of violating the NEC.

\begin{figure}[t]
\begin{center}
\includegraphics[width=\textwidth]{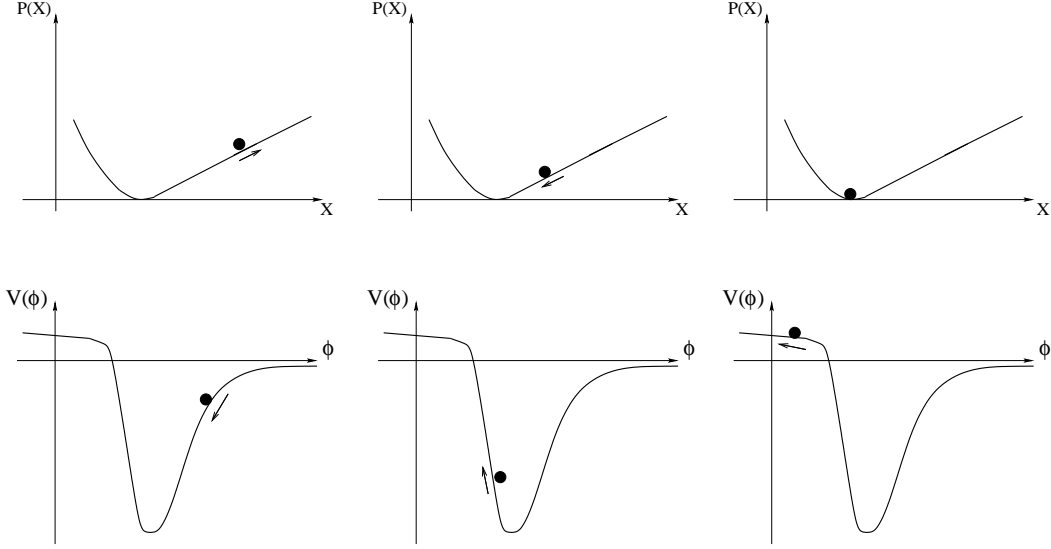}
\caption{\label{figure newekpyrotic} {\small This figure shows the evolution of the kinetic function $P(X)$ and of the potential $V(\phi)$ during the phases of ekpyrosis and bounce. Note that the field starts at approximately zero potential energy and moving very slowly, so that $\rho \sim -\dot{\pi} +V \approx 0.$ Then, during the ekpyrotic phase, the potential is negative, implying that $\dot{\pi}<0$ and hence $\dot{H}<0.$ Subsequently, during the ghost condensate phase, the potential shoots back up to positive values, implying that $\dot{\pi}>0$ and thus $\dot{H}>0,$ \ie the universe starts reverting from contraction to expansion. Figure reproduced with permission from \cite{Buchbinder:2007ad}.}}
\end{center}
\end{figure}

Nevertheless, it is still difficult to achieve a bounce because
NEC-violating modes scale with a positive power of the scale
factor, see equation (\ref{fluidscaling}). Hence they tend to
become subdominant in a contracting universe, and dominant in
an expanding one - exactly the opposite of what we want here!
This problem may be circumvented if it is the same field
driving both the ekpyrotic phase and the bounce. For this to
work though, we need the Lagrangian function $P(X)$ to be
canonical (linear) during ekpyrosis \cite{Buchbinder:2007ad}
(although see \cite{Creminelli:2007aq} for a different approach),
and then quadratic around the minimum; also, the potential $V$ has to become positive at the onset of the bounce phase in order to push the equation of state parameter $w$ to NEC-violating values below $-1.$ The complete evolution is sketched in Fig. \ref{figure
newekpyrotic}. After the bounce is complete, it is assumed that the
NEC-violating fluid decays by converting all its energy into
matter and radiation, and that it thereby reheats the universe
\cite{Buchbinder:2007ad}.

In these models, there are two sources of instability, which must be addressed. One is of the Jeans type and there is also a gradient instability, see \cite{Buchbinder:2007ad,Creminelli:2006xe,Creminelli:2007aq} for details. Both of these instabilities are harmless if the bounce phase is rather fast, \ie if it lasts only on the order of one Hubble time (as measured at the start of the bounce phase). In that case, there isn't enough time for the instabilities to grow. A further constraint that these models have to satisfy in order to obtain a successful bounce is that the value of the potential at the end of the ekpyrotic phase has to fall within a certain range, so that the ekpyrotic phase lasts long enough and so that the trajectory ends up near the ghost-condensate point at the end of ekpyrosis. As discussed in \cite{Buchbinder:2007tw}, this bound is \be m^4 e^{2\cN_{ek}} \ll |V_{end}| \ll \frac{M^4K}{p}, \ee where $\cN_{ek}=\frac{1}{2}\ln|V_{end}/V_{beg}|$ is the number of e-folds of the ekpyrotic phase; in other words, we must require that $M$ is exponentially larger than $m.$ Of course it would be satisfying to see how this large hierarchy in the effective theory could arise from an underlying, more fundamental, theory.

But this is not our direct concern here. The new ekpyrotic models rather aim to represent a proof-in-principle that it is possible to have a cosmological history that can be described entirely within the framework of a 4-dimensional effective theory, so that the evolution of all fields, and in particular of cosmological perturbations, can be followed unambiguously through the bounce. We will return to the issue of cosmological perturbations during the bounce phase in the next section.


\section{Cosmological Perturbations}
\label{section perturbations}

The ekpyrotic phase not only manages to address the standard
cosmological puzzles, but it also provides a new way of generating
cosmological perturbations. Both scalar and tensor perturbations
are generated, but, as we will see, they turn out to have very
different properties. The colliding branes picture provides a
heuristic way of understanding the origin of the perturbations \cite{Khoury:2001wf}:
although the ekpyrotic phase renders the branes very flat and
parallel classically, quantum fluctuations cause the branes to
ripple slightly. Thus, the branes do not collide everywhere at
exactly the same time, and therefore, in some places the big bang
happens slightly earlier than in others. In this way, some places
have a little bit more time to expand and cool than others, and
the result is a pattern of temperature fluctuations in the cosmic
microwave background (CMB). We will now make this cartoon quantitatively precise \cite{Khoury:2001wf,Khoury:2001zk,Gratton:2003pe,Boyle:2004gv}.

\subsection{Scalar perturbations}
\label{subsection scalar}

The detailed shape of the angular anisotropy power spectrum of the CMB radiation can be described by fluctuation modes that enter the horizon in succession, \ie in a timed, coordinated manner. This hints to having a period of cosmological evolution where the modes were in causal contact initially, and then went outside of the horizon, only to re-enter later in succession. There are, broadly speaking, two ways of producing this phenomenon: in inflation, the horizon is roughly constant and the modes get stretched rapidly beyond it, whereas in ekpyrosis, due to the increasing ultralocality in approaching the crunch, the horizon shrinks rapidly while the fluctuation modes themselves remain roughly constant in size. As we will see, for scalar, linear perturbations this leads to a kind of duality between the two theories. Tensor modes, on the other hand, are sensitive only to the evolution of the metric, and hence behave very differently in the two theories; this will be discussed in section \ref{subsection tensor}.

\subsubsection{Single Field}

We will start with a simple, heuristic argument for why one might expect to obtain scale-invariant perturbations in the presence of a scalar field with a steep negative potential \cite{Lehners:2007ac}. In fact, since the universe is contracting very slowly during the ekpyrotic phase, to a first approximation we can ignore gravity altogether. The validity of this approximation will be discussed shortly. So, for now consider a scalar field $\phi$ in Minkowski spacetime, with action
\be {\cal S} = \int \d^4 x \left(-{1\over 2} (\partial \phi)^2 +
V_0 e^{-c \phi}\right). \label{scl} \ee  We consider the case where $V_0$ is positive so the
potential energy is formally unbounded below and the scalar field
runs to $-\infty$ in a finite time. Note that the actual value of
$V_0$ is not a physical parameter, since by shifting the field
$\phi$ one can alter the value of $V_0$ arbitrarily. Furthermore,
in the ekpyrotic or cyclic models, the potential is expected to
turn up towards zero at large negative $\phi$ (because this
corresponds to the limit in heterotic M-theory in which the string
coupling approaches zero and the potential
disappears~\cite{Steinhardt:2001st}), but this detail is
irrelevant to the generation of perturbations on long wavelengths.

We now argue that, as the background scalar field rolls down the
exponential potential towards $-\infty$, then, to leading order in
$\hbar$, its quantum fluctuations acquire a scale-invariant
spectrum as the result of three features. First, the action
(\ref{scl}) is classically scale-invariant. Second, by re-scaling
$\phi \rightarrow \phi/c$ and re-defining $V_0$, the constant $c$
can be brought out in front of the action and absorbed into
Planck's constant $\hbar$, {\it i.e.}~$\hbar \rightarrow
\hbar/c^2$, in the expression $i {\cal S}/\hbar$ governing the
quantum theory. Finally, it shall be important that $\phi$ has
dimensions of mass in four spacetime dimensions.

To see the classical scale-invariance, note that shifting the
field $\phi \rightarrow \phi +\psi$, and re-scaling
coordinates $x^\mu \rightarrow x^\mu e^{c \psi/2}$, just
re-scales the action by $e^{c \psi}$ and hence is a symmetry
of the space of solutions of the classical field equations. Now we
consider a spatially homogeneous background solution corresponding
to zero energy density in the scalar field. In a cyclic model (see section \ref{section cyclic}), this ``zero energy''
condition is a reasonable initial state to assume for analyzing
perturbations because the phase
in which the perturbations are generated is preceded by a very low
energy density phase with an extended period of accelerated
expansion, like that of today's universe, which drives the
universe into a very low energy, homogeneous state. No new energy
scale enters and the solution for the scalar field is then
determined (up to a constant) by the scaling symmetry:
$\phi_b=(2/c) \ln (-A t)$. Next we consider quantum fluctuations
$\delta \phi$ in this background. The classical equations are
time-translation invariant, so a spatially homogeneous time-delay
is an allowed perturbation, $\phi = (2/c) \ln (-A(t+ \delta t))
\Rightarrow \delta \phi \propto t^{-1}$. On long wavelengths, for
modes whose evolution is effectively frozen by causality, {\it
i.e.} $|kt| \ll 1$, we can expect the perturbations to follow this
behavior. Hence, the quantum variance in the scalar field is \be
\langle \delta \phi^2 \rangle \propto \hbar t^{-2}. \label{qv} \ee
Restoring $c$ via $\phi \rightarrow c \phi$ and $\hbar \rightarrow
c^2 \hbar$ leaves the result unchanged. However, since $\delta
\phi$ has the same dimensions as $t^{-1}$ in four spacetime
dimensions, it follows that the constant of proportionality in
(\ref{qv}) is dimensionless, and therefore that $\delta \phi$ must
have a scale-invariant spectrum of spatial fluctuations.

It is straightforward to check this in detail. Setting
$\phi=\phi_b(t)+\delta \phi(t,{\bf x})$, to linear order in
$\delta \phi$ the field equation reads \be \ddot{\delta \phi} = -
V_{,\phi\phi} \,\delta \phi +{\bf \nabla}^2 \delta \phi \label{onef}
\ee Using the zero energy condition for the classical background,
we obtain $V_{,\phi \phi} = c^2 V = - c^2 \dot{\phi_b}^2/2 = -
2/t^2$. Next we set $\delta \phi(t,{\bf x}) = \sum_{\bf k} (a_{\bf
k}  \chi_{k}(t) e^{i {\bf k\cdot x}} + h.c.)$ with $a_{\bf k}$ the
annihilation operator and $\chi_k(t)$ the normalized positive
frequency modes. The mode functions $\chi_k$ obey \be
\ddot{\delta \chi}_{k} ={2\over t^2} \,\chi_{k} -k^2\chi_{k}, \label{chieq} \ee
and the incoming Minkowski vacuum state corresponds to \cite{Birrell:1982ix} \be \chi_k =
\frac{1}{\sqrt{2k}} e^{- i k t} \left(1-\frac{i}{k t}\right).\ee For large $|kt|$,
this solution tends to the usual Minkowski positive-frequency
mode. But as $|kt|$ tends to zero, each mode enters the growing
time-delay solution described above, with $\chi_k \propto t^{-1}$.
Computing the variance of the quantum fluctuation and subtracting
the usual Minkowski spacetime divergence, we obtain \be \langle
\delta \phi^2 \rangle = \hbar \int {k^2 \d k \over 4 \pi^2} {1 \over
k^3 t^2}, \label{a5} \ee where the integral is taken over $k$
modes which have ``frozen in'' to follow the time-delay mode. In
agreement with the general argument above, we have obtained a
scale-invariant spectrum of growing scalar field perturbations. To
recap, classical scale-invariance determines the $t$ dependence of
the perturbations, and dimensional analysis then gives a
scale-invariant spectrum in $k$, in three space dimensions.

However, it is important to realize that what is measured in the CMB is directly related to the curvature perturbation of the spacetime geometry. Hence it is imperative that we include gravity in the analysis, even though {\it a priori} its effects appear to be small corrections. As we will see, the inclusion of gravity entails some (perhaps unexpected) subtleties \cite{Creminelli:2004jg}. Thus, we now consider the Lagrangian \be {\cal L} =\sqrt{-g}[ \frac{1}{2}R-\frac{1}{2} (\partial \phi)^2 + V_0 e^{-c\phi}]. \ee The resulting equations of motion (the one-field versions of (\ref{Eomscalar}) and (\ref{EomFriedmann})) are solved by the one-field scaling solution (\ref{ekpyrosis-scaling}). It is convenient to define the parameter $\epsilon$ via \be \epsilon \equiv \frac{3}{2}(1+w). \ee In inflation this is the slow-roll parameter; here, since the potential is steep, $w \gg 1$ and $\epsilon$ rather represents ``fast-roll''. In the scaling solution, it is simply given by $\epsilon=1/p.$ It is also useful to continue the following analysis in terms of conformal time $\tau,$ defined via $\d t=a \d \tau.$ We will denote derivatives w.r.t conformal time by a prime, \ie $\frac{\d}{\d \tau}\equiv '.$ In terms of conformal time, the scaling solution (\ref{ekpyrosis-scaling}) reads (up to unimportant re-scalings) \be a = (-\tau)^{p/(1-p)}, \qquad \phi_i = {2\over c_i (1-p)} \ln
(-\tau), \qquad p=\sum_i {2 \over c_i^2}.
\label{ekpyrosis-scaling-tau}\ee

The single-field scalar perturbation theory was first discussed in \cite{Khoury:2001wf,Khoury:2001zk}.
In the next few paragraphs, we will closely follow the presentation of \cite{Boyle:2004gv}.
Including scalar perturbations, the metric can be written as
    \begin{eqnarray}
      \label{perturbed_metric}
      \textrm{d}s^{2}/a^{2}&=&-(1+2AY)\textrm{d}\tau^{2}
      -2BY_{i}\textrm{d}\tau\textrm{d}x^{i}\nonumber \\
      &&+\left[(1+2H_{L}Y)\delta_{ij}+2H_{T}Y_{ij}\right]
      \textrm{d}x^{i}\textrm{d}x^{j}
    \end{eqnarray}
    and the perturbed scalar field as
    \begin{equation}
      \label{perturbed_scalar_field}
      \phi=\phi_{0}(\tau)+\delta\phi(\tau)Y
    \end{equation}
  where $Y(\vec{x})$, $Y_{i}(\vec{x})$, and $Y_{ij}(\vec{x})$ are
  scalar harmonics. Here we work in Fourier space, so that we are not writing out the
 implicit subscripts $\vec{k}$ of perturbation modes. The
  perturbed Einstein equations $\delta G^{\mu}_{\;\;\nu}=\delta
  T^{\mu}_{\;\;\nu}$ then relate the metric and the matter perturbations to each other. As is
well known, scalar perturbations in a spatially-flat FRW
  universe with scalar field $\phi$ and potential $V(\phi)$ are
  completely characterized by a single gauge-invariant variable. Various choices for this variable are typically considered, including the ``Newtonian potential'' $\Phi$,
  and the ``curvature perturbation'' $\zeta$.
  The gauge-invariant Newtonian potential $\Phi$ is most easily
  understood in ``Newtonian gauge'' ($B\! =\! H_{T}\! =\! 0$), where
  it is related to the metric perturbations in a simple way:
  $\Phi=A=-H_{L}$.  It obeys the equation of motion
  \begin{equation}
    \label{Phi_EOM}
    \Phi''+2\left[\frac{a'}{a}-\frac{\phi_{0}''}{\phi_{0}'}\right]
    \Phi'+2\left[k^{2}+2\mathcal{H}'-2\mathcal{H}
    \frac{\phi_{0}''}{\phi_{0}'}\right]\Phi=0
  \end{equation}
  where $k=|\vec{k}|$ is the magnitude of the (comoving) Fourier
  3-vector.  On the other hand, the gauge-invariant perturbation
  variable $\zeta$ is most easily understood in ``comoving gauge''
  ($H_{T}=\delta T^{0}_{\;\; i}=0$), where it represents the curvature
  perturbation on spatial hypersurfaces, and is related to the spatial
  metric perturbation in a simple way: $\zeta=-H_{L}$.  The condition
  $\delta T^{0}_{\;\; i}=0$ also implies that $\delta\phi=0$ in this
  gauge.  $\zeta$ obeys the equation of motion
  \begin{equation}
    \label{zeta_EOM}
    \zeta''+2\frac{z'}{z}\zeta'+k^{2}\zeta=0,
  \end{equation}
  where we have defined $z\equiv a^{2}\phi_{0}'/a'$. $\Phi$ and $\zeta$ are related
  to each other by
    \label{zeta_phi_relations}
    \begin{eqnarray}
      \label{zeta_from_phi}
      \zeta & = & \Phi+\frac{1}{\epsilon}
      \left[\frac{a}{a'}\Phi'+\Phi\right]\\
      \label{Phi_from_zeta}
      \Phi & = & -\epsilon\frac{a'}{a k^2}\zeta'\;.
    \end{eqnarray}

  It is convenient to introduce new variables, $u$ and
  $v$ \cite{Mukhanov:1985rz} (these are of course unrelated to the coordinates used in section \ref{subsection Milne}), by multiplying $\Phi$ and $\zeta$ by
  $k$-independent functions of $\tau$:
  \begin{equation}
    \label{def_uv}
    u\equiv \frac{a}{\phi_{0}'}\Phi \qquad v\equiv z\zeta.
  \end{equation}
  Note that $u$ and $v$ have the same $k$-dependence as $\Phi$ and
  $\zeta$, respectively, and therefore the same spectral properties. $u$ and $v$ obey the simple equations of motion
    \begin{eqnarray}
      \label{u_EOM}
      u''+(k^{2}-\frac{(1/z)''}{(1/z)})u & = & 0\\
      \label{v_EOM}
      v''+(k^{2}-\frac{z''}{z})v & = & 0,
    \end{eqnarray}
  and are related to each other by
    \begin{eqnarray}
      \label{v_from_u}
      kv & = & 2k(u'+\frac{z'}{z}u)\\
      \label{u_from_v}
      -ku & = & \frac{1}{2k}
      (v'+\frac{(1/z)'}{1/z}v).
    \end{eqnarray}
     When $\epsilon$ is time-independent, we can use
  (\ref{ekpyrosis-scaling}) to find
    \begin{eqnarray}
      \label{theta_eps}
      \frac{(1/z)''}{1/z} & = &
      \frac{\epsilon}{(\epsilon-1)^{2}\tau^{2}}\\
      \label{z_eps}
      \frac{z''}{z} & = &
      \frac{2-\epsilon}{(\epsilon-1)^{2}\tau^{2}}
    \end{eqnarray}
Thus, at early times (when $|\tau|$ is large) the $k^2$ term dominates the $(1/z)''/(1/z)$ and $z''/z$ terms in the brackets in the equations of motion (\ref{u_EOM}) and (\ref{v_EOM}). Physically, this means that the relevant modes are well within the horizon, where the solutions are asymptotically oscillatory. On scales much smaller than the horizon, the curvature of spacetime is negligible, and thus we impose boundary conditions corresponding to the Minkowski vacuum seen by a coming observer \cite{Birrell:1982ix}:
    \begin{eqnarray}
      \label{u_bc}
      u & \rightarrow \frac{i}{(2k)^{3/2}}\textrm{e}^{-ik\tau}\\
      \label{v_bc}
      v & \rightarrow \frac{1}{\sqrt{2k}}\textrm{e}^{-ik\tau}
    \end{eqnarray}
  as $\tau\rightarrow-\infty$.

  The equations of motion are Bessel equations, and they can be solved exactly by
    \begin{eqnarray}
      u(x)=x^{1/2}\left[A^{(1)}H^{(1)}_{\alpha}(x)
    +A^{(2)}H^{(2)}_{\alpha}(x)\right]\\
      v(x)=x^{1/2}\left[B^{(1)}H^{(1)}_{\beta}(x)
    +B^{(2)}H^{(2)}_{\beta}(x)\right]
    \end{eqnarray}
  where $x\equiv k|\tau|$ is a dimensionless time variable,
  $A^{(1,2)}$ and $B^{(1,2)}$ are constants, $H^{(1,2)}_{s}(x)$ are
  Hankel functions, and we have defined
    \begin{eqnarray}
      \label{def_alpha}
      \alpha \equiv & \sqrt{\frac{(1/z)''}{1/z}\tau^{2}+1/4} &
      =\frac{1}{2}\left|\frac{\epsilon+1}{\epsilon-1}\right|\\
      \label{def_beta}
      \beta \equiv & \sqrt{\frac{z''}{z}\tau^{2}+1/4} &
      =\frac{1}{2}\left|\frac{\epsilon-3}{\epsilon-1}\right|
    \end{eqnarray}
  In the far past ($x\rightarrow\infty$) we use the asymptotic Hankel
  expression
  \begin{equation}
    \label{hankel_asymptotic}
      H^{(1,2)}_{s}(x)\rightarrow\sqrt{\frac{2}{\pi x}}
    \textrm{exp}\left[\pm\,
    i\left(x-\frac{s\,\pi}{2}-\frac{\pi}{4}\right)\right]
  \end{equation}
  so that the boundary conditions pick out the solutions
    \begin{eqnarray}
      \label{u_soln}
      u & = & \frac{\mathcal{P}_1}{2k}\sqrt{\frac{\pi x}{4k}}
      H_{\alpha}^{(1)}(x)\\
      \label{v_soln}
      v & = & \,\mathcal{P}_{2}\,\sqrt{\frac{\pi x}{4k}}
      H_{\beta}^{(1)}(x)
    \end{eqnarray}
  with the phase factors
    \begin{eqnarray}
      \label{P_1}
      \mathcal{P}_1=\textrm{exp}[i(2\alpha+3)\pi/4]\\
      \label{P_2}
      \mathcal{P}_{2}=\textrm{exp}[i(2\beta+1)\pi/4].
    \end{eqnarray}
In order to determine the power spectra of $\zeta$ and $\Phi,$ at late times when comoving scales are well outside Hubble radius, we need the asymptotic form of the Hankel functions as $x\rightarrow 0:$
\be
\label{H1asymptotics}
H^{(1)}_s (x) \rightarrow -\frac{i}{\pi}\,\Gamma(s)(\frac{x}{2})^{-s}\hspace{-4mm},
\ee
where $s>0$ and $\Gamma(s)$ is the Euler gamma function.
On large scales, the power spectrum of the Newtonian potential $\Phi$ is given by
\be
\label{phipowerspec}
P_\Phi (k) = \frac{k^3}{2\pi^2}\,\frac{|u|^2 \phi_0'^2}{a^2}\propto x^{1-2\alpha}.
\ee The spectral index for $\Phi$ is thus \be n_\Phi-1=1-2\alpha = 1-|\frac{\epsilon + 1}{\epsilon -1}|. \ee
Note that it is invariant under $\epsilon \rightarrow 1/
\epsilon,$ which is a manifestation (at the linear level) of a duality between inflation ($\epsilon \approx 0$) and ekpyrosis ($\epsilon \gg 1$) \cite{Boyle:2004gv}. In fact, this could have been anticipated on the basis of the invariance of both the equation of motion and the boundary condition for $u$ under the transformation $\epsilon \rightarrow 1/\epsilon.$ Note that in both limits of interest, namely $\epsilon \rightarrow 0$ (inflation) and $\epsilon \rightarrow \infty$ (ekpyrosis), the spectrum of the Newtonian potential turns out to be scale-invariant. If we had performed the same calculation, but with an equation of state that is slowly varying in time, then we would have obtained the spectral index \cite{Gratton:2003pe} \be
n_{\Phi}-1=-4(\bar{\epsilon}+\bar{\eta}),
\ee where \be
\bar{\epsilon}=(\frac{V}{V_{,\phi}})^2, \qquad
\bar{\eta}= 1-\frac{V V_{,\phi\phi}}{V_{,\phi}^2}.
\ee

On large scales, the power spectrum for $\zeta$ is given by
\be
\label{fullzetaspec}
P_\zeta (k) = \frac{k^3}{2\pi^2}\,\frac{|v|^2}{z^2} \propto x^{3-2\beta},
\ee
which means that the spectral index for $\zeta$ is \be n_{\zeta}-1 = 3-2\beta = 3-|\frac{\epsilon -3}{\epsilon -1}|. \ee In ekpyrotic models, $\epsilon$ is large, and thus the spectrum of the curvature perturbation $\zeta$ is very blue, meaning that there is much more power on smaller scales, in disagreement with observations.

This seems very puzzling at first, since one would have thought that it didn't matter whether one used $\Phi$ or $\zeta$ in order to perform the calculation. However, it is instructive to rewrite the relationship (\ref{zeta_from_phi}) between the two variables in the following form:
\be
\label{neatzeta}
\zeta= \frac{1}{\epsilon a^2}\,(\frac{a^3\Phi}{a'})'.
\ee
Expanding the Hankel function $H^{(1)}(x)$ again as $x\rightarrow 0$, but this time up to second order, one finds
\be
H^{(1)}_s (x)= -\frac{i}{\pi}\,\Gamma(s)(\frac{x}{2})^{-s}-\frac{i}{\pi}\,e^{-i\pi s}\,\Gamma(-s)(\frac{x}{2})^s + O(x^{2-s},\,x^{2+s}),
\ee
which means that $\Phi$ behaves roughly speaking as
\be
\label{Phiscale}
\Phi \sim k^{-3/2-1/\epsilon}(-\tau)^{-1-2/\epsilon}+ k^{-1/2+1/\epsilon},
\ee
where we have expanded all exponents up to linear order in $1/\epsilon.$ But using the same expansion, we have that \be \frac{a'}{a^3} \sim (-\tau)^{-1-2/\epsilon},\ee so that from (\ref{neatzeta}) we can see that the leading, scale-invariant contribution to $\Phi$ drops out of $\zeta !$

Here we see very clearly the importance of adding gravity to our analysis. In light of the discussion in section (\ref{subsection Milne}) this means that the single-field ekpyrotic model does not lead to a scale-invariant spectrum of cosmological perturbations unless there is a mixing of $\Phi$ and $\zeta$ at the bounce (in that case, the scale-invariant component of $\Phi$ would dominate on large scales over the blue intrinsic spectrum of $\zeta$). Such a mixing would depend on the precise dynamics at the bounce, and would therefore be a model-dependent prediction.  Examples of mixing due to higher-dimensional effects have been given in \cite{McFadden:2005mq,Tolley:2003nx}, and mixing can occur for example due to the breakdown of the 4-dimensional effective field theory. But what this means physically is that new degrees of freedom become relevant near the brane collision, and therefore it is rather natural to try to employ a more complete effective theory, containing more fields, in order to perform the above calculation. This will be the topic of the following section.

\subsubsection{Two Fields} \label{subsection twofields}

In a sense it is rather unnatural to consider only a single scalar field in the effective theory, since there are at least two ``universal'' fields that are always present in a higher-dimensional context: the radion field, determining the distance between the two end-of-the-world branes, and the volume modulus of the internal manifold. But as soon as there are more than one scalar field present, one can have entropy, or isocurvature, perturbations, which are growing mode perturbations in a collapsing universe \cite{Notari:2002yc}. Entropy perturbations can source the curvature perturbation, and hence (provided the entropy perturbations acquire a nearly scale-invariant spectrum), almost scale-free curvature perturbations can be generated just before the bounce. These then turn into growing mode perturbations in the ensuing expanding phase.

As will be shown in section \ref{section fundamental}, the effective theory arising from heterotic M-theory, which describes gravity and the two universal scalars mentioned above, is a 4-dimensional theory of gravity minimally coupled to two scalar fields (after a field redefinition). It is thus surprisingly simple. On top of that, we assume an attractive force between the boundary branes, modeled by two negative nearly exponential potentials for the scalar fields:
 \be V=
-V_1 e^{-\int c_1 \d \phi_1}-V_2 e^{-\int c_2 \d\phi_2},
\label{c1} \ee where $c_1=c_1(\phi_1)$, $c_2 = c_2(\phi_2)$, and
$V_1$ and $V_2$ are positive constants. We consider potentials in
which the $c_i$ are slowly varying and hence the potentials are
locally exponential in form. Furthermore, for simplicity, we focus
on scaling background solutions in which both fields {\it
simultaneously} diverge to $-\infty$. The background solution we are interested in is the two-field version of the  scaling solution (\ref{ekpyrosis-scaling}). We will find it useful to define the variable $\s$ via \cite{Gordon:2000hv}
\be \dot{\s} \equiv \sqrt{\dot{\phi}_1^2+\dot{\phi}_2^2}.
\label{c3} \ee $\s$ has the interpretation of being the path length along the background scalar field trajectory.
The angle $\theta$
of the background trajectory is defined by \cite{Gordon:2000hv} $
\cos(\theta)=\dot{\f}_1/\dot{\s} \, , \, \sin(\theta) =
\dot{\f}_2/\dot{\s}.$
Then the fast-roll parameter $\epsilon$ and the equation of state parameter $w$
are related to the background evolution via \be \epsilon \equiv
\frac{3}{2} (1+w) = \frac{\dot{\sigma}^2}{2H^2}. \label{epsil}\ee
During the phase in which the entropic perturbations are
generated, $\dot{\theta}=0$, which corresponds to a straight
background trajectory in scalar field space. In this case, we define \be
\dot{\phi}_2 \equiv \gamma \dot{\phi}_1, \label{phirel} \ee and,
with this notation, $c_1 = \gamma c_2$ and
 \be \epsilon_{ek}=
\frac{|\gamma  c_1 c_2| }{2(1+\gamma^2)}. \ee Thus we have an ekpyrotic phase as long as $c_1 c_2$ is large. $\gamma$ is typically of ${\cal O} (1);$ for the heterotic M-theory
colliding branes solution with empty branes, $\gamma =-
\frac{1}{\sqrt{3}}$ during the ekpyrotic phase, as will be derived in section \ref{section fundamental}. It turns out though that the results are not very sensitive to $\gamma.$

The angle of the background trajectory is related to the potential via \be \dot{\theta} = \frac{\dot{\phi_2} V_{,\phi_1}- \dot{\phi_1}
V_{,\phi_2}} {\dot{\phi_1}^2 +\dot{\phi_2}^2}.\ee
Hence, since we want a straight line trajectory, we have \be V = \tilde{V}(\phi_1)
+\gamma^2 \tilde{V}(\phi_2/\gamma), \label{gammarel} \ee for some function
$\tilde{V}(\phi)$.

\begin{figure}[t]
\begin{center}
\includegraphics[width=0.75\textwidth]{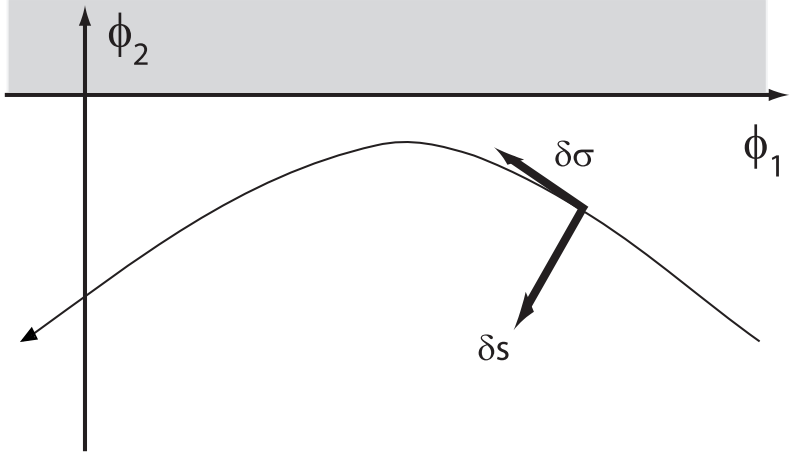}
\caption{\label{FigureSubdominant} {\small The trajectory in field
space reflects off a boundary at $\phi_2=0.$ The entropy
perturbation, denoted $\delta s$, is orthogonal to the trajectory.
The bending causes the conversion of entropy modes into adiabatic modes $\delta \sigma$,
which are perturbations tangential to the trajectory. }}
\end{center}
\end{figure}

In a contracting universe, a growing mode is given by
the entropy perturbation $\delta s$, namely the relative fluctuation in the
two fields, defined (at linear order) as follows
\cite{Gordon:2000hv} \be \delta s \equiv (\dot{\phi}_1\,\delta
\phi_2 - \dot{\phi}_2\,\delta \phi_1)/\dot{\s}. \ee The entropy perturbation is gauge-invariant and it
represents the perturbation orthogonal to the background scalar
field trajectory, as shown in Fig.~\ref{FigureSubdominant}; see
Ref.~\cite{Langlois:2006vv} for its definition to all orders.

The entropy perturbation equation of motion reads (see \eg \cite{Gordon:2000hv}) \be
\ddot{\delta s} + 3H\dot{\delta s} + \left(\frac{k^2}{a^2}
  + V_{ss} + 3\dot{\theta}^2 \right) \delta s =
\frac{4 k^2 \dot\theta}{a^2 \sqrt{\dot{\phi}_1^2 +
\dot{\phi}_2^2}}\, \Phi. \label{eq-entropylinear}\ee
Successive derivatives of the potential with respect to the
entropy field are given by \bea V_{s} &=&
\frac{1}{\dot{\s}}(\dot{\phi_1} V_{,\phi_2}- \dot{\phi_2}
V_{,\phi_1})\\ V_{ss} &=& \frac{1}{\dot{\s}^2} (\dot{\phi_1}^2
V_{,\phi_2 \phi_2}-2 \dot{\phi_1}\dot{\phi_2}
V_{,\phi_1 \phi_2} + \dot{\phi_2}^2 V_{,\phi_1 \phi_1}) \\
V_{sss} &=& \frac{1}{\dot{\s}^3}(\dot{\phi_1}^3 V_{,\phi_2 \phi_2
\phi_2}-3\dot{\phi_1}^2\dot{\phi_2} V_{,\phi_1 \phi_2 \phi_2} \nn \\
& & + 3\dot{\phi_1}\dot{\phi_2}^2 V_{,\phi_1 \phi_1 \phi_2} -
\dot{\phi_2}^3 V_{,\phi_1 \phi_1 \phi_1}). \eea Incidentally, note that this implies $\dot{\theta} = -V_{s}/\dot{\s}.$

Again, for simplicity
we will focus attention on straight line trajectories in scalar
field space. Since $\dot{\theta}=0$, the entropy perturbation is
not sourced by the Newtonian potential $\Phi$ and we can solve the
equations rather simply. On large scales, the linear entropy equation then reduces to
\be
\ddot{\delta s} + 3H\dot{\delta s} + V_{ss} \delta s =
0. \label{eq-entropylinear2}\ee
It is convenient at this point to continue the analysis in terms
of conformal time $\tau$. Introducing the re-scaled entropy field \be\delta S = a(\tau)
\,\delta s, \ee Eq. (\ref{eq-entropylinear2}) becomes \be {\delta S}'' +
\left(-\frac{a''}{a}
  + a^2 V_{,\phi_1 \phi_1}
 \right) \delta S = 0. \label{eq-entropy-S}
  \ee
The crucial term governing the spectrum of the
perturbations is then
\be \tau^2\,\Big(\,{a''\over a} - V_{,\phi_1 \phi_1} \,a^2\Big).
\label{potpert} \ee
When this quantity is approximately 2, we will
get nearly scale-invariant perturbations.
 We proceed \cite{Lehners:2007ac} by evaluating the quantity in (\ref{potpert}) in an
expansion in inverse powers of $\epsilon$ and its derivatives with
respect to $N$,  where $N = \ln (a/a_{end})$, where $a_{end}$ is
the value of $a$ at the end of the ekpyrotic phase. Note that $N$
decreases as the fields roll downhill and the contracting
ekpyrotic phase proceeds.

We obtain the first term in (\ref{potpert}) by differentiating
(\ref{eq-Hdot}), obtaining \be {a''\over a} = 2 H^2 a^2
\Big(1-{1\over 2} \,\epsilon\Big). \label{app} \ee The second term
in (\ref{potpert}) is found by differentiating (\ref{epsil}) twice
with respect to time and using the background equations and the
definition of $N$. We obtain \be a^2 V_{,\phi_1 \phi_1}= - a^2 H^2
\,\Big( 2 \epsilon^2  - 6 \epsilon - {5\over 2}
\,\epsilon_{,N}\Big) +O(\epsilon^0). \label{vpp} \ee Finally, need
to express ${\cal H} \equiv (a'/a) = a H$ in terms of the
conformal time $\tau$. From (\ref{app}) we obtain \be {\cal H}'=
{\cal H}^2(1-\epsilon), \label{hcp} \ee which integrates to \be
{\cal H}^{-1}  = \int_0^\tau \d \tau (\epsilon -1). \label{inthcp}
\ee Now, inserting $1= \d(\tau)/\d\tau$ under the integral and
using integration by parts we can re-write this as \be {\cal
H}^{-1} = \epsilon \tau \left(1 -{1\over \epsilon} - (\epsilon
\tau)^{-1} \int_0^\tau \epsilon' \tau \d \tau\right).
\label{integ} \ee Using the same procedure once more, the integral
in this expression can be written as \be (\epsilon \tau)^{-1}
\int_0^\tau \epsilon' \tau \d \tau = \frac{\epsilon'
\tau}{\epsilon} - (\epsilon \tau)^{-1} \int_0^\tau
\frac{\d}{\d\tau}(\epsilon' \tau)\,\tau \d \tau. \ee Now using the
fact that $\epsilon' = {\cal H} \epsilon_{,N}$, and that to
leading order in $1/\epsilon$, ${\cal H}$ can be replaced by its
value in the scaling solution (with constant $\epsilon$), ${\cal
H}\tau = \epsilon^{-1}$, we can re-write the second term on the
right-hand side as \be - (\epsilon \tau)^{-1}\int_0^\tau
\frac{\d}{\d\tau}(\epsilon' \tau)\,\tau \d \tau = - (\epsilon
\tau)^{-1}\int_0^\tau \d \tau \frac{1}{\epsilon}\,
\Big(\frac{\epsilon_{,N}}{\epsilon}\Big)_{\hspace{-0.5mm},N}, \ee
which shows that this term is of order ${1}/{\epsilon^2}$ and can
thus be neglected. Altogether we obtain \be {\cal H}^{-1}
=\int_0^\tau \d \tau (\epsilon-1) \approx \epsilon \tau \left(1-
{1\over \epsilon}- {\epsilon_{,N}\over \epsilon^2}\right).
\label{appepsint} \ee Using (\ref{app}) and (\ref{vpp}) with
(\ref{appepsint}) we can calculate the crucial term entering the
entropy perturbation equation, \be
\tau^2\,\Big(\hspace{0.5mm}{a''\over a} - V_{,\phi_1 \phi_1}
\,a^2\Big) = 2 \left(1 - \frac{3}{2 \epsilon} + \frac{3}{4}
\frac{\epsilon_{,N}}{\epsilon^2}\right). \ee Following the same steps as in the single field section above, it is straightforward to see that the deviation from
scale-invariance in the spectral index of the entropy perturbation
is now given by \be \label{tilt1} n_s -1 = \frac{2}{\epsilon } -
\frac{\epsilon_{,N}}{\epsilon^2}. \ee The first term on the
right-hand side is a gravitational contribution, which, being
positive, tends to make the spectrum blue. The second term is a
non-gravitational contribution, which tends to make the spectrum
red. We will return to this expression shortly. Before explaining how these entropic
perturbations are naturally converted to curvature perturbations, we must add an important comment: the growth of the entropy perturbations is intimately linked to an instability of the background trajectory \cite{Lehners:2007ac,Tolley:2007nq}. Indeed, the background trajectory follows a ridge in the potential \cite{Koyama:2007mg}, and in order to achieve a sufficiently long ekpyrotic phase, the trajectory has to remain near the crest of the ridge for long enough. This places important constraints on the two-field models, which we will discuss in section \ref{section cyclic}.

We have shown how an approximately scale-invariant spectrum of
entropy perturbations may be generated by scalar fields in a
contracting universe. We will now discuss how these
perturbations may be converted to curvature perturbations if the
scalar field undergoes a sudden acceleration, and we will estimate
the curvature perturbation amplitude.
At linear order, a
non-zero entropy perturbation combined with a bending
($\dot{\theta}\neq 0$) of the background trajectory sources the
curvature perturbation on large scales and results in a linear,
gaussian curvature perturbation (see \eg \cite{Gordon:2000hv}) \bea \zeta_L &=&
\int_{\Delta t} -\frac{2H}{\dot{\sigma}}\dot{\theta}\delta
s^{(1)} \\ &=& \int_{\Delta t}
\sqrt{\frac{2}{\epsilon_c}}\dot{\theta}\delta s^{(1)},
\label{curvaturelinear} \eea where we denote the duration of
the conversion by $\Delta t$ (\ie $\Delta t$ is the time during which $\dot{\theta} \neq 0$) and the sign corresponds to a contracting universe. Thus the strength of conversion is
proportional to $1/\sqrt{\epsilon_c}$ and this dependence on
the equation of state during the time of conversion will have repercussions for the magnitude
of the second order correction as well, as will be discussed in the next section. There are many ways in which such a bending of the background trajectory can occur. We will now discuss the various possibilities considered in the
literature, namely conversion after
the ekpyrotic phase during kinetic energy domination \cite{Lehners:2007ac}, conversion during the ekpyrotic phase \cite{Koyama:2007mg,Koyama:2007ag} or during the transition to a ghost condensate phase \cite{Buchbinder:2007ad,Creminelli:2007aq}, and
conversion after the big bang by modulated preheating \cite{Battefeld:2007st}.

\vspace{2cm}

{\noindent \it Conversion during kinetic energy domination}

In the original ekpyrotic and cyclic models,
 the phase dominated by the steep,
 ekpyrotic potential $V(\phi)$ comes to an end
(at $t=t_{end}<0$) before the big crunch/big bang transition ($t=0$),
and the universe becomes dominated by
the kinetic energy of the scalar fields, as we saw in section \ref{section crunch}. Consequently,
the equation of state at $t=t_{end}$
changes from $\epsilon_{ek} \gg 1$
to $\epsilon = 3$ (corresponding to $w \rightarrow 1$, the equation of state
for a kinetic energy dominated universe). The conversion from entropic to curvature
perturbations then takes place during this kinetic energy dominated phase \cite{Lehners:2007ac}.
The conversion occurs naturally
in the heterotic M-theory embedding of the cyclic
model because  the negative-tension brane bounces off a spacetime
singularity \cite{Lehners:2007nb}  --  creating
a bend in the trajectory in field space in the 4-dimensional effective theory --
before it collides with
the positive-tension brane (the big crunch/big bang transition). This will be explained in detail in section \ref{section fundamental}. Here, all we
need to know is that in the 4-dimensional effective description there are two
scalar field moduli, $\phi_1$ and $\phi_2$, living on the
half-plane $-\infty < \phi_1 < \infty$, $-\infty < \phi_2 < 0$. In other words, there is a boundary to moduli space at $\phi_2=0.$
Furthermore, the cosmological solution of interest is one in which
$\phi_2$ encounters this boundary $\phi_2=0$, and reflects off it
at time $t=t_{ref};$ see Figure \ref{FigureSubdominant}. This bending of the trajectory automatically induces the conversion of
entropy to curvature perturbations. Here
we will not restrict the analysis to this particular example,
but we will consider the general situation in which the scalar
field trajectory bends in a smooth way during the phase of
kinetic energy domination following an ekpyrotic phase.

In general, in the presence of $N$ scalar
fields with general K\"{a}hler metric $g_{ij}(\phi)$ on scalar
field space, the equation for the evolution of the curvature perturbation on large scales can be rewritten as \cite{Lehners:2007ac}
 \be \dot{\zeta} =  {H \over \dot{H}}
g_{ij} {D^2 \phi^i \over D t^2} s^j , \label{e1} \ee where the $N-1$ entropy perturbations
\be s^i= \delta \phi^i - \dot{\phi}^i \,\frac{g_{jk}(\phi)\,
\dot{\phi}^j \delta \phi^k }{ g_{lm}(\phi)\, \dot{\phi}^l
\dot{\phi^m}} \label{e2} \ee are just the components of $\delta
\phi^i$ orthogonal to the background trajectory, and the operator
$D^2/Dt^2$ is the geodesic operator on scalar field space. Here, things simplify because the scalar field space is flat,
so the metric is $g_{ij}=\delta_{ij}$, and $D/Dt$ reduces to an
ordinary time derivative. Considering only two scalar fields, we
have \be s^1 = - \dot\phi_2 \,\delta
s/\sqrt{\dot\phi_1^2+\dot\phi_2^2}, \qquad s^2 = + \dot\phi_1
\,\delta s/\sqrt{\dot\phi_1^2+\dot\phi_2^2}. \ee As a first approximation, we will assume that the scalar field trajectory simply reflects off the boundary at $\phi_2=0.$ The
scalar field trajectory is $\dot{\phi_2}=-\tilde{\gamma}
\dot{\phi_1}$, for $t<t_b$, and $\dot{\phi_2}=\tilde{\gamma}
\dot{\phi_1}$, for $t>t_b$, with $\dot{\phi_1}$ constant and
negative in the vicinity of the bounce. The bounce leads to a
delta function on the right-hand side of (\ref{e1}), \be
\frac{D^2\phi_2}{Dt^2} = \delta(t-t_{ref}) \,2 \dot\phi_2(t_{ref}^+), \label{e3} \ee
where from the higher-dimensional point of view $t_{ref}$ is the time of the bounce of the negative-tension
brane. As can be readily seen from (\ref{e1}), if the entropy
perturbations already have acquired a scale-invariant spectrum by
the time $t_{ref}$, then the bounce leads to their instantaneous
conversion into curvature perturbations with precisely the same
long wavelength spectrum.
We can estimate the amplitude of the resulting curvature
perturbation by integrating equation (\ref{e2}) using (\ref{e3}).
Since we have assumed the universe is kinetic-dominated at this
time, $H=1/(3t)$. As pointed out earlier, since the entropy
perturbation (\ref{c3}) is canonically normalized, its spectrum is
given by (\ref{a5}) up to non-scale-invariant corrections. This
expression only holds as long as the ekpyrotic behavior is still
underway: the ekpyrotic phase ends at a time $t_{end}$
approximately given by $|V_{end}|=2/(c^2 t_{end}^2)$. After
$t_{end}$, during the kinetic phase, the entropy perturbation obeys $\ddot{\delta s} +
t^{-1} \dot{\delta s} = 0$, which has the solution $\delta s = A +
B \ln(-t).$ Matching this solution to the growing mode solution
$t^{-1}$ in the ekpyrotic phase, one finds that by $t_{ref}$ the
entropy grows by an additional factor of $1+ \ln(t_{end}/t_{ref})$.
Employing the Friedmann equation to relate $\dot{\phi_2}=
\tilde{\gamma} \dot{\phi_1}$ to $H$, putting everything together
and restoring both the Planck constant and the Planck mass, we find for the variance of the
spatial curvature perturbation in the scale-invariant case, \be
\langle {\zeta}^2 \rangle = \hbar \,{c_1^2 |V_{end}|\over 3 \pi^2
M_{Pl}^2} \,{ \tilde{\gamma}^2 \over (1+\tilde{\gamma}^2)^2}
\,\left(1+\ln(t_{end}/t_{ref})\right)^2 \int {\d k \over k} \equiv
\int {\d k\over k}\, \Delta^2_{\zeta}(k) \label{r5} \ee for the
perfectly scale-invariant case. Notice that the result depends
only logarithmically on $t_{ref}$: the main dependence is on the
minimum value of the effective potential and the parameter $c_1$.
Observations on the current Hubble horizon indicate $
\Delta^2_{\zeta}(k_0) \approx 2.4\times 10^{-9} $ with $k_0=0.002 \, Mpc^{-1}$ \cite{Komatsu:2008hk}. Ignoring the
logarithm in (\ref{r5}), this requires \be c_1 \sqrt{|V_{end}|}
\approx 10^{-3} M_{Pl}, \label{amplitudeconstraint}\ee or approximately the GUT scale. This is
of course entirely consistent with the heterotic M-theory setting
\cite{Banks:1996ss}. Note also that this means that the ekpyrotic phase ends at a time of about $10^3$ Planck times before the big crunch.

In fact, our approximation of a perfect reflection should really be improved upon, especially since eventually we want to consider second-order corrections. Specific examples \cite{Lehners:2007nb} show that in the presence of matter on the branes, one expects a smooth repulsive potential and a gradual bending of the trajectory. We will discuss this in more detail shortly in section \ref{subsection non-gaussianity}. It does turn out however that for gradual reflections, and taking into account the evolution of the entropy perturbation during the process of conversion, the efficiency of conversion is only reduced by a factor of about $5$ or so \cite{Lehners:2008my}, and hence the above rough estimate for the amplitude of the perturbations is rather good. Here we simply note that because of the evolution of the entropy perturbation during the time of conversion, sharper reflections in fact lead to less efficient conversions. Turning this the other way around, one can also say that sharper conversions require steeper ekpyrotic potentials (larger values of $c_1$) to yield an amplitude of cosmological fluctuations compatible with observations.

{\noindent \it Conversion during the ekpyrotic phase}

In the ``new ekpyrotic'' models
\cite{Buchbinder:2007ad,Creminelli:2007aq,Koyama:2007mg}, typically the
conversion of entropy to curvature perturbations is assumed to take place
because the background trajectory switches from the two-field
unstable scaling solution to a late-time single-field attractor solution; in
other words, the trajectory starts out close to the ridge of the
two-field potential and then falls off one of the steep sides,
either because of the initial conditions or due to an additional
feature in the potential (for example, one field could see a sharp rise in its potential if it enters the ghost condensate phase before the second one). Adding a feature to the potential makes
little difference to the results as long as the reflection remains
gradual \cite{Lehners:2008my}, so in fact we will not consider an additional potential
here. Compared to conversion during the kinetic phase, the entropy perturbation falls off by rather less (see section \ref{subsection non-gaussianity}), but the value of $\epsilon_c=\epsilon_{ek}$ is larger. Altogether this implies from (\ref{curvaturelinear}) that the resulting amplitude of curvature perturbations obeys roughly the same constraints as those discussed above, see equation (\ref{amplitudeconstraint}).

{\noindent \it Conversion via modulated preheating}

Finally, it has been proposed
\cite{Battefeld:2007st} that, instead of
converting entropy perturbations into curvature
perturbations before the big crunch/big bang transition, the
 conversion could occur during the phase shortly
following the bang through modulated reheating. The concept is
that massive matter fields are produced copiously at the brane
collision and dominate the energy density immediately after the
bang.  The massive fields are assumed to couple to ordinary matter
with a strength proportional to $h(\delta s)$, so that their decay
into ordinary matter occurs at slightly different times depending
on the value of $\delta s.$  In this way, the ordinary matter
perturbations inherit the entropic perturbation spectrum. Since
the conversion happens while $\epsilon_c \approx 3,$ the resulting amplitude of the perturbations should be of the same magnitude as for conversion during the kinetic phase.
Note however that a detailed prediction is made difficult by
the fact that $h$ is an unknown function of $\delta s.$ Indeed,
the entropy perturbations are converted with an efficiency
\cite{Battefeld:2007st} \be e = \frac{3}{2} \frac{|h_{,s}|}{h}.
\ee In the absence of a concrete model predicting the shape of the function $h$, it is therefore impossible to go beyond the present order-of-magnitude estimate.

{\noindent \it Comparing Predictions for the Spectral Index}

Now that we have seen how curvature perturbations with the right amplitude can be generated from a multiple-field ekpyrotic phase, we should add some comments about the spectral index of the perturbations, since this is one of the quantities that can be measured with great precision. As seen above, the curvature perturbations inherit their spectral index from the entropy perturbations, and therefore it is given by (\ref{tilt1}). In this section, we
analyze this relation using several techniques and compare the prediction
to those for curvature perturbations in inflation and
for a cyclic model in which
 the single-field time-delay fluctuations
are converted to curvature perturbations before/at the bang \cite{Lehners:2007ac}.

As a first approach, let us consider the model-independent
estimating procedure used in Ref.~\cite{Khoury:2003vb}.  We begin
by re-expressing Eq.~(\ref{tilt1}) in terms of ${\cal N}$, the
number of e-folds before the end of the ekpyrotic phase (where $\d
{\cal N} = (\epsilon-1) \d N$ and
 $\epsilon \gg 1$):
\be n_s -1 = \frac{2}{\epsilon} -  \frac{\d \ln \epsilon}{\d {\cal
N}}. \ee This expression is identical to the case of the Newtonian
potential perturbations derived in \cite{Khoury:2003vb}, except
that the first term has the opposite sign. In this expression,
$\epsilon({\cal N})$ measures the equation of state during the
ekpyrotic phase, which decreases from a value much greater
than unity to a value of order unity in the last ${\cal N}$
e-folds. If we estimate  $\epsilon \approx {\cal N}^{\alpha}$,
then  the spectral tilt is \be n_s-1 \approx \frac{2}{{\cal
N^{\alpha}}} - \frac{\alpha}{{\cal N}}. \label{tilt2} \ee Here we see that the
sign of the tilt is sensitive to $\alpha$. For nearly exponential
potentials ($\alpha \approx 1$),  the spectral tilt is $n_s
\approx 1+ 1/{{\cal N}} \approx 1.02$, slightly blue, because the
first term dominates.  However, there are well-motivated examples
(see below) in which the equation of state does not decrease
linearly with ${\cal N}$. We have introduced $\alpha$ to
parameterize these cases.  If $\alpha > 1.14$,
 the spectral tilt is red.  For example, $n_s = 0.97$ for
$\alpha \approx 2$.   These examples
represent  the range that can be
achieved for the entropically-induced curvature perturbations in
the simplest models, roughly
\be 0.97 < n_s < 1.02. \ee

For comparison, if we use the same estimating procedure for the
Newtonian potential fluctuations in the cyclic model (assuming
they converted to curvature fluctuations before the bounce), we obtain $0.95 < n_s < 0.97$. This range agrees with
the estimate obtained  by an independent analysis based on
studying  inflaton potentials directly \cite{Boyle:2005ug}.
Furthermore, as shown in Ref.~\cite{Khoury:2003vb}, the same range
is obtained for time-delay (Newtonian potential) perturbations in
the cyclic model,
 due to the duality in the linear perturbation equations discussed above~\cite{Boyle:2004gv}.
Hence, all estimates are consistent with one another, and
we can conclude that the range of spectral tilt obtained from
entropically-induced curvature perturbations is typically bluer by
a few percent.

We note that it is possible that
curvature perturbations are created both by the entropic mechanism
and by converting Newtonian potential perturbations into curvature
perturbations through higher-dimensional effects.  In this case, the
cosmologically relevant contribution is the one with the bigger
amplitude.  In particular, the conversion of Newtonian potential
perturbations is sensitive to the brane collision velocity
\cite{Tolley:2003nx,McFadden:2005mq}, whereas the entropic mechanism is
not. So, conceivably, either contribution could dominate.

A second way of analyzing the spectral tilt is to assume a form
for the scalar field potential.
Consider the case
where the two fields have
 steep potentials that can be modeled as
 $V(\phi_1) = -V_0\, e^{-\int c \,\d\phi}$
and $\dot{\phi_2} = \gamma \dot{\phi_1}$. Then
Eq.~(\ref{tilt1}) becomes
\be n_s -1 =
{4(1+\gamma^2) \over c^2 M_{Pl}^2} - { 4 c_{,\phi} \over c^2},
\label{f5} \ee
where we have used the fact that $c(\phi)$ has the dimensions of inverse
mass and restored the factors of Planck mass.  The presence of $M_{Pl}$
 clearly indicates that the first term on
the right is a gravitational term.  It is also the piece that
makes a blue contribution
to the spectral tilt.  The second term is a non-gravitational
term. For a pure exponential potential, which has $c_{,\phi}=0$, the
non-gravitational contribution is zero, and the spectrum is slightly
blue, as our model-independent analysis suggested.  For plausible
values of $c = 20$ and $\gamma = 1/2$, say, the gravitational piece is
about one percent and the spectral tilt is $n_s \approx 1.01$,
also consistent with our earlier estimate.
However, this case with $c_{,\phi}$ precisely equal to
zero is unrealistic.  In the cyclic
model, for example, the steepness of the potential
must decrease as the field rolls downhill in order that the ekpyrotic
phase comes to an end,  which corresponds to $c_{,\phi} >0$.
If $c(\phi)$ changes from some initial value $\bar{c} \gg 1$
to
some value of order unity at the end of the ekpyrotic phase after $\phi$
changes by an amount $\Delta \phi $, then
$c_{,\phi} \sim \bar{c}/\Delta{\phi}$.  When $c$ is large,
the non-gravitational term in
Eq.~(\ref{f5}) typically
dominates and the spectral tilt is a few per cent towards
the red.
For example, suppose $c \propto
\phi^\beta$ and $\int c(\phi)\, \d \phi \approx 125$; then, the
spectral tilt is \be
 n_s-1 = -0.03 {\beta \over 1+\beta}, \label{spectralindex}
\ee which corresponds to $0.97 < n_s < 1$ for positive $0<\beta
<\infty$, in agreement with our earlier estimate.

Finally, we can estimate the running of the spectral index, given by $\d n_s/\d \ln k$ \cite{Kosowsky:1995aa}. Again, we can use the model-independent estimating technique used above. Then, starting with equation (\ref{tilt2}) and using $\cN= \ln(a_{end}H_{end}/aH)$ as well as $k=aH,$ we obtain \be \frac{\d n_s}{\d \ln k} = -\frac{\d n_s}{\d \cN} = \frac{2 \a}{\cN^{\a+1}} - \frac{\a}{\cN^2}. \ee Since we have $\a > 1,$ we can see that the running of the spectral index is of order ${\cal O}(\cN^{-2}),$ which is typically $\approx 10^{-3}$ or less. The current WMAP bounds on the running are $|\d n_s/\d \ln k| \lesssim 0.03$ \cite{Komatsu:2008hk} and are thus easily satisfied.

\subsubsection{Non-Gaussianity}
\label{subsection non-gaussianity}

Cosmological measurements, as performed for example by the WMAP satellite, are just now becoming sensitive enough to probe the second order corrections to the background evolution of the universe. If we stop at linear order in perturbation theory, then the equations of motion can be derived from a quadratic action $S.$ Hence, the fluctuation modes obey gaussian statistics (as the probability $\sim e^{-S}$). In this sense the second order corrections give us a measure of non-gaussianity in the distribution of cosmological perturbations, and, in conjunction with precision measurements of the spectral index, measures of non-gaussianity provide very sensitive tests of cosmological models. As we will see, ekpyrotic models predict a substantial amount of non-gaussianity, in contrast to simple single-field inflationary models. This is principally due to the fact that the potential is steep, and hence self-interactions of the ekpyrotic field are important, whereas in inflation, the potential is very flat and self-interactions of the inflaton are negligible. The fact that non-gaussianity will be measured with high precision in the near future (with the upcoming Planck satellite) and that hints of it have already been seen \cite{Yadav:2007yy} justifies a detailed examination of the ekpyrotic predictions. For conversion during the ekpyrotic phase, the order-of-magnitude of non-gaussianity has been estimated in \cite{Creminelli:2007aq,Buchbinder:2007tw} and more detailed calculations were performed in \cite{Koyama:2007if,Buchbinder:2007at}; for conversion during the kinetic phase, the non-gaussianity was calculated in \cite{Lehners:2007wc}. The various models have been compared in \cite{Lehners:2008my}.

What we have to do is rather clear: namely, we have to repeat the analysis of the previous paragraphs, but this time we need to include all the second order terms as well. To this effect, we will decompose the entropy perturbation into a
linear, gaussian part and a second-order perturbation by
writing $\delta s = \delta s^{(1)} + \delta s^{(2)}.$ Its
equation of motion, on large scales and up to second order in
field perturbations is then given by \cite{Langlois:2006vv}
\bea && \ddot{\delta s} +3H \dot{\delta s}+
\left(V_{ss}+3\dot{\theta}^2 \right) \delta s \nn \\
&&+\frac{\dot{\theta}}{\dot{\sigma}}(\dot{\delta s}^{(1)})^2
+\frac{2}{\dot{\sigma}}\left( \ddot{\theta}+
\dot{\theta}\frac{V_{\sigma}}{\dot{\sigma}} -
\frac{3}{2}H\dot{\theta}\right)\delta s^{(1)} \dot{\delta
s}^{(1)} \nn \\ && +\left( \frac{1}{2}
V_{sss}-\frac{5\dot{\theta}}{\dot{\sigma}}V_{ss}-
\frac{9\dot{\theta}^3}{\dot{\sigma}} \right)(\delta s^{(1)})^2
+\frac{2\dot{\theta}}{\dot{\sigma}}\delta \epsilon^{(2)}= 0.
\label{eq-entropy} \eea   The last term in equation
(\ref{eq-entropy}) is a non-local term proportional to the
difference in spatial gradients between the linear entropy
perturbation and its time derivative. This difference evolves
as $a^{-3}$ \cite{Langlois:2006vv}, so that it remains
approximately constant during the ekpyrotic phase when $a$ is
very slowly varying. This ends up being exponentially
suppressed compared to the entropy perturbation itself, which
grows by a factor of $10^{30}$ or more during this same period
\cite{Lehners:2007ac}.  After the ekpyrotic phase has ended,
the non-local term grows because $a \propto (-t)^{-1/3}$, but
this growth is negligible compared to the exponential
suppression during the ekpyrotic phase.  Hence, the non-local
term can be safely neglected. Note that this is a closed equation for
the entropy perturbation. Here $V_{\sigma}$ denotes a
derivative of the potential along the background trajectory.

It is useful to recast the evolution in terms of the adiabatic
and entropic variables $\s$ and $s$. Up to unimportant additive
constants (that we will fix below), they can be defined by \bea
\s &\equiv& \frac{\dot\phi_1 \phi_1 + \dot\phi_2
\phi_2}{\dot\s}
\\ s &\equiv& \frac{\dot\phi_1 \phi_2 - \dot\phi_2
\phi_1}{\dot\s}. \eea Then we can expand the potential up to third order as follows
\cite{Buchbinder:2007tw}: \be V_{ek}=-V_0 e^{\sqrt{2\e}\s}[1+\e
s^2+\frac{\k_3}{3!}\e^{3/2} s^3],
\label{potentialParameterized}\ee where $\k_3$ is of ${\cal O}(1)$
 for typical potentials (the case of exact exponentials corresponds to $\k_3=-4\sqrt{2/3}$).
The scaling
solution (\ref{ekpyrosis-scaling}) can be rewritten as \be a(t)=(-t)^{1/\e} \qquad
\s=-\sqrt{\frac{2}{\e}}\ln \left(-\sqrt{\e V_0} t\right) \qquad
s=0. \label{ScalingSolution}\ee  The intrinsic non-gaussianity
in the entropy perturbation is produced during the ekpyrotic
phase and can be determined from the equation of motion for the
entropy field (\ref{eq-entropy}), which reduces to \be
\ddot{\delta s}+3H\dot{\delta s}+V_{ss}\delta s +
\frac{1}{2}V_{sss} (\delta s^{(1)})^2 = 0.
\label{EomGeneration}\ee The last term is \be
\frac{1}{2}V_{sss} =-\frac{\k_3}{2t^2} \sqrt{\e}
 \ee and thus the solution, at long wavelengths and up to second order
in field perturbations, is given by
\cite{Koyama:2007if,Lehners:2007wc} \bea \delta s(t) &=& \delta
s^{(1)}(t) + \delta s^{(2)}(t) \nn
\\ &=& \delta s_{end} \frac{t_{end}}{t} + \tilde{c} (\delta
s_{end} \frac{t_{end}}{t})^2, \label{entropys2ekpyrosis}\eea
where \be \tilde{c} = \frac{\k_3\sqrt{\e}}{8}.
\label{ctildedefinition}\ee Thus, the intrinsic non-gaussianity
present in ekpyrotic models (second term in
(\ref{entropys2ekpyrosis}) above) is of order ${\cal{O}}
(\sqrt{\epsilon_{ek}})$ and it is due to the steepness of the
potentials and the resulting self-interactions of the scalar
fields. This is in sharp contrast with inflation, where the
intrinsic non-gaussianity is extremely small due to the
flatness of the potential (or, equivalently, $\epsilon_{inf}\ll
1$).

What is measured is not directly the non-gaussianity present in
the entropy perturbation, but the non-gaussianity it imprints
on the curvature perturbation. Thus, it is important to know
the strength with which this intrinsic non-gaussianity gets
transferred to the curvature perturbation. The time evolution
of the curvature perturbation to second order in field
perturbations and at long wavelengths is given in FRW time by
\cite{Langlois:2006vv} \bea \dot{\zeta} &=&
-\frac{2H}{\dot{\s}}\dot{\theta}\delta s +
\frac{H}{\dot{\sigma}^2} [(V_{ss} + 4 \dot{\theta}^2) (\delta
s^{(1)})^2-\frac{V_{,\sigma}}{\dot\sigma}\delta s \dot{\delta s}], \label{zetadotquadratic} \eea where we have omitted a non-local term which can be neglected in ekpyrotic models for the same reason as the non-local term omitted from (\ref{eq-entropy}) above. As we saw in equation (\ref{curvaturelinear}), at linear order, a
non-zero entropy perturbation combined with a bending
($\dot{\theta}\neq 0$) of the background trajectory sources the
curvature perturbation on large scales and results in a linear,
gaussian curvature perturbation $ {\zeta}_L = \int_{\Delta t}
\sqrt{\frac{2}{\epsilon_c}}\dot{\theta}\delta s^{(1)}.$ Thus the strength of conversion is
proportional to $1/\sqrt{\epsilon_c}$ and this dependence on
the equation of state significantly influences the magnitude
of the second order correction as well, as we will see presently.

As in most inflationary models, the fluctuations are generated by
scalar fields with canonical kinetic energy density, so they
generate non-gaussianity of the ``local'' type, as considered in
Refs.~\cite{Verde:1999ij,Komatsu:2000vy,Maldacena:2002vr,Babich:2004gb}. The local
wavelength-independent non-gaussian contribution to ${\zeta}$
can then be characterized in terms of the leading linear,
gaussian curvature perturbation ${\zeta}_{L}$ according to \be
{\zeta}={\zeta}_L+\frac{3}{5}f_{NL}{\zeta}_L^{2}, \label{fNLdefinition} \ee
using the sign convention for wavelength-independent
non-gaussianity parameter $f_{NL}$
 in \cite{Komatsu:2000vy}. Current observational constraints on $f_{NL}$ have been reported by the WMAP team \cite{Komatsu:2008hk} to lie in the range \be -9 < f_{NL} < 111 \ee at the $2\s$ level.

For the two-field models that we are interested in here, Eq. (\ref{zetadotquadratic}) implies that the contributions to $f_{NL}$ can be divided into three
parts \cite{Lehners:2007wc}  \bea
f_{NL}^{intrinsic}&=&-\frac{5}{3{\zeta}_L^2}\int_{\Delta t}
\frac{2H}{\dot{\s}}\dot{\theta}\delta s^{(2)} \label{fNLintrinsic}\\
f_{NL}^{reflection}&=&\frac{5}{3{\zeta}_L^2}\int_{\Delta t}
\frac{H}{\dot{\s}^2} [(V_{ss} + 4 \dot{\theta}^2) (\delta
s^{(1)})^2-\frac{V_{,\sigma}}{\dot\sigma}\delta s \dot{\delta s}]\label{fNLreflection}\\
f_{NL}^{integrated}&=&\frac{5}{6 {\zeta}_L^2}(\delta
s^{(1)}(t_{end}))^2, \label{fNLintegrated}\eea with the total $f_{NL}$ being the sum of all three contributions.
$f_{NL}^{intrinsic}$ arises from the direct translation of the
intrinsic non-linearity present in the entropy perturbation
into a corresponding non-linearity in the curvature
perturbation. By contrast, $f_{NL}^{reflection}$ and
$f_{NL}^{integrated}$ would be non-zero even if the entropy
perturbation were exactly gaussian. Both contributions are due
to the non-linear relationship between the curvature
perturbation and the entropy perturbation, as expressed in
equation (\ref{zetadotquadratic}) - the difference is that
$f_{NL}^{integrated}$ gets generated during the ekpyrotic phase
and $f_{NL}^{reflection}$ during the conversion.

The intrinsic non-gaussianity can be estimated by combining
equations (\ref{curvaturelinear}), (\ref{entropys2ekpyrosis}), (\ref{ctildedefinition}) and (\ref{fNLintrinsic}):
\bea f_{NL}^{intrinsic} &\approx& \pm
\frac{5}{3{\zeta}_L^2}\int_{\Delta
t} \sqrt{\frac{2}{\epsilon_c}} \dot{\theta} \tilde{c}{\delta s^{(1)}}^2 \nn \\
&\sim& \sqrt{\epsilon_c \epsilon_{ek}}. \label{fNLestimate}
\eea It is given by the geometric mean of the $\epsilon$
parameters during the phases of generation and conversion of
the perturbations. This is perhaps the most important result
because it fixes a rough magnitude for $|f_{NL}|$ which can
only be significantly reduced through accidental cancelations
due to the  other terms or other effects not included in the
present analysis. It explains in a nutshell why the
non-gaussianity in ekpyrotic/cyclic models is necessarily
more than an order of magnitude greater than in simple inflation
models, and it explains why the equation of state during
conversion can have a significant quantitative effect on the
prediction.

To go beyond this qualitative estimate to a more precise one, we need
to take account of
the details of the conversion mechanism.
We will again discuss the various possibilities considered above in turn.

{\noindent \it Conversion during kinetic energy domination}

We can immediately perform an
order-of-magnitude estimate of the non-gaussianity generated by this conversion mechanism (described in section \ref{subsection twofields}) since during
the kinetic phase $\epsilon_c=3,$ and so (\ref{fNLestimate})
would lead us to expect $f_{NL}$ to be of order \be f_{NL} \sim
\sqrt{\epsilon_{ek}} \sim {\cal{O}}(c_1). \ee

We will treat the bend as if only one field reflects
($\phi_2$), and we will consider the case where $\dot\theta >
0,$ see Fig. \ref{FigureSubdominant}. Other cases can be
related to these representative examples by changing the
coordinate system in field space appropriately. In the
heterotic M-theory example, the reflection occurs because
$\phi_2$ comes close to a boundary of moduli space ($\phi_2=0$)
and is forced to bounce \cite{Lehners:2007nb}. For the purposes
of this study, we will treat the reflection as being due to a
potential, $V^R(\phi_2),$ an additional contribution unrelated
to the exponential potentials that were dominant during the
ekpyrotic phase (but which are negligible during the kinetic
energy dominated phase).

It is important to know the evolution of the entropy
perturbation during the process of conversion. If there is a
phase of pure kinetic energy domination before the conversion,
then the background scalar field trajectory is also a straight
line during this phase, but with the potentials being
irrelevant. The equation of motion reduces to \be \ddot{\delta
s}+\frac{1}{t}\dot{\delta s} = 0, \label{EomKE}\ee and by
matching onto the ekpyrotic solution and its first time
derivative at $t=t_{end}$ we find \bea
\delta s(t) &=& \delta s^{(1)}(t) + \delta s^{(2)}(t) \nn \\
&=& \delta s_{end}(1+ \ln{\frac{ t_{end}}{t}}) + \tilde{c}
\delta s_{end}^2 (1+2 \ln{\frac{t_{end}}{t}}). \eea
Incidentally, note that \be t\frac{ \dot{\delta s}}{\delta s}
\sim \frac{1}{\ln(-t)} \label{c4} \ee during the kinetic phase,
while during the ekpyrotic phase $t\dot{\delta s} \sim \delta
s.$ This observation will simplify our analysis later on.

\begin{figure}[t]
\begin{center}
\includegraphics[width=0.75\textwidth]{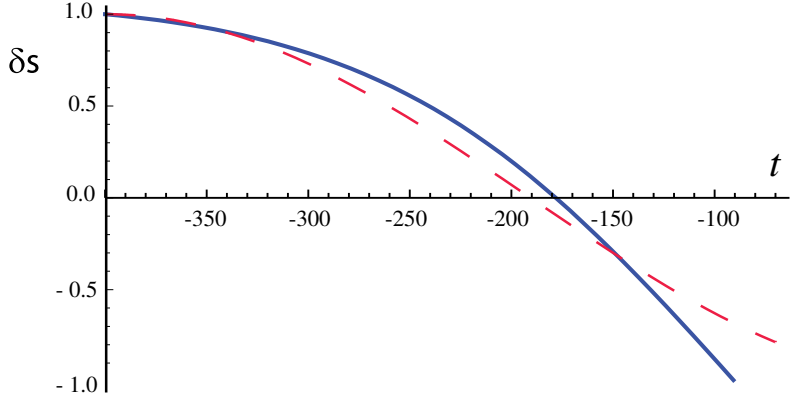}
\caption{\label{Figure3} {\small
The evolution of the linear entropy perturbation during conversion: the solid
(blue)
line shows the actual evolution calculated numerically, while the dashed
(purple) line
shows the approximate solution (\ref{entropys1conversion}) with $\omega=3,$
$\dot{\theta}=-1/t_{ref}$ and $t_{ref} =-400 M_{Pl}^{-1}$. For the
purposes of illustration, $\delta
s^{(1)}(t_{ref})$ has  been normalized to $1$.}}
\end{center}
\end{figure}

During the conversion, even though the kinetic energy of the
scalar fields is still the dominant contribution to the total
energy, the potential $V^R(\phi_2)$ that causes the bending has
a significant influence on the evolution of the entropy
perturbation. To analyze this, we will approximate (as in
\cite{Buchbinder:2007at}) the bending of the trajectory to be
gradual by taking $\dot{\theta}$ constant and non-zero for a
period of time $\Delta t,$ starting from $t=t_{ref}.$ Note
that, assuming the total angle of bending is ${\cal O}(1)$
radian, we have $|\dot{\theta}| \approx 1/|t_{ref}|$ in this
case. Then one can relate the derivatives of the potential to
expressions involving $\dot{\theta},$ for example \be
V^R_{ss}=-2\dot{\theta}^2+\frac{\dot{\theta}}{\gamma
t}.\label{Vssapproximation} \ee Assuming a gradual conversion
($\Delta t \sim t_{ref}$), we can ignore higher derivatives of
$\theta$. (In \cite{Lehners:2007wc} it was shown that sharp
transitions lead to unacceptably large values of $f_{NL};$
hence these cases are of less phenomenological interest.) To
satisfy the constraint on the amplitude of the curvature
perturbation obtained by the WMAP observations, we set $t_{ref}
\approx - 10^3 M_{Pl}^{-1}$ \cite{Lehners:2007ac}. At linear
order, the equation of motion (\ref{eq-entropy}) then reads \be
\ddot{\delta s}^{(1)}+3H\dot{\delta
s}^{(1)}+(\dot{\theta}^2+\frac{1}{\gamma t}\dot{\theta})\delta
s^{(1)} = 0, \label{EomReflectionLinear}\ee As indicated by
equation (\ref{c4}), we can set $\dot{\delta s}^{(1)} = 0$ as a
first approximation and, thus,
 neglect the damping term in
the equation of motion. Also, we will simply evaluate the
coefficient of the last term midway through the reflection, and
define \be \omega \equiv \dot{\theta}
\sqrt{1+\frac{1}{\dot{\theta}\gamma (t_{ref}+\Delta t/2)}}. \ee
For a gradual reflection $\omega \approx (2-3) \dot{\theta}.$
Then the solution for the linear entropy perturbation is \be
\delta s^{(1)} = \delta
s^{(1)}(t_{ref})\cos{\omega(t-t_{ref})}.
\label{entropys1conversion}\ee Instead of continuing to grow
logarithmically, the entropy perturbation actually falls off
during the conversion, see Fig.~\ref{Figure3}. This has the
consequence that the conversion from entropy to curvature
perturbations is less efficient than one might have naively
thought. From (\ref{curvaturelinear}), we can estimate \be
\zeta_L= \sqrt{\frac{2}{3}}\dot{\theta}\int_{\Delta t}
\delta s^{(1)}= \sqrt{\frac{2}{3}} \frac{\dot{\theta}}{\omega}
\delta s^{(1)}(t_{ref})\sin{\omega \Delta t},
\label{curvaturelinestimate}\ee where we have used
$\epsilon_c\approx 3$,  which is a  good approximation for
subdominant reflections (and acceptable for the estimating
purposes for dominant ones).
Since $\omega \Delta t \approx 3,$ the entropy perturbation
evolves over nearly a half-cycle and consequently $\sin \omega
\Delta t$ is a small factor which we will take to be about
$1/3$ in our estimates, which fits well with numerical results.

We also need to know the evolution of the second order entropy
perturbation during the time of conversion. Using the same
approximations as above (implying that we can neglect $V_{sss}$
compared to $\dot\theta V_{ss}/\dot\sigma$ and
$\dot\theta^3/\dot\sigma$ by the time the conversion is
underway), the equation of motion (\ref{eq-entropy}) simplifies
to \be \ddot{\delta s}^{(2)}+ \omega^2 \delta s^{(2)} -
\frac{\dot{\theta}}{\dot{\sigma}} \omega^2 (\delta s^{(1)})^2 =
0, \label{entropys2reflection} \ee where $\delta s^{(1)}$ is
given in (\ref{entropys1conversion}). Putting
$\dot{\theta}\approx \dot{\sigma}$ at the start of the
reflection and keeping in mind that we impose the boundary
condition $\dot{\delta s}^{(2)}\approx 0$, the solution for the
second order entropy perturbation is given by \bea  \delta
s^{(2)} &=& \delta
s^{(2)}(t_{ref})\cos[\omega(t-t_{ref})] \nn \\
& & +\frac{1}{12}(\delta s^{(1)}(t_{ref}))^2 \Big(
-4\cos[\omega(t-t_{ref})] \nn \\ && +4\cos^4[\omega(t-t_{ref})]
+9\sin^2[\omega(t-t_{ref})] \nn
\\ & & +\sin[\omega(t-t_{ref})]\sin[3\omega(t-t_{ref})] \Big). \label{s2}
\eea At large $\epsilon_{ek}$ (or, equivalently, large
$|\tilde{c}|$), the second order entropy perturbation falls off
in the same way as the linear perturbation, but at small
$|\tilde{c}|$ there are significant corrections to this
behavior. We will also need the integral \bea C^{-2}
\int_{\Delta t}\delta s^{(2)}   &=&
\frac{(1+2\ln(t_{end}/t_{ref}))}{(1+\ln(t_{end}/t_{ref}))^2}
\frac{\sin(\omega \Delta t)}{ \omega}\tilde{c} \nn \\
&& +\frac{\Delta t}{2} -\frac{\sin(\omega \Delta
t)}{3\omega}-\frac{\sin(2\omega \Delta t)}{12\omega}, \eea
where $C= \delta s^{(1)}(t_{ref})$. In all cases of interest,
the last two terms are negligible.

We are finally in a position to evaluate the various
contributions to the non-linearity parameter $f_{NL}.$ The
intrinsic contribution, defined in (\ref{fNLintrinsic}),
becomes \bea f_{NL}^{intrinsic} \approx A \k_3
\sqrt{\epsilon_{ek}}+B, \label{simp} \eea with \bea A &=&
\frac{5\omega(1+2\ln{(t_{end}/t_{ref})})}{8\sqrt{6}\dot{\theta}
(1+\ln{(t_{end}/t_{ref})})^2\sin(\omega \Delta t)} \nn \\ B &=&
\frac{5\omega^2}{2\sqrt{6}\dot{\theta}^2\sin^2(\omega \Delta
t)}. \eea Eq.~(\ref{simp}) is one of our key results because it
shows that the essential contribution scales in a simple way
with $\epsilon_{ek}$ and has a value that exceeds the total
$f_{NL}$ for simple inflationary models by more than an
 order of magnitude.
As suggested by Eq.~(\ref{fNLestimate}), the first term in
$f_{NL}^{intrinsic}$ can be re-expressed as ${\cal O}
\,(\sqrt{\epsilon_c \epsilon_{ek}})$ with $\epsilon_c=3$. This
contribution comes directly from the non-zero $\delta s^{(2)}$
generated during the ekpyrotic phase, which is due to the third
derivative w.r.t. $s$ of the ekpyrotic potential.  It is
interesting to note that, in the case where $V_{sss}=0$ or
$\k_3=0$, $f_{NL}^{intrinsic}$ is nevertheless non-negligible
because of the positive offset $B$ generated by the linear
entropy perturbation  $\delta s^{(1)}$ during conversion, the
piece proportional to  $(\delta s^{(1)})^2$ in (\ref{s2}).

Using the definition (\ref{fNLreflection}) together with
(\ref{Vssapproximation}) it is straightforward to see that for
conversion during the kinetic phase, $f_{NL}^{reflection}$ is
always negative, and it can be estimated as \bea
f_{NL}^{reflection} &\approx&
\frac{5}{6\zeta_L^2}\int_{\Delta t} (2 \dot{\theta}^2 t +
\frac{\dot{\theta}}{\gamma}) (\delta s^{(1)})^2 \nn \\
&\approx& -\frac{15\omega^2}{8\dot{\theta}^2\sin^2(\omega
\Delta t)}\frac{|t_{ref}+\Delta t/2|}{\Delta t}, \eea  where we
have used $\gamma t_{ref} \dot{\theta} \approx 1$ and \be
\int_{\Delta t} t \sin^2 \omega (t-t_{ref})\approx
\frac{1}{2}(t_{ref}+\frac{\Delta t}{2})\Delta t, \ee which is a
valid approximation since $\delta s^{(1)}$ evolves over
approximately a half-cycle. We have neglected the contribution
due to the term proportional to $\d s \dot{\d s},$ since it is
suppressed on account of (\ref{c4}). The integrated
contribution to $f_{NL}$ generated during the ekpyrotic phase
gives an additional contribution of
 \bea
f_{NL}^{integrated}&=&\frac{5}{12 \zeta_L^2}(\delta
s^{(1)}(t_{end}))^2 \nn \\ &\approx&
\frac{5\omega^2}{8\dot{\theta}^2(1+\ln{(t_{end}/t_{ref})})^2\sin^2(\omega
\Delta t)}. \eea Note that neither the reflected nor the
integrated contributions depend on $\epsilon_{ek};$ they both
simply shift the final result by a number depending on the
sharpness of the transition and the duration of the purely
kinetic phase respectively.

The total $f_{NL}$ is the sum of all the above contributions.
Since we are only considering gradual conversions, we expect
the ratio $|t_{ref}+\Delta t/2| /\Delta t$ to lie between 1 and
2. Also, the duration of the pure kinetic phase between the end
of the ekpyrotic phase and the conversion is necessarily rather
short, so that we expect $\ln(t_{end}/t_{ref}) \lesssim {\cal
O}(1).$

\begin{figure}[t]
\begin{center}
\includegraphics[width=0.75\textwidth]{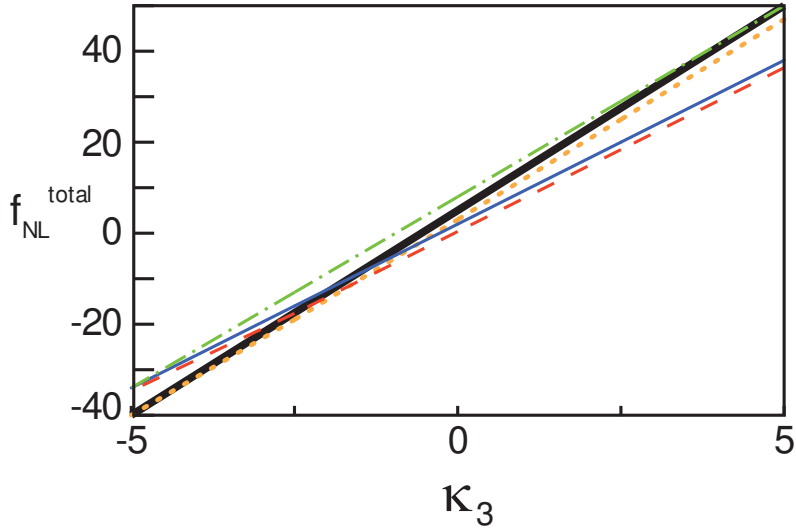}
\caption{\label{Figure4} {\small A comparison of the results from
numerical calculations with the fitting formula given
in Eq.~(\ref{fNLapproximation}) and indicated by the thick black line.
Here we have fixed the value of $\epsilon_{ek} = 36.$
The plot confirms that $f_{NL}$ then grows linearly with $\k_3,$ in good agreement with (\ref{fNLapproximation}). The
sample models (dashed and dotted lines) are representative of the range of models and
parameters shown in \cite{Lehners:2007wc}. We have similarly
checked the dependence on $\sqrt{\e}$ when the value of $\k_3$ is kept fixed.}}
\end{center}
\end{figure}

Putting everything together, we get the following estimates:
\bea f_{NL}^{intrinsic} &\approx& 10\tilde{c} + 50 \\
f_{NL}^{reflection} &\approx& -50 \label{fNLrefestimate} \\
f_{NL}^{integrated} &\approx& 5. \eea The analytic estimate for
$f_{NL}^{reflection}$ suggests a range that could extend to
$-100$, but the value above is more consistent with the exact
numerical calculations for a wide range of parameters.
Altogether, we end up with the following fitting
formula \bea f_{NL}^{total} &\approx&  10 \tilde{c} +5 \\
&\approx& \frac{3}{2}\k_3 \sqrt{\epsilon_{ek}} +5.
\label{fNLapproximation}\eea Fig. \ref{Figure4} shows that this
formula agrees well with the results from exact numerical
calculations. In particular, we have been able to estimate the
slope, and hence the dependence on $\k_3\sqrt{\e_{ek}},$ rather
accurately.

\vspace{1cm}

{\noindent \it Conversion during the ekpyrotic phase}

If the conversion happens during the
ekpyrotic phase, {\it i.e.} while the ekpyrotic potentials are
relevant, $\epsilon_c=\epsilon_{ek},$ and from (\ref{fNLestimate})
we expect the non-gaussianity to be of order \be f_{NL} \sim
\epsilon_{ek} \sim {\cal{O}}(c_1^2), \ee \ie we expect the non-gaussianity to be larger by a factor of $c_1$ compared to the case of conversion during the kinetic phase.

As shown by Koyama {\it et al.} \cite{Koyama:2007if}, the $\delta \cN$ formalism is well
suited to treating this case, and the derivation surprisingly quick. According to the $\delta \cN$ formalism \cite{Starobinsky:1986fxa,Sasaki:1995aw}, the comoving curvature perturbation is identified with a perturbation in the number of e-folds of ekpyrosis (the perturbations are identified on a uniform density hypersurface at $t=t_f$, starting the evolution from a flat initial hypersurface at $t=t_i$)
\be \zeta = \delta \cN = \cN_{,\a_g}\delta \a_g + \frac{1}{2} \cN_{,\a_g \a_g} (\delta \a_g)^2, \label{deltaN}\ee where we have expanded $\delta \cN$ up to second order in terms of a gaussian variable $\a_g.$ We will specify the meaning of $\a_g$ shortly. Then, comparing with the definition (\ref{fNLdefinition}), we immediately see that the non-gaussianity parameter is given by \be f_{NL}=\frac{5}{6}\frac{\cN_{,\a_g \a_g}}{(\cN_{,\a_g})^2}. \ee We can evaluate $\cN$ by using the one and two-field versions of the scaling solution (\ref{ekpyrosis-scaling}), with the result that \cite{Koyama:2007if} \be \cN = \int_{t_i}^{t_{ref}} H dt + \int_{t_{ref}}^{t_f} H dt = -\frac{2}{c_j^2} \ln |H_{ref}| + constant, \label{ekpyrosisefolds}\ee where the index $j$ corresponds to the field that becomes frozen in the late-time
single-field solution. Here we have assumed that at time $t_{ref}$ the trajectory instantaneously switches from the unstable two-field scaling solution to a single-field attractor scaling solution \cite{Koyama:2007mg}. Now, using that $\delta s \propto t^{-1} \propto H,$ we can rewrite equation (\ref{entropys2ekpyrosis}) as \be \delta s =\a_g H + \tilde{c} (\a_g H)^2, \ee whence $\a_g$ is seen to be a parameter distinguishing different trajectories. Turning this relation around, we have that \be \a_g = \frac{\delta s}{H}(1-\tilde{c} \delta s). \ee Hence at the time of transition (reflection) $t_{ref},$ when $\delta s=\delta s_{ref}$ is fixed, we have that $\a_g \propto H_{ref}^{-1}.$ Thus, using (\ref{ekpyrosisefolds}), we can rewrite (\ref{deltaN}) as \be \delta \cN = \frac{2}{c_j^2 \a_g} \delta \a_g -\frac{1}{c_j^2 \a_g^2}(\delta \a_g)^2. \ee It immediately follows that the non-gaussianity parameter $f_{NL}$ is given by \cite{Koyama:2007if} \be f_{NL} = \frac{5}{6}(\frac{-2}{c_j^2 \a_g^2})(\frac{c_j^2 \a_g}{2})^2=
-\frac{5}{12}c_j^2, \label{fNLKoyama}\ee where, again, the index $j$
corresponds to the field that becomes frozen in the late-time
single-field solution. This expression has been checked numerically, both in \cite{Koyama:2007if} and in \cite{Lehners:2008my}, by different methods, and the numerics have been found to agree well with the analytic formula above. Note the particular feature that the sign of $f_{NL}$ is always positive for
these cases.

A qualitative understanding of this result can be achieved in
the present context as follows: from (\ref{fNLintrinsic}) we
can see that, since $H/\dot{\sigma}$ is negative and
approximately constant, the sign of $f_{NL}^{intrinsic}$ is
given by the sign of $\int \dot{\theta} \delta s^{(2)}.$ In
practice, numerical simulations indicate that, for conversion
during the ekpyrotic phase, we have \be |\dot{\theta}| \approx
\frac{1}{10|t|} \label{thetadotekpyrotic} \ee during most of
the conversion, but with $\dot{\theta}$ growing faster towards
the end. We will use this numerical input to guide our
analysis. The sign of $f_{NL}^{intrinsic}$ is essentially
determined by the sign of $\dot{\theta} \delta s^{(2)}$ towards
the end of the period of conversion. Naively it is difficult to
perform a purely analytic estimate of $f_{NL}$ in the current
scheme, since all the terms in the equation of motion
(\ref{eq-entropy}) go as $c_1 t^{-4};$ however, equation
(\ref{thetadotekpyrotic}) tells us that \be |V_{ss}| = 2/t^2
\gg \dot{\theta}^2, \label{Vssdominant}\ee and so there are
surprisingly few terms in the equation of motion for $\delta
s^{(2)}$ that are actually important during (most of) the time
of conversion. In fact, to a first approximation, we are simply
left with \be \ddot{\delta s^{(2)}} = (-\frac{1}{2}V_{sss} +
\frac{\dot{\theta}}{\dot{\sigma}} V_{ss}) (\delta s^{(1)})^2.
\ee $V_{sss}$ decreases in importance as the single-field
scaling solution is reached. Thus, even though the $V_{sss}$
term determines the initial evolution of $\delta s^{(2)},$
eventually the $V_{ss}$ term dominates. Then, since $V_{ss}<0$
it is easy to see that the sign of $\delta s^{(2)}$ is always
driven to be opposite to that of $\dot{\theta},$ and
consequently $f_{NL}^{intrinsic}$ is negative in all cases.

Since $f_{NL}^{reflection}$ is proportional to
$\zeta_L^{-2},$ it provides a contribution of the same order
${\cal{O}}(\epsilon_{ek}).$ Using (\ref{Vssdominant}), it is
straightforward to see that the part of $f_{NL}^{reflection}$
that is proportional to $(\d s)^2$ is always positive, while
the part proportional to $\d s \dot{\d s}$ is always negative.
This implies that there will be a competition between the two
contributions. Numerical integration then shows that the part
proportional to $(\d s)^2$ approximately cancels out
$f_{NL}^{intrinsic},$ while the $\d s \dot{\d s}$ part by
itself is very close in numerical value to the final answer.
In all cases $f_{NL}^{integrated}$ is completely negligible.

\begin{table}
\begin{tabular}{|c|c||c||c|}
  \hline
  $c_1$ & $c_2$ & $f_{NL,\delta N}$ & $f_{NL}$ \\ \hline
  10 & 10 & -41.67  & -39.95   \\
  10 & 15 & -41.67  & -40.45  \\
  10 & 20 & -41.67  & -40.62 \\
  15 & 10 & -93.75  & -91.01 \\
  15 & 15 & -93.75  & -92.11  \\
  15 & 20 & -93.75  & -92.49   \\
  20 & 10 & -166.7  & -162.5   \\
  20 & 15 & -166.7  & -164.4  \\
  20 & 20 & -166.7  & -165.1  \\
  \hline
\end{tabular}
\caption{\small Ekpyrotic conversion: the values of $f_{NL}$
estimated by the $\d N$ formalism compared to the numerical
results obtained by directly integrating the equations of
motion.} \label{Table1}
\end{table}

Consequently, the total $f_{NL}$ turns out to be negative and
moreover in good agreement with the $\delta N$ result
(\ref{fNLKoyama}) \cite{Koyama:2007if}, as shown in Table
\ref{Table1}. Clearly, the $\delta N$ formalism provides a fast
and elegant derivation of non-gaussianity for conversion during
the ekpyrotic phase that agrees well with direct integration of
the equations of motion.  (Note, however, no analogous $\delta
N$ approach applies to conversion in the kinetic energy
dominated phase.)

The conclusion that $f_{NL}$ is substantially less than zero
differs from Ref. \cite{Buchbinder:2007at}; their approximation
method was incomplete though, since the term proportional to
$\d s \dot{\d s}$ in (\ref{zetadotquadratic}) was not included.
As shown above, this term typically contributes significantly
to the final result. Also, the evolution of the entropy
perturbation during conversion was neglected, which meant, for
example, that values of $f_{NL}^{intrinsic}$ of either sign
were obtained. Our results here show that their approximations
were too crude, and that $f_{NL}$ is generally negative with
$f_{NL} \lesssim -20$. This is inconsistent with current limits
by roughly 3 $\sigma$ \cite{Komatsu:2008hk}.

Hence,  we conclude that models with conversion during the
ekpyrotic phase ($\epsilon_c = \epsilon_{ek} \gg 1$) are
difficult to accommodate with current observations in contrast
to models in which conversion occurs during the kinetic energy
dominated phase ($\epsilon_c=3$).

{\noindent \it Conversion after the crunch/bang transition}

If the conversion of entropy to curvature perturbations occurs via the scenario of modulated preheating after the big bang
\cite{Battefeld:2007st}, then the conversion happens while $\epsilon_c \approx 3,$ as described in section \ref{subsection twofields}. Thus we can expect
an intrinsic contribution to $f_{NL}$ of \be f_{NL} \sim
\sqrt{\epsilon_{ek}} \sim {\cal{O}}(c_1), \ee {\it i.e.} $f_{NL}$
is of the same order as in the case of conversion during the
kinetic phase preceding the big crunch. Moreover, as shown in
\cite{Battefeld:2007st}, no significant additional contributions
to $f_{NL}$ are expected. Therefore, models of this type are
subject to roughly the same observational constraints as the
models where the conversion happens during the kinetic phase
before the bang and where the intrinsic contribution to the
non-gaussianity is dominant.
Note however that
the entropy perturbations are converted with an unknown efficiency
\cite{Battefeld:2007st} \be e = \frac{3}{2} \frac{|h_{,s}|}{h}.
\ee Since the non-linearity in the entropy field is of
magnitude $\tilde{c},$ this implies a non-linearity in the
curvature perturbation given by \be f_{NL} \approx \pm
\frac{\tilde{c}}{e}, \ee where the $\pm$ sign reflects our
ignorance of the sign of $h_{,s}$ and of the direction of
bending of the scalar field trajectory. Thus, in the absence of
a more detailed model, we cannot go beyond this rough
order-of-magnitude estimate at present.

\vspace{1cm}

The analysis of non-gaussianity can be summarized in a few rules of thumb:
\begin{itemize}
\item The intrinsic contribution to $|f_{NL}|$ is
    proportional to the geometric mean of $\epsilon_{ek}$
    and $\epsilon_c$, which is at least two orders of
    magnitude greater than in simple inflationary models
    (where $|f_{NL}^{intrinsic}| = {\cal O}(0.1)$). When
    all contributions are considered, the total $f_{NL}$ is
    generically more than an order of magnitude greater
    than in simple inflationary models, (where
    $f_{NL}^{total} = {\cal O}(1)$).
\item  For a fixed value of $\k_3,$ the value of
    $|f_{NL}^{total}|$ is correlated with the spectral tilt: smaller
    $|f_{NL}^{total}|$ implies smaller $\epsilon_{ek}$, which tends to
    make the spectral tilt bluer. Current limits on $f_{NL}$ fit well
    with limits on the spectral tilt for the simplest models with
    conversion during the kinetic energy driven phase before the bang
    or during reheating after the big bang.
\item Models in which the conversion occurs during a phase
    with larger $\epsilon_c $ produce a  larger intrinsic
    $|f_{NL}|$ and are more difficult to fit with the
    current observed limits on $f_{NL}$ and spectral tilt. In particular, models with conversion in the ekpyrotic phase favor $|f_{NL}|$ large and {\it negative}.
\item Cases in which the intrinsic contribution to $f_{NL}$
    is much smaller than the reflection plus integrated
    contributions produces very large values of $|f_{NL}|
    \gg 100$ that are inconsistent with current
    observational bounds. This includes all cases where the
    conversion is sharp.
\end{itemize}

\begin{figure}[t]
\begin{center}
\includegraphics[width=0.75\textwidth]{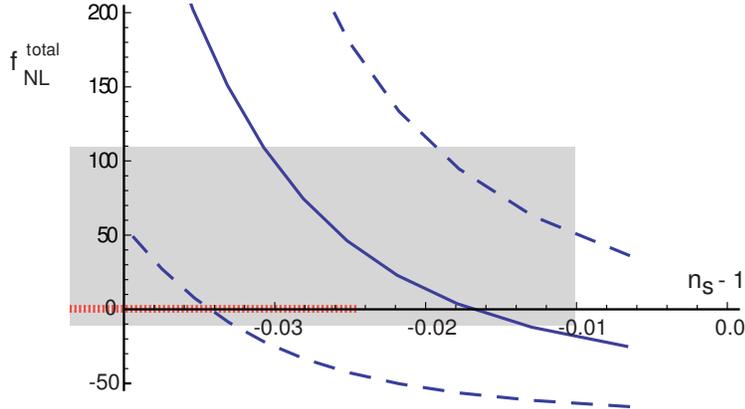}
\caption{\label{fnlvtilt} {\small
A plot for characterizing the correlation
between $|f_{NL}|$ and scalar spectral tilt, $n_s -1$,
here illustrated for the case of  the cyclic model
 in which the conversion from entropic to
curvature perturbations occurs during the kinetic energy dominated
phase just before the big crunch/big bang transition. Different curves correspond to different fixed amounts of skewness $\k_3$ in the potential (the central curve corresponds to $\k_3=4\sqrt{2/3}$), while we vary the steepness of the potential $\epsilon_{ek}.$
The curves  show
the general trend that $|f_{NL}|$ increases as the spectrum becomes redder.
 Simple inflationary models correspond to the narrow
 horizontal hashed (red) strip  with $|f_{NL}| \lesssim 1$.
 The shaded rectangle represents the current observational constraints
 on $f_{NL}$ and tilt (95\% confidence) from WMAP5 \cite{Komatsu:2008hk}.
 }}
\end{center}
\end{figure}

The analysis suggests  a  useful characteristic plot
 for differentiating
cosmological models: $f_{NL}$ versus tilt.
Figure~\ref{fnlvtilt} illustrates the prediction for a cyclic
model in which the conversion from
entropic to curvature perturbations occurs in the kinetic energy
dominated epoch following the ekpyrotic phase. Here we keep the skewness $\k_3$ of the potential fixed, as we vary its steepness $\epsilon.$ The prediction is
a swath whose width is largely due to the uncertainty in the trajectory,
parameterized by $\gamma$. Although the swath includes $f_{NL}$
near zero, positive $f_{NL}$ between 10 and 100 is preferred for tilts
in the range suggested by WMAP5. The prediction for simple inflationary
models is confined to the narrow
band $\Delta f_{NL} \lesssim 1$ around zero.

The results are surprisingly predictive. If observations of
$f_{NL}$ lie in the range predicted by the intrinsic
contribution of either inflationary or ekpyrotic/cyclic models,
it is reasonable to apply Occam's razor and Bayesian analysis
to favor one cosmological model over the other. Combining with
measurements of the spectral tilt significantly sharpens the
test. The current observational bounds obtained by the WMAP
satellite are still inconclusive, but it is clear from the
estimates presented here that non-gaussianity should be
detected by the Planck satellite if the ekpyrotic/cyclic model
is correct. At the same time, this provides a strong incentive
to further refine other methods of measuring non-gaussianity,
such as looking for evidence in measurements of the large scale
structure of the universe \cite{Slosar:2008hx}. As an example, a positive $f_{NL}$ would mean that the largest structures formed slightly earlier, and thus would have drawn more mass out of the voids, and hence one might for example expect slightly emptier voids, as compared to a perfectly gaussian distribution of primordial density fluctuations.

\subsection{Tensor perturbations}
\label{subsection tensor}

The ekpyrotic phase also generates gravitational waves, but with a spectrum radically different from the inflationary prediction. Thus, gravitational waves provide another useful test for differentiating between models of the early universe. Unfortunately, it is unclear at present how soon measurement devices will be sensitive enough to probe the gravitational wave background.

Tensor perturbations are simpler to analyze than scalar perturbations since there are no issues of gauge variance.
In the context of ekpyrotic models, the analysis was first performed in \cite{Khoury:2001wf,Boyle:2003km}. Here, we will follow the treatment of \cite{Boyle:2004gv}.  The perturbed metric is given by
  \begin{equation}
    \label{tensor_perturbed_metric}
    \textrm{d}s^{2}/a^{2}=-\textrm{d}\tau^{2}
    +[\delta_{ij}+2h_{T}Y^{(2)}_{ij}]
    \textrm{d}x^{i}\textrm{d}x^{j}
  \end{equation}
  where $Y^{(2)}_{ij}$ is a tensor harmonic.  The tensor perturbation $h_{T}$ is
  gauge-invariant and obeys the equation of motion
  \begin{equation}
    \label{tensor_EOM}
    h_{T}''+2(a'/a)h_{T}'+k^{2}h_{T}=0.
  \end{equation}
  It is useful to define the re-scaled variable $f_{T}\equiv a h_{T}$ which
  obeys
  \begin{equation}
    \label{f_EOM}
    f_{T}''+(k^{2}-a''/a)f_{T}=0.
  \end{equation}
  Again, the standard vacuum choice is the Minkowski vacuum of a
  comoving observer in the far past, corresponding to the boundary
  condition
  \begin{equation}
    \label{tensor_bc}
    f_{T}\rightarrow \frac{1}{\sqrt{2k}}e^{-ik\tau}
    \qquad\textrm{as}\quad\tau\rightarrow -\infty.
  \end{equation}
  We can now solve for $f_{T}$, just as in the scalar case.  But it is
  simpler to notice that (\ref{ekpyrosis-scaling-tau})
  implies $z(\tau)\propto a(\tau)$, and hence $z''/z=a''/a$ when
  $\epsilon$ is constant.  Thus, since $v$ and $f_{T}$ obey identical
  equations of motion (compare (\ref{v_EOM}) with (\ref{f_EOM})) and
  boundary conditions (compare (\ref{v_bc}) with (\ref{tensor_bc})),
  we find
  \begin{equation}
    \label{f_soln}
    f_{T}=v=\mathcal{P}_{2}\sqrt{\frac{\pi x}{4k}} H_{\beta}^{(1)}(x).
  \end{equation}

  The tensor spectral index is defined in the long-wavelength limit by
  $k^{3}|f_{T}|^{2}\propto k^{n_{T}}$.  Using the small $x$ expansion of the Hankel function
  and (\ref{f_soln}) we find
  \begin{equation}
  n_{T}=3-2\beta=3-\left|\frac{\epsilon-3}{\epsilon-1}\right|.
  \end{equation}
Thus, in ekpyrotic models, the tensor spectrum is very blue, $n_T \approx 2.$ There is a cut-off of the spectrum at high frequencies \cite{Boyle:2003km}, due to a combination of two facts: first, gravitational waves are not confined to the brane - they see the whole spacetime, as they are excitations of the metric. And secondly, after the ekpyrotic phase has ended, the kinetic dominated phase starts on the brane worldvolume. However, as described in section \ref{subsection Milne}, the corresponding higher-dimensional spacetime near the brane collision is compactified Milne spacetime, which is locally Minkowski space. Thus, gravitational waves with wavelengths smaller than about the GUT scale, which leave the horizon after the end of the ekpyrotic phase, behave as in Minkowski space and are not amplified (see also section \ref{section fundamental} for more details on the higher-dimensional spacetime near the collision). The high-frequency modes just below the cut-off lead to the strongest constraint on the gravitational wave spectrum, as they are susceptible of interfering with big bang nucleosynthesis. The corresponding constraint requires a minimum temperature at reheating, and will be presented in section \ref{section cyclic} along with other constraints that ekpyrotic and cyclic models have to satisfy.

The ekpyrotic spectrum is in marked contrast with the model of inflation, which predicts a nearly scale-invariant spectrum of gravitational waves, essentially because there, the gravitational waves are produced in the same way as the scalar perturbations. Because of the very blue spectrum, gravity waves are just about unobservable in ekpyrotic models \cite{Khoury:2001wf}: if we assume the same Hubble parameter $H_b$ at reheating in a model of inflation, and at the brane collision in an ekpyrotic model, then modes of wavelength $H_b^{-1}$ will have approximately the same amplitude, namely $H_b/M_{Pl}.$ But now consider a mode of wavelength comparable to the present-day horizon, $\sim H_0^{-1}.$ Then, for inflation, we would expect an amplitude of similar magnitude due to the approximate scale-invariance of the spectrum. But for ekpyrosis, the current horizon is of the order of 60 e-folds larger than $H_b^{-1},$ and hence, due to the blue spectrum, the amplitude of a horizon-sized gravity wave today would be exponentially small and thus unobservable. In fact, on large scales, one would expect the leading contribution to the gravitational wave background to be induced by the backreaction on the geometry of the first-order scalar perturbations \cite{Baumann:2007zm}.

Hence gravitational waves provide one of the crucial observational tests that are capable of unambiguously ruling out ekpyrotic models. If gravitational waves are detected, and it can be established that their spectrum is close to scale-invariant, this will constitute very strong evidence in favor of inflation. If, on the other hand, primordial gravitational waves remain unobserved, ekpyrotic models will deserve our attention and we should look for other significant tests (such as non-gaussianity) in order to improve our understanding of the early universe.


\section{The End is the Beginning: a Cyclic Model of the Universe}
\label{section cyclic}

\begin{figure}[t]
\begin{center}
\includegraphics[width=0.85\textwidth]{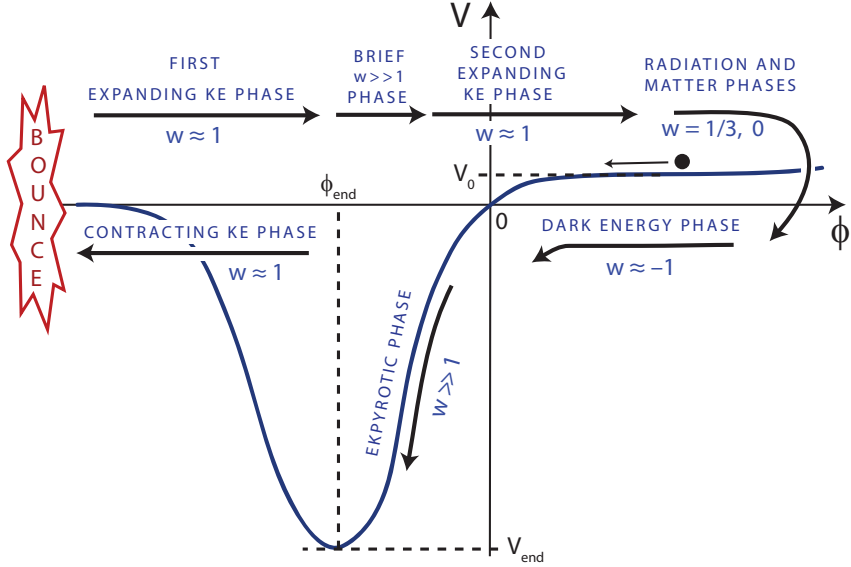}
\caption{\label{figure cycpotential} {\small
The potential for the cyclic universe integrates the ekpyrotic part and a quintessence epoch, but is irrelevant at the brane collision. A possible form for the potential is $V(\f)=V_0(e^{b\f}-e^{-c\f})F(\f),$ with $b\ll 1, \, c\gg 1$ and $F(\f)$ tends to unity for $\f>\f_{end}$ and to zero for $\f < \f_{end}.$ Reproduced with permission from \cite{Erickson:2006wc}.}}
\end{center}
\end{figure}

We have seen how the ekpyrotic phase can resolve the standard
puzzles of big bang cosmology, and how it can generate scalar
and tensor fluctuations with interesting properties. However,
this leaves many aspects of our universe unaddressed, and
foremost the rather mysterious dark energy. The cyclic model
\cite{Steinhardt:2001vw,Steinhardt:2001st,Steinhardt:2002kw,Steinhardt:2002ih,Steinhardt:2004gk,Khoury:2003rt,Erickson:2006wc} is an ambitious attempt at providing a complete
history of the universe, within the framework of a braneworld
view of the universe, while incorporating both the ekpyrotic
mechanism and dark energy in an essential way. We will first
outline the main ideas, then discuss the constraints that the
cyclic model has to satisfy and finally describe the global
view and new possibilities that the model offers.
The cyclic model relies on having a scalar field potential of
the general shape depicted in figure \ref{figure cycpotential}.
The plausibility of such a potential arising within the context
of string theory will be discussed in section \ref{section
fundamental}. The idea is that the ekpyrotic phase, which precedes
the big crunch/big bang transition, is itself preceded by a
phase where the potential is positive and flat. During this
phase, the radion field rolls down the potential very slowly,
and this in fact provides a period of dark energy domination
(in the form of quintessence \cite{Peebles:1987ek}). But this could be our
current universe! The dark energy that we see today is
reinterpreted as a small attractive force between our brane and
the second, parallel, end-of-the-world brane. At some point in the future, this force becomes stronger;
the potential energy becomes negative, and the universe reverts
from expansion to contraction. We then enter an ekpyrotic
phase, which locally flattens and homogenizes the branes. After
a brief phase of kinetic energy domination, the branes collide,
matter and radiation are produced and brane dwellers experience
a big bang. After the bang, the scalar field acquires a small
boost and quickly rolls back across the potential well and onto
the positive energy plateau, where the scalar gets
quasi-stabilized. Now the universe undergoes the usual periods
of radiation and matter domination, while the scalar field
starts rolling slowly back. Eventually, the resulting dark
energy comes to dominate the energy density of the universe,
and the whole cycle starts again. In this way, physical
processes in the universe today provide the initial conditions
for the next cycle. What should be noted is the amazing economy
of ingredients: only branes and their motion are used to
address the general evolution of the universe over its entire
history! Of course, aesthetic appeal is nice in a physical
model, but we must ask the question: can this work? In order to
answer this, we must have a closer look at the details.

First, there is the question of entropy \cite{Steinhardt:2001st}. The second law of thermodynamics implies that the entropy of each cycle should be larger than that of the previous cycle; but this seems to preclude the possibility of a truly cyclic universe (and historically, Tolman used this reasoning to show the impossibility of cyclic models in closed universes with zero cosmological constant \cite{Tolman}). The resolution is that, as we will see shortly, there is a huge net expansion every cycle, and hence, even though the total entropy always increases, the entropy gets diluted during the dark energy phase, and the entropy density can return to approximately the same value cycle after cycle. Note that the ekpyrotic phase does not undo the dilution, since the scale factor shrinks very modestly during ekpyrosis.

It is useful to keep track of various scales during the evolution of a given cycle. We are now at the start of a dark energy phase, during which the scale factor $a$ grows by an amount $e^{\cN_{DE}},$ where, by definition, $\cN_{DE}$ is the number of e-folds of dark energy domination. During this phase, the Hubble parameter remains approximately constant. Then, during the ekpyrotic phase, the scale factor shrinks very modestly, by $|V_{end}/V_{beg}|^{-p/2} \sim {\cal O}(1),$ where $V_{beg}$ is the value of the potential at the beginning of the ekpyrotic phase; this value coincides with the value of the present day cosmological constant $V_{beg} \approx V_0$. During the same time, the Hubble parameter grows immensely, by an amount $|V_{end}/V_{beg}|^{1/2} \equiv e^{\cN_{ek}},$ where $\cN_{ek}$ denotes the number of e-folds of ekpyrosis. As the branes collide, radiation and matter is produced and the branes ``reheat'' with a finite temperature, where we define $T_r$ to be the temperature at the time that the radiation density equals the scalar kinetic energy density. There is a contracting kinetic energy phase just before the brane collision, followed by an expanding kinetic phase after the bang. These two phases are almost exactly symmetrical, except for a slight boost that the scalar field obtains at the bounce. This boost must occur in order to achieve a cyclic model, because, due to extra Hubble damping from the radiation, the scalar needs a slight excess of kinetic energy to cross the potential well onto the plateau, see figure \ref{figure cycpotential}. There are two known ways by which the scalar field can obtain this slight excess energy \cite{Steinhardt:2001st,Turok:2004gb}: either due to the coupling of the matter fields to the radion, or if there is slightly more matter produced on the negative-tension brane than on the positive-tension one. The latter result makes sense heuristically, as more negative than positive inertia is produced in that case. The first expanding kinetic phase is followed by a brief, and unimportant \cite{Erickson:2006wc}, $w\gg 1$ phase, which in turn is followed by another expanding kinetic phase. We can treat all of these phases as a single kinetic phase, as in \cite{Erickson:2006wc}. Then, since $\rho \propto a^{-6} \propto t^{-2}$ during kinetic energy domination, the net expansion of the scale factor is given by $(|V_{end}|^{1/4}/T_r)^{2/3} = e^{2\gamma_{KE}/3},$ where we define the parameter \be \gamma_{KE} \equiv \ln \left( \frac{|V_{end}|^{1/4}}{T_r} \right). \ee Meanwhile, the Hubble parameter has shrunk by $(|V_{end}|^{1/4}/T_r)^{-2}=e^{-2\gamma_{KE}}.$ Finally, during radiation and matter domination, the universe grows by a factor of $T_r/T_0 \equiv e^{N_{rad}},$ where $T_0$ is the temperature of the CMB today; at the same time, the Hubble parameter shrinks by $(T_r/T_0)^{-2}=e^{-2N_{rad}}.$ Therefore, to recapitulate, the scale factor $a$ grows by a total of \be \cN_{DE}+\frac{2\gamma_{KE}}{3}+N_{rad} \label{growth}\ee e-folds over the course of a single cycle. However, because \be \frac{(-V_{end})^{1/2}}{V_{beg}^{1/2}}\frac{T_r^2}{(-V_{end})^{1/2}}\frac{T_0^2}{T_r^2} \approx 1,\ee due to $V_{beg} \approx T_0^4,$ the Hubble parameter actually returns to its original value after one cycle. This result can also be re-written as the statement that \cite{Erickson:2006wc} \be \cN_{ek} \approx 2(\gamma_{KE} + N_{rad}). \ee There is no constraint on the number of e-folds of dark energy domination $\cN_{DE}.$

In order for the cyclic model to work, the temperature $T_r$ has to satisfy a number of constraints. The most obvious one is that the reheat temperature has to be high enough for nucleosynthesis to take place. Also, it should be below the GUT scale, so that no topological defects are formed. Hence the first constraint is \be 1 MeV < T_r < 10^{16} GeV. \ee Next, in order for the successful predictions of nucleosynthesis not to be adversely affected by the overproduction of gravitational waves \cite{Allen:1996vm,Smith:2006nka}, the temperature has to satisfy \cite{Boyle:2003km} \be T_r \geqslant \frac{p}{20} |V_{end}|, \ee where $p$ is defined in (\ref{ekpyrosis-scaling}) and is typically of ${\cal O}(10^{-3})$ for realistic models. Also, the radiation component should not dominate the cosmological evolution until the scalar field has crossed the potential well and gotten onto the plateau of the potential. Now, combining the evolution of the scalar field, equation (\ref{kinetic solution}), with the form of the ekpyrotic potential $V \sim -e^{-c\phi},$ we have that $V \propto t^{-\sqrt{2/3}c}.$ Hence, the time $t_{cross}$ it takes for the scalar field to cross the well (\ie to go from $-V_{end}$ to $V_0$) is $t_{cross}=t_{end}|V_{end}/V_0|^{\sqrt{3/2}/c}.$ We need the time of radiation domination $t_r$ to be larger than $t_{cross},$ or, using that $t_r \approx H_r^{-1} \approx T_r^{-2},$ we get the constraint (this bound was derived in \cite{Khoury:2003rt}, although there is a typo regarding the sign of the inequality in the paper) \be T_r \leqslant (-V_{end})^{1/4}|\frac{V_0}{V_{end}}|^{\sqrt{3p/16}}. \ee It should be noted that a broad range of parameters satisfies all the constraints for cycling \cite{Khoury:2003rt}. Typical values for the various parameters are for example \be |V_{end}|=(10^{15} GeV)^4, \quad T_r=10^{12} GeV, \quad p=10^{-3}. \ee Then, with $T_0=10^{-3} eV,$ we have $\gamma_{KE} \approx 7$ and $N_{rad} \approx 55.$ Hence this leads to more than $120$ e-folds of ekpyrosis, largely sufficient to solve the standard cosmological puzzles. And assuming zero e-folds of dark energy domination, according to (\ref{growth}), the universe grows by a factor of about $e^{60}\approx 10^{26}$ per cycle. Note that this means that our currently observable universe measured only about a cubic meter one cycle ago!

The enormous time scales involved in the cyclic model (a single cycle can already easily last a trillion years) mean that new approaches are possible for addressing some long-standing problems, such as for example the cosmological constant problem. In trying to explain the smallness of the cosmological constant by a dynamical relaxation mechanism, the problem has typically been that there simply isn't enough time between the big bang and now in order to achieve the required level of tuning. This has been an issue with the model of Abbott \cite{Abbott:1984qf} for example. However, in the cyclic model, the same model can work, as shown in \cite{Steinhardt:2006bf}. The idea is that the plateau in the cyclic potential is really part of a tilted washboard-shaped potential for an axion-type scalar. That scalar can then tunnel down the potential, spending more and more time at lower and lower values of the cosmological constant. It spends an exponentially longer time at the lowest positive local minimum of the potential (before tunneling to negative values, where the universe then collapses), and hence that value of the cosmological constant is the most probable one. This is just one example, but the general idea is that there could be evolutionary time scales vastly larger than the ones that are usually considered, and which can lead to qualitatively new explanations.

Let us turn to the question of ``initial'' conditions. The single-field cyclic model is an attractor, while at the same time managing to generate scalar perturbations that are nearly scale-invariant \cite{Steinhardt:2001st,Heard:2002dr,Gratton:2003pe}. This is really quite remarkable, and it means that the issue of initial conditions is essentially circumvented. Evidently, avoiding the issue of initial conditions is one of the main attractions of cyclic models. However, for two-field models \cite{Lehners:2007ac}, the situation is more complicated. As discussed in section \ref{subsection twofields}, in order to achieve a nearly scale-invariant spectrum of scalar perturbations, the background trajectory must fall off along a ridge of the potential, \ie in the presence of an unstable transverse direction. This instability is essential to obtain perturbations that are compatible with observations \cite{Tolley:2007nq}, and hence there is an issue of initial conditions. This is an open problem at present. The question of how the field returns to a starting position very close to the ridge after each cycle remains to be addressed. Of course, to some extent this problem is linked to the general problem of moduli stabilization (or quasi-stabilization in the cosmological context). In heterotic string theory, this problem is less well understood than for example in the type IIB theory at present. More work is certainly needed in this area. In the more phenomenological context of the new ekpyrotic models, Buchbinder {\it et al.} \cite{Buchbinder:2007tw} have proposed to include a stabilizing positive mass potential at the top of the ekpyrotic ridge, while coupling the entropy field to light fermions. Then, during this pre-ekpyrotic phase, the entropy field oscillates around the minimum of the potential. The field gets localized near the top of the ridge, because the oscillations are damped as energy is dissipated in the form of thermal radiation due to the coupling of the entropy field to light fermions. Something similar could also work in the cyclic model, and in fact, as the entropy field rolls slightly up the stabilizing potential, this increased potential energy could be identified with today's cosmological constant\footnote{This idea was suggested to the author by Paul Steinhardt.}. Of course, one would like to see such a potential arise from a fundamental theory, and avoid putting it in by hand.

\begin{figure}[t]
\begin{center}
\includegraphics[width=0.5\textwidth]{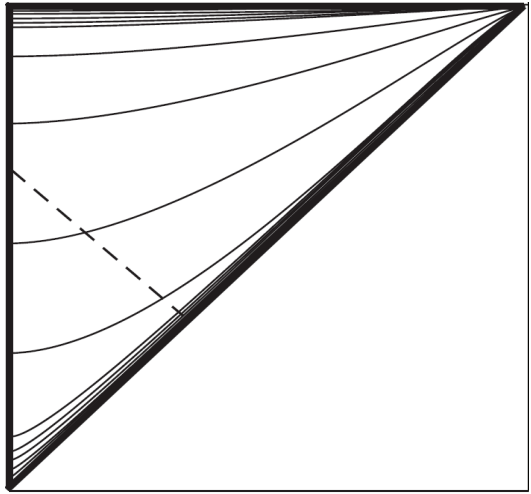}
\caption{\label{figure Penrose} {\small
On average, the Hubble parameter and the energy density are positive in the cyclic universe, and hence it can be represented approximately by the conformal diagram of de Sitter space, but with singular hypersurfaces corresponding to the big crunch/big bang transitions (here these are drawn only in the region covered by the usual flat coordinates on de Sitter space). In this diagram, time is vertical, and space horizontal, with each point representing a 2-sphere. A representative light signal is represented by the dashed line - it travels through many crunches, which makes it unlikely that one could see back through many cycles, even in principle.}}
\end{center}
\end{figure}

Finally, we will address the global structure of the cyclic universe and some more speculative questions. Since the universe undergoes an exponentially large net expansion every cycle, over large patches the universe will approximately have the structure of de Sitter space, but with singular hypersurfaces corresponding to the big crunch/big bang transitions \cite{Steinhardt:2004gk}. Hence, starting from a given patch of spacetime, the Penrose diagram will look approximately like the one shown in Fig. \ref{figure Penrose}. The cycling can however not be synchronous over super-horizon scales, as the ekpyrotic phase only homogenizes the branes over large, but finite-sized patches. Occasionally, there will be large quantum fluctuations as the brane collision is approached, leading to the formation of a large black hole at the collision. In this region, the cycling will stop \cite{Erickson:2006wc}; but since there is such a huge expansion of the universe during each cycle, these regions will be exponentially rare. In the same vein, one might ask whether an advanced civilization, with sufficient control over the spacetime metric, could stop the cycling in order to prolong its lifetime. Presumably so, but due to causality, only the region in its vicinity would be affected, and again, due to the vast net expansion, there would be plenty of room left for the cycling to continue unaffected. Thus, during each cycle, vast new regions of space are created, but within the same universe; in contrast to eternal inflation \cite{Linde:1986fc}, there is no proliferation of universes. Note also that it seems likely that only information from the previous cycle is remembered, and not from the ones before \cite{Erickson:2006wc}. This is essentially due to the effectiveness of the ekpyrotic phase in washing out all classical inhomogeneities, while new inhomogeneities are produced by random quantum processes. Nevertheless, could an advanced civilization have left us a message from the previous cycle? This seems extremely unlikely, as it would need to have had access to precisely the one cubic meter of space that has now become our currently observable universe. Could there have been a beginning to the cycling, for example if the two branes tunneled into existence? This is certainly possible, but if one really cannot see back more than one cycle prior to the current one, then this question might be rather meaningless. Finally, we note that, according to the present understanding, there seems to be no limit to the number of cycles allowed\footnote{Theorems about past-incompleteness \cite{Borde:2001nh} as well as entropy bounds \cite{Goheer:2002vf} do not apply, because of the singular crunch/bang transitions \cite{Steinhardt:2004gk}.}. Hence the cycles can continue indefinitely, creating more and more space, while injecting new matter and life into the universe periodically.


\section{The Link to Fundamental Theory: The Embedding in Heterotic M-theory}
\label{section fundamental}

It is the link between 11-dimensional supergravity and heterotic string theory which led to the colliding branes picture in the first place, and even though many of the relevant calculations are done in a 4-dimensional effective field theory, the higher-dimensional picture is of crucial importance for a deeper understanding of ekpyrotic/cyclic cosmology. Moreover, it is helpful in trying to visualize many of the processes described in the rest of this review, as it provides a geometrical interpretation of these\footnote{Many early studies dealing with the embedding of ekpyrotic/cyclic models in string theory/supergravity have considered the space in between the branes to be a slice of Anti de Sitter spacetime, see \eg \cite{Khoury:2001wf,Steinhardt:2001st,Tolley:2003nx,McFadden:2005mq}. Obtaining 5-dimensional AdS in supergravity is straightforward \cite{Schwarz:1983qr}, but as soon as the branes are added, the story becomes much more complicated, and not all issues have been resolved at present \cite{Chan:2000ms,Duff:2000az,Cvetic:2000id,Lehners:2007xa}. This is why we only present the theoretically better-motivated embedding into heterotic M-theory here.}.

The starting point for both ekpyrotic and cyclic cosmology is the brane picture of our universe as suggested by string theory. The various string theories and 11-dimensional supergravity are related to each other by a web of dualities \cite{Witten:1995ex}, and so any particular one of these theories could be used as a model for the early universe. However, certain theories have proved more intuitive than others in tackling particular problems, and it is the duality between 10-dimensional heterotic string theory with gauge group $E_8 \times E_8$ and 11-dimensional supergravity \cite{Horava:1995qa,Horava:1996ma} that has inspired the early universe models described in this review. Neglecting fermions, the action of the maximal, 11-dimensional supergravity is given by \cite{Cremmer:1978km} \bea S_{11} =
\frac{1}{2\k_{11}^2}\int_{M^{11}}\sqrt{-\hat{g}} && \bigg[
                    \hat{R}-\frac{1}{24}\hat{G}_{MNPQ}\hat{G}^{MNPQ}
                    \nn \\
          && -\frac{\sqrt{2}}{1728}\e^{M_1...M_{11}}
               \hat{A}_{M_1M_2M_3}\hat{G}_{M_4...M_7}\hat{G}_{M_8...M_{11}}\bigg],
\eea where we denote 11-dimensional quantities with hats. $\k_{11}$ determines the strength of gravity in 11 dimensions. Here
$\hat{G}_{MNPQ} = 4!\pt_{[M}\hat{A}_{NPQ]}$ is the 4-form field
strength associated with the 3-form $\hat{A}_{MNP}.$ The $\hat{A}_{(3)}\hat{G}_{(4)}\hat{G}_{(4)}$ term is a
Chern-Simons term and it is topological in nature. It will be unimportant in what follows. 11-dimensional supergravity is a classical theory, which nobody knows how to quantize. However, we can calculate anomalies (which do not depend on the microscopic structure of spacetime), and these allow us a glimpse of quantum M-theory. If we consider the 11-dimensional spacetime to take the form of a 10-dimensional spacetime plus a line segment, then anomaly cancelation requires us to add the following boundary terms \cite{Horava:1995qa,Horava:1996ma}
\bea S_{boundary} =
                - \frac{1}{8\pi\k_{11}^2}\left(\frac{\k_{11}}{4\pi}\right)^{2/3}
        && \!\!\! \int_{M_{10}^{(1)}}\sqrt{-g}\;\left[
           \tr(F^{(1)})^2 - \frac{1}{2}\tr R^2\right] \nn \\
        - \frac{1}{8\pi\k_{11}^2}\left(\frac{\k_{11}}{4\pi}\right)^{2/3}
          && \!\!\! \int_{M_{10}^{(2)}}\sqrt{-g}\;\left[
               \tr(F^{(2)})^2 - \frac{1}{2}\tr R^2\right]. \label{action-boundary}
\eea
The gauge fields correspond to the group $E_8$ and the $\tr R^2$ terms are required by supersymmetry. In the limit that the line segment becomes vanishingly small, the two boundaries sit on top of one another, and we recover the supergravity approximation to the 10-dimensional $E_8 \times E_8$ heterotic string theory. In fact, the size of the line segment is related to the coupling constant of the heterotic theory (roughly because the size of the line segment is reinterpreted as the dilaton field in 10 dimensions), via the famous relation \cite{Witten:1995ex}: \be R_{11} = g_s^{\frac{2}{3}}
\ee where $R_{11}$ is the radius of the 11th dimension and
$g_s$ is the string coupling constant. Thus, at strong coupling, the world looks 11-dimensional. As the boundary branes collide, the 11th dimension disappears, but at the same time the coupling goes to zero. Hence one might expect the collision to be rather mild.

In order to get a realistic model of particle physics, we can now consider curling up 6 of the spatial dimensions into a Calabi-Yau manifold. Then it turns out that in order to obtain the right magnitude of Newton's constant in 4 dimensions, the line segment must be larger than the radius of the Calabi-Yau manifold by a factor of about 30 or so \cite{Banks:1996ss,Witten:1996mz}. Hence, going up in energy, the world looks 4-dimensional, then 5-dimensional (as the line segment becomes visible), and only above the GUT scale 11-dimensional. This is the reason for choosing a 5-dimensional braneworld setup in describing the dynamics of the universe close to the big bang. The relevant 5-dimensional theory is called heterotic M-theory, and it can be obtained by dimensionally reducing the 11-dimensional theory described above on a 6-dimensional Calabi-Yau manifold, as follows \cite{Lukas:1998yy,Lukas:1998tt}.
The metric is reduced according to \be \d s_{11}^2 =
V_{CY}^{-2/3}g_{mn}\d x^m \d x^n +V_{CY}^{\frac{1}{3}} g_{ab}\d x^a \d x^b \; .\ee
where $V_{CY}$ denotes the volume of the Calabi-Yau, $g_{ab}$ being its
metric. Indices $m,n,...$ run over 5 spacetime dimensions (including the line segment direction, labeled $11,$ but where we also use the coordinate $x^{11}=y$), and $a,b,...$ denote Calabi-Yau directions. Due to the boundary terms (\ref{action-boundary}), there is a modified Bianchi identity for the 4-form mode, which has to satisfy \be (\d\hat{G})_{11MNPQ} = -\frac{1}{2\sqrt{2}\pi}
    \left(\frac{\k_{11}}{4\pi}\right)^{2/3} \left\{
       J^{(1)}\delta (y-1) + J^{(2)}\delta (y+1)
       \right\}_{MNPQ} \label{Bianchi} \ee where the currents are given by \be J^{(i)}
    = {\rm tr}F^{(i)}\wedge F^{(i)}
      - \frac{1}{2}{\rm tr}R\wedge R \; \ee and where we have chosen the coordinate positions $y=\pm 1$ for the location of the boundary branes. In the weakly coupled heterotic string theory the Bianchi
identity reads \be \d G \sim \rm{tr} R \wedge R - \rm{tr} F^{(1)}
\wedge F^{(1)} - \rm{tr} F^{(2)} \wedge F^{(2)} \; .\ee The
so-called ``standard embedding'' of the gauge connection into the
spin connection ensures that \be \rm{tr} F^{(1)} \wedge F^{(1)} =
\rm{tr} R \wedge R \label{stand emb} \ee and together with setting
$F^{(2)} = 0$ we see that we can have $\d G = 0$ and thus set $G$ to
zero consistently.
In our case, there appears to be no way of achieving $\rm{tr}
F^{(i)} \wedge F^{(i)} = \frac{1}{2} \rm{tr} R \wedge R$, and thus
there will always be some components of $\hat{G}$ that are
required to be non-zero. This mode is called the non-zero mode of
the reduction. The standard embedding (\ref{stand emb}) then leads to
\be (\d \hat{G})_{11MNPQ} = -\frac{1}{4\sqrt{2}\pi}
\left(\frac{\k_{11}}{4\pi}\right)^{2/3} (\delta(y-1) - \delta(y+1))\left\{ {\rm
tr}R\wedge R \right\}_{MNPQ}. \label{Bianchi2} \ee The
interpretation of this equation is that the $\pm {\rm
tr}R\wedge R$ terms are sources for oppositely charged branes sitting at the endpoints ($y=\pm 1$) of the line segment.
We can solve the modified Bianchi identity (\ref{Bianchi2}) above
(and at the same time satisfy the equation of motion for $\hat{G}$) to lowest order by setting\footnote{Compared to
\cite{Lukas:1998yy,Lukas:1998tt}, we have re-scaled $\a$ such that $\a =
\a_{\mathrm{LOSW}}/3\sqrt{2}$. Note that the parameters $\a$ and $\b$ used in this section are unrelated to the use of the same letters as indices of Hankel functions in section \ref{section perturbations}.}
\begin{equation}
 \hat{G}_{abcd} = \frac{\a}{\sqrt{2}}{\e_{abcd}}^{ef}\,\o_{ef}\,\theta(y)\; ,
            \label{nonzero}
\end{equation} where \bea \theta(y) &=&
\begin{cases} +1 & \mbox{for } -1 \le y < +1 \quad {\rm mod} \; 4 \\ -1 & \mbox{for }
-3 < y < -1 \quad {\rm mod} \; 4
\end{cases}
\eea and \be \a = \frac{\pi}{6}\left( \frac{\k_{11}}{4
\pi}\right)^{\frac{2}{3}}\b \ee where $\b = - \frac{1}{8
\pi^2}\int_{\mathcal{C}}{\rm tr}R \wedge R$ is the first
Pontryagin class of the Calabi-Yau (the integration being done
over the (2,2)-cycle $\mathcal{C}$ corresponding to the K\"{a}hler
form $\o_{ab}$). Note that $\hat{G}_{abcd}$ is related to the
integer $\b$ and thus $\hat{G}$ is quantized; hence $\a$ is also quantized. We are only considering the lowest mode
in the (Fourier) expansion of the full solution to the modified
Bianchi identity, as all higher order modes are massive (their
mass being of the order of the Calabi-Yau volume, {\it i.e.} the
GUT scale) and decouple in the approximation considered here. In contrast
to the other modes arising in the dimensional reduction of
$\hat{G}$, the non-zero mode cannot be set to zero.

Keeping only gravity, the non-zero mode and the Calabi-Yau volume modulus in 5 dimensions (which is a consistent truncation), we end up with the action \bea S_5 &=&
\frac{1}{2\k_5^2}\int_{5d}\sqrt{-g}\left[
                  R-\frac{1}{2}V_{CY}^{-2}\partial_m V_{CY}\partial^m V_{CY}
                   -6 \a^2 V_{CY}^{-2}\right] \nn
                   \\ && +\frac{1}{2\k_5^2}\left\{-12\a \int_{4d,y=+1}\sqrt{-g}
                   \, V_{CY}^{-1} +12 \a \int_{4d,y=-1}\sqrt{-g}\,
                   V_{CY}^{-1} \right\} \; .  \label{5dAction}
                   \eea
One should especially note the presence of the potential term
$6\a^2/V_{CY}^2$ which arose from the non-zero mode. This
potential means that (1,4)-dimensional Minkowski space is not a
solution to our theory, and hence not the vacuum. However such a
``cosmological'' term is dual in 5 dimensions to a 5-form field
strength, which can support a 3-brane solution. In 5 dimensions a
3-brane is of codimension 1 and thus represents a domain wall.
Note that in general for a 3-brane solution to be completely
consistent, one needs a (4-dimensional) source action. But we
actually have two 4-dimensional sources in the action
(\ref{5dAction}) located at $y=\pm 1.$ This double domain wall
vacuum was found in \cite{Lukas:1998yy}. It is given by
\bea
\d s^2 &=& h^{2/5}(y)\,\big[A^2 \,(-\d \t^2 + \d \vec{x}^2) + B^2 \,\d y^2\big], \nn \\
 V_{CY} &=& B\, h^{6/5}(y), \label{domainwall} \nn \\
h(y) &=& 5\a\, y+C,
\eea
where $A$, $B$ and $C$ are arbitrary constants.
The $y$ coordinate is taken to span
the orbifold (line segment) $S^1/\mathbb{Z}_2$ with fixed points at $y=\pm 1$.
In an `extended' picture of the solution, obtained by $\mathbb{Z}_2$-reflecting the solution across the branes,
there is a downward-pointing kink at $y=-1$ and an upward-pointing kink at $y=+1$.
These ensure the Israel conditions are satisfied, with the negative-tension brane being located at $y=-1$ and
the positive-tension brane at $y=+1$.
In the static domain wall solution above, the volume of the
Calabi-Yau manifold and the distance between the boundary branes
are determined in terms of the moduli $B$ and $C$, while the scale
factors on the branes are determined in terms of $A$ and $C$.  The
modulus $C$ additionally determines the height of the harmonic
function $h$ at a given position in $y$.
Before continuing by describing how we can extend this solution to allow for motion of the branes, we should add some remarks: by considering only the boundary branes, we are looking at a universal, but simplified version of heterotic M-theory. Typically, there are many more fields that can play a role, and there can be additional branes in the bulk. Such extra ingredients are usually necessary in order to obtain a realistic model of particle physics on the branes, see \eg \cite{Bouchard:2005ag,Braun:2005ux}. Here we focus on the simplest case of having only boundary branes, since we are interested in the collision of these boundary branes. Also, the braneworld picture suggest a natural place for dark matter to reside, namely on the second boundary brane. Any form of matter residing on the ``other'' brane will appear as dark to us, since we would only detect its effects via gravity. However, there is no quantitative argument known for how much of the dark matter should be ordinary matter on second brane, or weakly-interacting matter on the visible brane.

We are interested in finding the effective 4-dimensional theory describing the motion of the boundary branes that we have just discussed. We will obtain this by the method of the moduli space approximation \cite{Lehners:2006ir}: we promote the moduli $A,B,C$ in the solution above to arbitrary
functions of the brane conformal time $\t$, yielding the ansatz:
\bea \d s^2 &=&
h^{2/5}(\t,y)\left[A^2(\t)\,(-\d \t^2 + \d \vec{x}^2) + B^2(\t) \d y^2\right], \nn \\
V_{CY} &=& B(\t)\, h^{6/5}(\t,y), \label{TimeMod} \nn \\
h(\t,y) &=& 5\a y +C(\t), \qquad -1 \le y \le +1.
\eea
Let us give a brief justification for this ansatz:
firstly, we note that the ansatz satisfies the $\tau y$ Einstein
equation identically. This is important, since otherwise the $\tau
y$ equation would act as a constraint, see {\it e.g.}
\cite{Gray:2003vw}. Secondly, there is no $g_{\tau y}$ modulus,
since this metric component is odd under the $\mathbb{Z}_2$
symmetry, and therefore has to vanish at the location of the
branes. Any such component which is zero at the location
of the branes, but non-zero in the bulk, is necessarily
massive. In fact, from the work of \cite{Lehners:2005su}, we know
that, apart from the above moduli, all other perturbations have a
positive mass squared.
Having defined the time-dependent moduli, we would now like to
derive the action summarizing their equations of motion. This is
achieved by simply plugging the ansatz (\ref{TimeMod}) into the
action (\ref{5dAction}), yielding the result
\bea
S_{\mathrm{mod}} = -6 \int_{4d}
A^2BI_{\frac{3}{5}} && [ \Big(\frac{{A'}}{A}\Big)^2-\frac{1}{12}\Big(\frac{{B'}}{B}\Big)^2
 +\frac{{A'}{B'}}{AB} \nn \\ &&
-\frac{1}{25}\frac{I_{-\frac{7}{5}}}{I_{\frac{3}{5}}}\,{C'}^2
+\frac{3}{5}\,\frac{I_{-\frac{2}{5}}{A'}{C'}}{I_{\frac{3}{5}}A}],
\label{ActionMSA1}
\eea
where we have defined
\bea
I_n &=& \int_{-1}^{1} dy \ h^n = \frac{1}{5\a(n+1)}[(C+5\a)^{(n+1)}-(C-5\a)^{(n+1)}].
\eea
This action can be greatly simplified by
introducing the field redefinitions
\bea
a^2 &\equiv& A^2\,B\,I_{\frac{3}{5}}, \\
e^{\f_1/\sqrt{2}} &\equiv& B\,(I_{\frac{3}{5}})^{3/4}, \\
\f_2 &\equiv& -\frac{\sqrt{6}}{20}\int \d C\, (I_{\frac{3}{5}})^{-1}\,
\big[9\,(I_{-\frac{2}{5}})^2+16\,I_{-\frac{7}{5}}I_{\frac{3}{5}}\big]^{1/2}.
\label{DefinitionChi}
\eea
Note that $a$ has the interpretation of
being roughly the four-dimensional scale factor, whereas $\f_1$ and
$\f_2$ are four-dimensional scalars. The definition
(\ref{DefinitionChi}) can be rewritten as stating that
\be
\d\f_2 = -\frac{\sqrt{6}\d C}{2\,(C+5\a)^{1/5}\,(C-5\a)^{1/5}\, I_{\frac{3}{5}}}.
\ee This expression can be integrated to yield \be C = 5\a \left[
\frac{(1+e^{2\sqrt{2/3}\f_2})^{5/4}
+(1-e^{2\sqrt{2/3}\f_2})^{5/4}}{(1+e^{2\sqrt{2/3}\f_2})^{5/4}
-(1-e^{2\sqrt{2/3}\f_2})^{5/4}}\right]. \label{relCmoduli} \ee In terms of
$a$, $\f_1$ and $\f_2,$ the moduli space action (\ref{ActionMSA1})
then reduces to the remarkably simple form \cite{Lehners:2006ir} \be S_{\mathrm{mod}} =
 \int_{4d} [-6{a'}^2 + a^2 (\f_1^{'2} + \f_2^{'2})].
\label{ActionMSA2} \ee The minus sign in front of the kinetic term
for $a$ is characteristic of gravity, and in fact this is the
action for gravity with scale factor $a$ and two minimally coupled
scalar fields that is used for the calculations of cosmological perturbations in section \ref{section perturbations}. There is also a manifest $O(2)$ rotation symmetry
for the scalar fields. The equation of motion for $a$ reads \be
6{a''} = - a\,(\f_1^{'2} + \f_2^{'2}), \ee while the
equations of motion for $\f_1$ and $\f_2$ immediately lead to the
conserved charges $Q_{1}$ and $Q_{2}$, according to \be
a^2\, {\f'_1} = Q_{1}, \qquad a^2 \,{\f'_2} = Q_{2} .
\ee The solutions to these equations are given by  \bea
a^2 &=& 2\,\sqrt{Q_{1}^2+Q_{2}^2}\,(\t-\t_a), \\
\psi &=& \frac{\sqrt{\frac{3}{2}}Q_{1}}{\sqrt{Q_{1}^2+Q_{2}^2}}\,\ln [\f_{1,0}(\t-\t_a)], \\
\chi &=& \frac{\sqrt{\frac{3}{2}}Q_{2}}{\sqrt{Q_{1}^2+Q_{2}^2}}\,\ln [\f_{2,0}(\t-\t_a)],
\eea
where $Q_{1}$, $Q_{2}$, $\t_a$, $\f_{1,0}$, and
$\f_{2,0}$ are constants of integration.

We can now return to the ansatz (\ref{TimeMod}) and relate
physical quantities in five dimensions to the moduli fields $a$,
$\f_1$ and $\f_2$: if we denote the distance between the branes by
$d$, and the volume of the Calabi-Yau and the brane scale factors at the locations $y = \pm1$ by
$V_{CY\pm}$ and $a_{\pm}$ respectively,
then we have the relations
\bea
d &=& \frac{1}{3(2\a)^{1/4}}\, e^{\f_1/\sqrt{2}-\sqrt{3/2}\f_2}\,
[(1+e^{2\sqrt{2/3}\f_2})^{3/2}-|1-e^{2\sqrt{2/3}\f_2}|^{3/2}], \label{relModDistance} \\
V_{CY\pm} &=& (2\a)^{3/4} \, e^{\f_1/\sqrt{2}} \,\left\{
\begin{array}{lll}
                     \left( \cosh \sqrt{2/3}\f_2 \right)^{3/2}  & \\
                     \left(-\sinh \sqrt{2/3}\f_2 \right)^{3/2}, & \\
                    \end{array} \right.  \\
a_{\pm} &=& (2\a)^{1/8} \,a \,e^{-\f_1/2\sqrt{2}} \,\left\{
\begin{array}{lll}
                     \left( \cosh \sqrt{2/3}\f_2 \right)^{1/4}  & \\
                     \left(-\sinh \sqrt{2/3}\f_2 \right)^{1/4}. & \\
                    \end{array} \right. \label{relModScalefactor}
\eea
These relations are useful in
interpreting particular solutions to the moduli equations of
motion. Note in particular that the 4-dimensional effective theory describes the motion of {\it both} branes at the same time, as well as the geometry of the 5-dimensional bulk. In the limit that $\f_2 \rightarrow + \infty$, we have \bea
d &\simeq& (2\a)^{-1/4}\, e^{\f_1/\sqrt{2}-\f_2/\sqrt{6}}, \\
V_{CY\pm} &\simeq& (2\a)^{3/4}\, e^{\f_1/\sqrt{2}+\sqrt{3/2}\f_2},
\eea
whereas for $\f_2 \rightarrow - \infty$, we have
\bea
d &\simeq& (2\a)^{-1/4}\, e^{\f_1/\sqrt{2}+\f_2/\sqrt{6}}, \\
V_{CY\pm} &\simeq& (2\a)^{3/4}\, e^{\f_1/\sqrt{2}-\sqrt{3/2}\f_2}.
\eea Thus, in both limits, $\ln{d}$ and $\ln V_{CY\pm}$ are orthogonal
variables. This means that, sufficiently far away from the $\f_2 =
0$ axis, the fields $\f_1$ and $\f_2$ are, up to a re-scaling,
simply related to $\ln{d}$ and ${\ln V_{CY\pm}}$ by a rotation in field
space. Since the moduli space trajectories in terms of $\f_1$ and
$\f_2$ are straight lines, far from the $\f_2 =
0$ axis, the trajectories will also be approximately straight lines in terms
of $\ln{d}$ and $\ln V_{CY\pm}$.

The $\f_2=0$ axis plays a special role, in that it represents a boundary to moduli space. Positive values of $\f_2$ would lead to an unphysical negative volume of the Calabi-Yau manifold at the location of the negative-tension brane, and are forbidden. Physically, the boundary corresponds to a naked singularity in the bulk spacetime \cite{Lehners:2007nb}. In fact, it can be shown that in the presence of a wide variety of matter fields on the negative-tension brane, the negative-tension brane bounces off this singularity in a smooth way and without encountering the singularity \cite{Lehners:2007nb}. This is only possible because of the special properties of gravity on an object of negative tension. In comparing the moduli to
higher-dimensional quantities via equations
(\ref{relModDistance})-(\ref{relModScalefactor}), it is apparent
that this bounce of the negative-tension brane is equivalent to
flipping the sign of $\f_2$ and thus ${\f'_2}$ also, and hence
the otherwise straight trajectories in scalar field space get reflected off the
$\f_2 = 0$ boundary, as shown in Fig. \ref{FigureSubdominant}. This reflection has the important consequence of converting entropy perturbations around the background evolution into curvature perturbation, as discussed in section \ref{subsection twofields}.

As we have just seen, the general solution to the 4-dimensional effective theory (\ref{ActionMSA2}) is a straight line of arbitrary slope in scalar field space, but bending at the boundary $\f_2=0.$ In general, from the higher-dimensional point of view, the solutions correspond to a Kasner spacetime near the brane collision. However, one particular slope is special in that it corresponds to the least singular colliding branes solution studied in \cite{Lehners:2006pu}. It is given by \be a= |2 y_0 \t|^{1/2},
\qquad e^{\f_1}= |2 y_0 \t|^{3/2\sqrt{2}}, \qquad e^{\f_2}= |4 \a
y_0 \t|^{\sqrt{3}/2\sqrt{2}}, \label{IntConst2}\ee where $2 y_0$ has the interpretation of being the velocity of the branes at the collision, as, for small $\tau,$ we have from (\ref{relModDistance}) that $d \approx 2 y_0 \tau$. For this solution, the brane scale factors and the Calabi-Yau volume are finite at the brane collision (as can be verified directly using (\ref{relModDistance})-(\ref{relModScalefactor})), and hence the spacetime metric asymptotes to Milne near the collision. It was argued in section \ref{section crunch} that we can take this solution as a model spacetime for the crunch/bang transition. As discussed there, the ekpyrotic phase ensures that the curvatures remain small until the quantum gravity regime is reached. In particular, higher-derivative corrections involving the Riemann tensor will also be small near the collision for the quasi-Milne spacetime, and hence we should be able to trust the above solution up to about a Planck time (or so) on either side of the collision.

\begin{figure}[t]
\begin{center}
\hspace{-1cm}
\includegraphics[width=10cm]{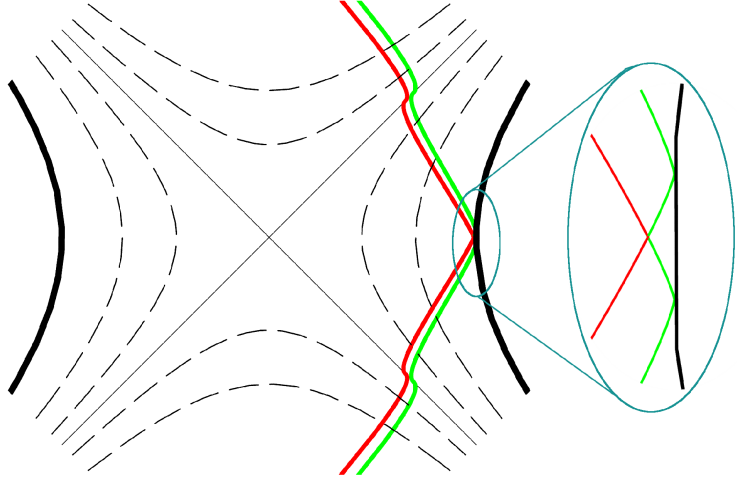}
\vspace{0.5cm} \caption{ \label{figure Kruskal} {\small
A Kruskal-type spacetime diagram in which the exact trajectories of the positive- and negative-tension
branes are plotted. The negative-tension brane (in green) is always to the right of the positive-tension brane (in red), thereby shielding it from the naked singularity (thick black line). The
bounces of the negative-tension brane off the naked singularity,
as well as the collision of the branes themselves, are shown at a
magnified scale in the inset. Note that the physical spacetime is restricted to the region bounded by the two branes.} }
\end{center}
\end{figure}

Note that the solution above is simply a solution to the moduli space effective action, and hence is only the leading order solution in an expansion in the brane velocity. However, the corresponding exact solution may actually be found directly in 5-dimensional heterotic M-theory (\ref{5dAction}) \cite{Lehners:2006pu}. Above, we have used a gauge in which the bulk is dynamical and the branes are kept at fixed coordinate positions. Here, it will be more useful to use a gauge where we have a static bulk through which the branes are moving. Then, subject to the boundary condition that the spacetime asymptotes to Milne at the collision, one can derive a Birkhoff-type theorem which fixes the bulk metric to be given by \cite{Lehners:2006pu} \bea \label{staticbulk}
\d s^2 &=& -f\,\d T^2 + \frac{r^{12}}{f}\,\d r^2 + r^2\, \d\vec{x}^2, \nn \\[1ex]
f(r) &=& \a^2 r^2 - \mu r^4, \qquad V_{CY} = r^6, \eea where $0\le
r\le \sqrt{\mu}/\a$, and the coordinate $T$ is unrelated to the
time coordinate appearing in the section \ref{subsection Milne}. Physically, this
solution describes a timelike naked singularity located at $r=0.$ The coordinates above do not cover the whole spacetime manifold, but
the maximal extension may easily be constructed following the
usual Kruskal procedure \cite{Lehners:2006pu}. The branes then move in this background spacetime according to the Israel junction conditions, and this motion is shown in Fig. \ref{figure Kruskal} (for a detailed derivation see \cite{Lehners:2006pu}).
One finds that the induced brane metrics
are indeed cosmological, with scale factors given by
\be
\label{bscalefactors} a_\pm = (1\pm 4\sqrt{\mu}\, \t_\pm)^{1/4},
\ee
where $\tau_\pm$ is the brane conformal time, and we have re-scaled
$a_\pm$ to unity at the collision, which is taken to occur at
$\tau_\pm = 0$.
Upon setting $\mu=0$, we immediately recover the static domain
wall solution (\ref{domainwall}), after a suitable change of
coordinates. More generally, we require $\mu\ge 0$ to avoid the
appearance of imaginary scale factors. $\mu$ is directly related to the brane velocity $y_0,$ and one can verify that the scaling solution (\ref{IntConst2}) is recovered as an approximation at low velocity \cite{Lehners:2006ir}.
For $\mu >0$, starting from the brane collision, the two branes
proceed to separate.  However, while the positive-tension brane
travels out to large radii unchecked, the negative-tension brane
reaches the naked singularity (at which $a_-$ and $V_{CY-}$ tend to
zero) in a finite brane conformal time $\tau_- =
(4\sqrt{\mu})^{-1}$. The resulting
singularity is simple to regularize. As stated above, if almost
any type of well-behaved matter is present on the negative-tension
brane -- even in only vanishing quantities -- then, rather than
hitting the singularity, the negative-tension brane will instead
undergo a bounce at some small finite value of the scale factor
and move away from the singularity. We can make this a little bit more precise by looking at the Friedmann equation on the negative-tension brane. For the case in which a time-dependent scalar field (with a
$V_{CY}$ coupling to the Calabi-Yau volume scalar) is present on the brane, this
equation takes the form (see \cite{Lehners:2006pu})
\be
H_-^2 = \frac{\dot{a}_-^2}{a_-^2}= -
2 \a \,\frac{\rho_-}{a_-^{18}}+ \frac{\rho_-^2}{a_-^{24}}+\frac{\mu}{a_-^{10}},
\ee
where the constant $\rho_-$ parameterizes the scalar kinetic energy
density. The key feature of this equation is the negative sign in front of
the first term on the right-hand side, reflecting the fact that
matter on the negative-tension brane couples to gravity with the
wrong sign.
For sufficiently large values of the scale factor the right-hand
side is dominated by the $\mu\, a_-^{-10}$ term. If we further
assume that the matter density on the branes is small compared to
the brane tension (as is in any case necessary for the
existence of a four-dimensional effective description, \ie $\rho_-
\ll \a$),
so that the term linear in $\rho_-$ dominates over the quadratic term, then
it follows that at some small value of the scale factor the entire
right-hand side must vanish. Thus a negative-tension brane,
initially traveling towards the singularity, will generically
undergo a smooth bounce at some small value of the scale factor.
After this bounce the brane travels away from the singularity back
towards large values of the scale factor. This behavior is
specific to the negative-tension brane, and moreover, persists
even in the limit in which $\rho_-$ (and hence the initial matter
density) is negligibly small. It can also be generalized to a wide class of negative-tension branes and matter types \cite{Lehners:2007nb}.

In order for the cyclic model and the ekpyrotic mechanism to work, we must assume a force between the boundary branes expressed by a potential of the form shown in Fig. \ref{figure cycpotential}. Could a potential of this shape arise from the heterotic M-theory setting under discussion here? First of all, we should note that a negative potential is completely natural in supergravity (hence all the problems in trying to explain the observed dark energy). Further, one can imagine that an attractive force between the branes could arise from virtual exchange of membranes stretching between the boundary branes (see \cite{Lehners:2002tw} for an example of such membrane solutions). Preliminary studies do indicate that an attractive force can arise from open membrane instantons \cite{Moore:2000fs,Lima:2001jc} in a certain parameter range. Unfortunately, a complete calculation of the potential involving all non-perturbative effects has been impossible to do so far. The vanishing of the potential near the brane collision seems natural, as the string coupling constant $g_s$ goes to zero there. Hence one would expect non-perturbative effects to vanish also as $\sim e^{-1/g_s}$ \cite{Steinhardt:2001st}. Certainly, one would eventually like to see a complete derivation of the potential in heterotic M-theory or a related setup.


\section{Conclusions and Outlook}
\label{section outlook}

Postulating a slowly contracting phase of the universe with equation of state $w \gg 1$ (preceding the standard expanding phase of big bang cosmology) has extremely powerful consequences: it renders the universe flat and homogeneous over large patches, it delays the onset of chaos until the quantum gravity regime is reached, it prepares the universe ideally for a big crunch/big bang transition, and, on top of that, it generates small-scale scalar perturbations with a nearly scale-invariant spectrum. Hence, even if the higher-dimensional picture and the link to string theory seem unappealing to you, at the very least the ekpyrotic phase represents a new mechanism for solving the standard cosmological puzzles and a new way of generating cosmological perturbations. Moreover, these perturbations contain a significant non-gaussian component, so that they can be tested observationally and compared to simple, gaussian, inflationary models. Further into the future, the prediction of a blue small-amplitude gravitational spectrum will also be tested against observations.

But more generally, ekpyrotic and cyclic models present a very broad conceptual framework, in which the big bang is not the beginning of time, but rather a physical event amenable to a testable physical description. Cyclic models allow for evolutionary timescales much larger than the time since the last big bang, and can provide qualitatively new explanations in this way. The embedding of ekpyrotic/cyclic models in string theory provides a geometric interpretation for many cosmological processes, and can eventually provide observational tests of the theory. Many open issues remain, which should be tractable within string theory: in particular, the issue of moduli stabilization requires further work, as does the closely related question of ``initial'' conditions in the presence of many moduli. Also, the all-important cyclic potential has not yet been derived from first principles. Most promising is the recent realization that the brane collision, and the corresponding 4-dimensional big crunch/big bang transition, can perhaps be understood within the context of generalizations of the non-perturbative gauge/gravity correspondence. Much work remains to be done here before a physically realistic case can be treated; but this is a topic of such importance to cosmology that we should see this as an opportunity, rather than a drawback. It would be truly magnificent if we could find observational signatures of higher-dimensional physics, and of a time before the big bang.

\section*{Acknowledgments}

It is a pleasure to thank Thorsten Battefeld, Ben Craps, Thomas Hertog, Justin Khoury, Kazuya Koyama, Paul McFadden, Burt Ovrut, S\'{e}bastien Renaux-Petel, Daniel Wesley and especially Paul Steinhardt and Neil Turok for many stimulating and valuable discussions. I would also like to thank my editor, Marc Kamionkowski, for encouragement and many helpful suggestions regarding the manuscript.



\begin{thebibliography}{999}

\bibitem{Wiki}
Explanation taken from $http://en.wikipedia.org/wiki/Ekpyrotic\_universe$.

\bibitem{Abbott:1984qf}
  L.~F.~Abbott,
  ``A Mechanism For Reducing The Value Of The Cosmological Constant,''
  Phys.\ Lett.\  B {\bf 150}, 427 (1985).

\bibitem{Adams:2006sv}
  A.~Adams, N.~Arkani-Hamed, S.~Dubovsky, A.~Nicolis and R.~Rattazzi,
  ``Causality, analyticity and an IR obstruction to UV completion,''
  JHEP {\bf 0610}, 014 (2006)
  [arXiv:hep-th/0602178].

\bibitem{Albrecht:1982wi}
  A.~Albrecht and P.~J.~Steinhardt,
  ``Cosmology For Grand Unified Theories With Radiatively Induced Symmetry
  Breaking,''
  Phys.\ Rev.\ Lett.\  {\bf 48}, 1220 (1982).

\bibitem{Allen:1996vm}
  B.~Allen,
  ``The stochastic gravity-wave background: Sources and detection,''
  arXiv:gr-qc/9604033.

\bibitem{ArkaniHamed:2003uy}
  N.~Arkani-Hamed, H.~C.~Cheng, M.~A.~Luty and S.~Mukohyama,
  ``Ghost condensation and a consistent infrared modification of gravity,''
  JHEP {\bf 0405}, 074 (2004)
  [arXiv:hep-th/0312099].

\bibitem{Babich:2004gb}
  D.~Babich, P.~Creminelli and M.~Zaldarriaga,
  ``The shape of non-Gaussianities,''
  JCAP {\bf 0408}, 009 (2004)
  [arXiv:astro-ph/0405356].

\bibitem{Banks:1996ss}
  T.~Banks and M.~Dine,
  ``Couplings and Scales in Strongly Coupled Heterotic String Theory,''
  Nucl.\ Phys.\  B {\bf 479}, 173 (1996)
  [arXiv:hep-th/9605136].

\bibitem{Battefeld:2007st}
  T.~Battefeld,
  ``Modulated Perturbations from Instant Preheating after new Ekpyrosis,''
  Phys.\ Rev.\  D {\bf 77}, 063503 (2008)
  [arXiv:0710.2540 [hep-th]].

\bibitem{Battefeld:2004mn}
  T.~J.~Battefeld, S.~P.~Patil and R.~Brandenberger,
  ``Perturbations in a bouncing brane model,''
  Phys.\ Rev.\  D {\bf 70}, 066006 (2004)
  [arXiv:hep-th/0401010].

\bibitem{Baumann:2007zm}
  D.~Baumann, P.~J.~Steinhardt, K.~Takahashi and K.~Ichiki,
  ``Gravitational Wave Spectrum Induced by Primordial Scalar Perturbations,''
  Phys.\ Rev.\  D {\bf 76}, 084019 (2007)
  [arXiv:hep-th/0703290].

\bibitem{Belinsky:1970ew}
  V.~A.~Belinsky, I.~M.~Khalatnikov and E.~M.~Lifshitz,
  ``Oscillatory approach to a singular point in the relativistic cosmology,''
  Adv.\ Phys.\  {\bf 19}, 525 (1970).

\bibitem{Berkooz:2002je}
  M.~Berkooz, B.~Craps, D.~Kutasov and G.~Rajesh,
  ``Comments on cosmological singularities in string theory,''
  JHEP {\bf 0303}, 031 (2003)
  [arXiv:hep-th/0212215].

\bibitem{Birrell:1982ix}
  N.~D.~Birrell and P.~C.~W.~Davies,
  ``Quantum Fields In Curved Space,''
{\it  Cambridge, Uk: Univ. Pr. ( 1982) 340p}

\bibitem{Borde:2001nh}
  A.~Borde, A.~H.~Guth and A.~Vilenkin,
  ``Inflationary space-times are incompletein past directions,''
  Phys.\ Rev.\ Lett.\  {\bf 90}, 151301 (2003)
  [arXiv:gr-qc/0110012].

\bibitem{Bouchard:2005ag}
  V.~Bouchard and R.~Donagi,
  ``An SU(5) heterotic standard model,''
  Phys.\ Lett.\  B {\bf 633}, 783 (2006)
  [arXiv:hep-th/0512149].

\bibitem{Boyle:2003km}
  L.~A.~Boyle, P.~J.~Steinhardt and N.~Turok,
  ``The cosmic gravitational wave background in a cyclic universe,''
  Phys.\ Rev.\  D {\bf 69}, 127302 (2004)
  [arXiv:hep-th/0307170].

\bibitem{Boyle:2004gv}
  L.~A.~Boyle, P.~J.~Steinhardt and N.~Turok,
  ``A new duality relating density perturbations in expanding and  contracting
  Friedmann cosmologies,''
  Phys.\ Rev.\  D {\bf 70}, 023504 (2004)
  [arXiv:hep-th/0403026].

\bibitem{Boyle:2005ug}
  L.~A.~Boyle, P.~J.~Steinhardt and N.~Turok,
  ``Inflationary predictions reconsidered,''
  Phys.\ Rev.\ Lett.\  {\bf 96}, 111301 (2006)
  [arXiv:astro-ph/0507455].

\bibitem{Brandenberger:2001bs}
  R.~Brandenberger and F.~Finelli,
  ``On the spectrum of fluctuations in an effective field theory of the
  ekpyrotic universe,''
  JHEP {\bf 0111}, 056 (2001)
  [arXiv:hep-th/0109004].

\bibitem{Braun:2005ux}
  V.~Braun, Y.~H.~He, B.~A.~Ovrut and T.~Pantev,
  ``A heterotic standard model,''
  Phys.\ Lett.\  B {\bf 618}, 252 (2005)
  [arXiv:hep-th/0501070].

\bibitem{Buchbinder:2007ad}
  E.~I.~Buchbinder, J.~Khoury and B.~A.~Ovrut,
  ``New Ekpyrotic Cosmology,''
  Phys.\ Rev.\  D {\bf 76}, 123503 (2007)
  [arXiv:hep-th/0702154].

\bibitem{Buchbinder:2007tw}
  E.~I.~Buchbinder, J.~Khoury and B.~A.~Ovrut,
  ``On the Initial Conditions in New Ekpyrotic Cosmology,''
  JHEP {\bf 0711}, 076 (2007)
  [arXiv:0706.3903 [hep-th]].

\bibitem{Buchbinder:2007at}
  E.~I.~Buchbinder, J.~Khoury and B.~A.~Ovrut,
  ``Non-Gaussianities in New Ekpyrotic Cosmology,''
  arXiv:0710.5172 [hep-th].

\bibitem{Chan:2000ms}
  C.~S.~Chan, P.~L.~Paul and H.~L.~Verlinde,
  ``A note on warped string compactification,''
  Nucl.\ Phys.\  B {\bf 581}, 156 (2000)
  [arXiv:hep-th/0003236].

\bibitem{Craps:2007ch}
  B.~Craps, T.~Hertog and N.~Turok,
  ``Quantum Resolution of Cosmological Singularities using AdS/CFT,''
  arXiv:0712.4180 [hep-th].

\bibitem{Craps:2003ai}
  B.~Craps and B.~A.~Ovrut,
  ``Global fluctuation spectra in big crunch / big bang string vacua,''
  Phys.\ Rev.\  D {\bf 69}, 066001 (2004)
  [arXiv:hep-th/0308057].

\bibitem{Creminelli:2006xe}
  P.~Creminelli, M.~A.~Luty, A.~Nicolis and L.~Senatore,
  ``Starting the universe: Stable violation of the null energy condition and
  non-standard cosmologies,''
  JHEP {\bf 0612}, 080 (2006)
  [arXiv:hep-th/0606090].

\bibitem{Creminelli:2004jg}
  P.~Creminelli, A.~Nicolis and M.~Zaldarriaga,
  ``Perturbations in bouncing cosmologies: Dynamical attractor vs scale
  invariance,''
  Phys.\ Rev.\  D {\bf 71}, 063505 (2005)
  [arXiv:hep-th/0411270].

\bibitem{Creminelli:2007aq}
  P.~Creminelli and L.~Senatore,
  ``A smooth bouncing cosmology with scale invariant spectrum,''
  JCAP {\bf 0711}, 010 (2007)
  [arXiv:hep-th/0702165].

\bibitem{Cremmer:1978km}
  E.~Cremmer, B.~Julia and J.~Scherk,
  ``Supergravity theory in 11 dimensions,''
  Phys.\ Lett.\  B {\bf 76}, 409 (1978).

\bibitem{Cvetic:2000id}
  M.~Cvetic, M.~J.~Duff, J.~T.~Liu, H.~Lu, C.~N.~Pope and K.~S.~Stelle,
  ``Randall-Sundrum brane tensions,''
  Nucl.\ Phys.\  B {\bf 605}, 141 (2001)
  [arXiv:hep-th/0011167].

\bibitem{Duff:2000az}
  M.~J.~Duff, J.~T.~Liu and K.~S.~Stelle,
  ``A supersymmetric type IIB Randall-Sundrum realization,''
  J.\ Math.\ Phys.\  {\bf 42}, 3027 (2001)
  [arXiv:hep-th/0007120].

\bibitem{Durrer:2002jn}
  R.~Durrer and F.~Vernizzi,
   ``Adiabatic perturbations in pre big bang models: Matching conditions and
  scale invariance,''
  Phys.\ Rev.\  D {\bf 66}, 083503 (2002)
  [arXiv:hep-ph/0203275].

\bibitem{Erickson:2006wc}
  J.~K.~Erickson, S.~Gratton, P.~J.~Steinhardt and N.~Turok,
  ``Cosmic perturbations through the cyclic ages,''
  Phys.\ Rev.\  D {\bf 75}, 123507 (2007)
  [arXiv:hep-th/0607164].

\bibitem{Erickson:2003zm}
  J.~K.~Erickson, D.~H.~Wesley, P.~J.~Steinhardt and N.~Turok,
  ``Kasner and mixmaster behavior in universes with equation of state $w >=1$,''
  Phys.\ Rev.\  D {\bf 69}, 063514 (2004)
  [arXiv:hep-th/0312009].

\bibitem{Goheer:2002vf}
  N.~Goheer, M.~Kleban and L.~Susskind,
  ``The trouble with de Sitter space,''
  JHEP {\bf 0307}, 056 (2003)
  [arXiv:hep-th/0212209].

\bibitem{Gordon:2000hv}
  C.~Gordon, D.~Wands, B.~A.~Bassett and R.~Maartens,
  ``Adiabatic and entropy perturbations from inflation,''
  Phys.\ Rev.\  D {\bf 63}, 023506 (2001)
  [arXiv:astro-ph/0009131].

\bibitem{Gratton:2003pe}
  S.~Gratton, J.~Khoury, P.~J.~Steinhardt and N.~Turok,
  ``Conditions for generating scale-invariant density perturbations,''
  Phys.\ Rev.\  D {\bf 69}, 103505 (2004)
  [arXiv:astro-ph/0301395].

\bibitem{Gray:2003vw}
  J.~Gray and A.~Lukas,
  ``Gauge five brane moduli in four-dimensional heterotic models,''
  Phys.\ Rev.\  D {\bf 70}, 086003 (2004)
  [arXiv:hep-th/0309096].

\bibitem{Guth:1980zm}
  A.~H.~Guth,
  ``The Inflationary Universe: A Possible Solution To The Horizon And Flatness
  Problems,''
  Phys.\ Rev.\  D {\bf 23}, 347 (1981).

\bibitem{Heard:2002dr}
  I.~P.~C.~Heard and D.~Wands,
  ``Cosmology with positive and negative exponential potentials,''
  Class.\ Quant.\ Grav.\  {\bf 19}, 5435 (2002)
  [arXiv:gr-qc/0206085].

\bibitem{Horava:1995qa}
  P.~Horava and E.~Witten,
  ``Heterotic and type I string dynamics from eleven dimensions,''
  Nucl.\ Phys.\  B {\bf 460}, 506 (1996)
  [arXiv:hep-th/9510209].

\bibitem{Horava:1996ma}
  P.~Horava and E.~Witten,
  ``Eleven-Dimensional Supergravity on a Manifold with Boundary,''
  Nucl.\ Phys.\  B {\bf 475}, 94 (1996)
  [arXiv:hep-th/9603142].

\bibitem{Hwang:2001zt}
  J.~c.~Hwang and H.~Noh,
  ``Non-singular big-bounces and evolution of linear fluctuations,''
  Phys.\ Rev.\  D {\bf 65}, 124010 (2002)
  [arXiv:astro-ph/0112079].

\bibitem{Kallosh:2007ad}
  R.~Kallosh, J.~U.~Kang, A.~Linde and V.~Mukhanov,
  ``The New Ekpyrotic Ghost,''
  JCAP {\bf 0804}, 018 (2008)
  [arXiv:0712.2040 [hep-th]].

\bibitem{Khoury:2001bz}
  J.~Khoury, B.~A.~Ovrut, N.~Seiberg, P.~J.~Steinhardt and N.~Turok,
  ``From big crunch to big bang,''
  Phys.\ Rev.\  D {\bf 65}, 086007 (2002)
  [arXiv:hep-th/0108187].

\bibitem{Khoury:2001wf}
  J.~Khoury, B.~A.~Ovrut, P.~J.~Steinhardt and N.~Turok,
  ``The ekpyrotic universe: Colliding branes and the origin of the hot big
  bang,''
  Phys.\ Rev.\  D {\bf 64}, 123522 (2001)
  [arXiv:hep-th/0103239].

\bibitem{Khoury:2001zk}
  J.~Khoury, B.~A.~Ovrut, P.~J.~Steinhardt and N.~Turok,
  ``Density perturbations in the ekpyrotic scenario,''
  Phys.\ Rev.\  D {\bf 66}, 046005 (2002)
  [arXiv:hep-th/0109050].

\bibitem{Khoury:2003vb}
  J.~Khoury, P.~J.~Steinhardt and N.~Turok,
  ``Great expectations: Inflation versus cyclic predictions for spectral
  tilt,''
  Phys.\ Rev.\ Lett.\  {\bf 91}, 161301 (2003)
  [arXiv:astro-ph/0302012].

\bibitem{Khoury:2003rt}
  J.~Khoury, P.~J.~Steinhardt and N.~Turok,
  ``Designing cyclic universe models,''
  Phys.\ Rev.\ Lett.\  {\bf 92}, 031302 (2004)
  [arXiv:hep-th/0307132].

\bibitem{Komatsu:2008hk}
  E.~Komatsu {\it et al.}  [WMAP Collaboration],
  ``Five-Year Wilkinson Microwave Anisotropy Probe (WMAP)
  Observations: Cosmological Interpretation,''
  arXiv:0803.0547 [astro-ph].

\bibitem{Komatsu:2000vy}
  E.~Komatsu and D.~N.~Spergel,
  ``Acoustic signatures in the primary microwave
  background bispectrum,''
  Phys.\ Rev.\  D {\bf 63}, 063002 (2001).

\bibitem{Kosowsky:1995aa}
  A.~Kosowsky and M.~S.~Turner,
  ``CBR anisotropy and the running of the scalar spectral index,''
  Phys.\ Rev.\  D {\bf 52}, 1739 (1995)
  [arXiv:astro-ph/9504071].

\bibitem{Koyama:2007if}
  K.~Koyama, S.~Mizuno, F.~Vernizzi and D.~Wands,
  ``Non-Gaussianities from ekpyrotic collapse with multiple fields,''
  JCAP {\bf 0711}, 024 (2007)
  [arXiv:0708.4321 [hep-th]].

\bibitem{Koyama:2007ag}
  K.~Koyama, S.~Mizuno and D.~Wands,
  ``Curvature perturbations from ekpyrotic collapse with multiple fields,''
  Class.\ Quant.\ Grav.\  {\bf 24}, 3919 (2007)
  [arXiv:0704.1152 [hep-th]].

\bibitem{Koyama:2007mg}
  K.~Koyama and D.~Wands,
  ``Ekpyrotic collapse with multiple fields,''
  JCAP {\bf 0704}, 008 (2007)
  [arXiv:hep-th/0703040].

\bibitem{Langlois:2006vv}
  D.~Langlois and F.~Vernizzi,
  ``Nonlinear perturbations of cosmological scalar fields,''
  JCAP {\bf 0702}, 017 (2007)
  [arXiv:astro-ph/0610064].

\bibitem{Lehners:2006pu}
  J.~L.~Lehners, P.~McFadden and N.~Turok,
  ``Colliding Branes in Heterotic M-theory,''
  Phys.\ Rev.\  D {\bf 75}, 103510 (2007)
  [arXiv:hep-th/0611259].

\bibitem{Lehners:2006ir}
  J.~L.~Lehners, P.~McFadden and N.~Turok,
  ``Effective Actions for Heterotic M-Theory,''
  Phys.\ Rev.\  D {\bf 76}, 023501 (2007)
  [arXiv:hep-th/0612026].

\bibitem{Lehners:2007ac}
  J.~L.~Lehners, P.~McFadden, N.~Turok and P.~J.~Steinhardt,
  ``Generating ekpyrotic curvature perturbations before the big bang,''
  Phys.\ Rev.\  D {\bf 76}, 103501 (2007)
  [arXiv:hep-th/0702153].

\bibitem{Lehners:2005su}
  J.~L.~Lehners, P.~Smyth and K.~S.~Stelle,
  ``Stability of Horava-Witten spacetimes,''
  Class.\ Quant.\ Grav.\  {\bf 22}, 2589 (2005)
  [arXiv:hep-th/0501212].

\bibitem{Lehners:2007xa}
  J.~L.~Lehners, P.~Smyth and K.~S.~Stelle,
  ``Kaluza-Klein Induced Supersymmetry Breaking for Braneworlds in Type IIB
  Supergravity,''
  Nucl.\ Phys.\  B {\bf 790}, 89 (2008)
  [arXiv:0704.3343 [hep-th]].

\bibitem{Lehners:2007wc}
  J.~L.~Lehners and P.~J.~Steinhardt,
  ``Non-Gaussian Density Fluctuations from Entropically Generated Curvature
  Perturbations in Ekpyrotic Models,''
  Phys.\ Rev.\  D {\bf 77}, 063533 (2008)
  [arXiv:0712.3779 [hep-th]].

\bibitem{Lehners:2008my}
  J.~L.~Lehners and P.~J.~Steinhardt,
  ``Intuitive understanding of non-gaussianity in ekpyrotic and cyclic
  models,''
  arXiv:0804.1293 [hep-th].

\bibitem{Lehners:2002tw}
  J.~L.~Lehners and K.~S.~Stelle,
  ``D = 5 M-theory radion supermultiplet dynamics,''
  Nucl.\ Phys.\  B {\bf 661}, 273 (2003)
  [arXiv:hep-th/0210228].

\bibitem{Lehners:2007nb}
  J.~L.~Lehners and N.~Turok,
  ``Bouncing Negative-Tension Branes,''
  Phys.\ Rev.\  D {\bf 77}, 023516 (2008)
  [arXiv:0708.0743 [hep-th]].

\bibitem{Lima:2001jc}
  E.~Lima, B.~A.~Ovrut, J.~Park and R.~Reinbacher,
  ``Non-perturbative superpotential from membrane instantons in heterotic
  M-theory,''
  Nucl.\ Phys.\  B {\bf 614}, 117 (2001)
  [arXiv:hep-th/0101049].

\bibitem{Linde:1981mu}
  A.~D.~Linde,
  ``A New Inflationary Universe Scenario: A Possible Solution Of The Horizon,
  Flatness, Homogeneity, Isotropy And Primordial Monopole Problems,''
  Phys.\ Lett.\  B {\bf 108}, 389 (1982).

\bibitem{Linde:1986fc}
  A.~D.~Linde,
  ``Eternal Chaotic Inflation,''
  Mod.\ Phys.\ Lett.\  A {\bf 1}, 81 (1986).

\bibitem{Liu:2002kb}
  H.~Liu, G.~W.~Moore and N.~Seiberg,
  ``Strings in time-dependent orbifolds,''
  JHEP {\bf 0210}, 031 (2002)
  [arXiv:hep-th/0206182].

\bibitem{Lukas:1998yy}
  A.~Lukas, B.~A.~Ovrut, K.~S.~Stelle and D.~Waldram,
  ``The universe as a domain wall,''
  Phys.\ Rev.\  D {\bf 59}, 086001 (1999)
  [arXiv:hep-th/9803235].

\bibitem{Lukas:1998tt}
  A.~Lukas, B.~A.~Ovrut, K.~S.~Stelle and D.~Waldram,
  ``Heterotic M-theory in five dimensions,''
  Nucl.\ Phys.\  B {\bf 552}, 246 (1999)
  [arXiv:hep-th/9806051].

\bibitem{Lyth:2001pf}
  D.~H.~Lyth,
  ``The primordial curvature perturbation in the ekpyrotic universe,''
  Phys.\ Lett.\  B {\bf 524}, 1 (2002)
  [arXiv:hep-ph/0106153].

\bibitem{Maldacena:2002vr}
  J.~M.~Maldacena,
  ``Non-Gaussian features of primordial fluctuations in single field
  inflationary models,''
  JHEP {\bf 0305}, 013 (2003)
  [arXiv:astro-ph/0210603].

\bibitem{Martin:2004pm}
  J.~Martin and P.~Peter,
  ``On the properties of the transition matrix in bouncing cosmologies,''
  Phys.\ Rev.\  D {\bf 69}, 107301 (2004)
  [arXiv:hep-th/0403173].

\bibitem{Martin:2001ue}
  J.~Martin, P.~Peter, N.~Pinto Neto and D.~J.~Schwarz,
  ``Passing through the bounce in the ekpyrotic models,''
  Phys.\ Rev.\  D {\bf 65}, 123513 (2002)
  [arXiv:hep-th/0112128].

\bibitem{McFadden:2005mq}
  P.~L.~McFadden, N.~Turok and P.~J.~Steinhardt,
  ``Solution of a braneworld big crunch / big bang cosmology,''
  Phys.\ Rev.\  D {\bf 76}, 104038 (2007)
  [arXiv:hep-th/0512123].

\bibitem{Misner:1974qy}
  C.~W.~Misner, K.~S.~Thorne and J.~A.~Wheeler,
  ``Gravitation,''
{\it  San Francisco 1973, 1279p}

\bibitem{Moore:2000fs}
  G.~W.~Moore, G.~Peradze and N.~Saulina,
  ``Instabilities in heterotic M-theory induced by open membrane  instantons,''
  Nucl.\ Phys.\  B {\bf 607}, 117 (2001)
  [arXiv:hep-th/0012104].

\bibitem{Mukhanov:1985rz}
  V.~F.~Mukhanov,
  ``Gravitational Instability Of The Universe Filled With A Scalar Field,''
  JETP Lett.\  {\bf 41}, 493 (1985)
  [Pisma Zh.\ Eksp.\ Teor.\ Fiz.\  {\bf 41}, 402 (1985)].

\bibitem{Niz:2006ef}
  G.~Niz and N.~Turok,
  ``Classical propagation of strings across a big crunch / big bang
  singularity,''
  Phys.\ Rev.\  D {\bf 75}, 026001 (2007)
  [arXiv:hep-th/0601007].

\bibitem{Notari:2002yc}
  A.~Notari and A.~Riotto,
  ``Isocurvature perturbations in the ekpyrotic universe,''
  Nucl.\ Phys.\  B {\bf 644}, 371 (2002)
  [arXiv:hep-th/0205019].

\bibitem{Peebles:1987ek}
  P.~J.~E.~Peebles and B.~Ratra,
  ``Cosmology with a Time Variable Cosmological Constant,''
  Astrophys.\ J.\  {\bf 325}, L17 (1988).

\bibitem{Perlmutter:1998np}
  S.~Perlmutter {\it et al.}  [Supernova Cosmology Project Collaboration],
  ``Measurements of Omega and Lambda from 42 High-Redshift Supernovae,''
  Astrophys.\ J.\  {\bf 517}, 565 (1999)
  [arXiv:astro-ph/9812133].

\bibitem{Riess:1998cb}
  A.~G.~Riess {\it et al.}  [Supernova Search Team Collaboration],
  ``Observational Evidence from Supernovae for an Accelerating Universe and a
  Cosmological Constant,''
  Astron.\ J.\  {\bf 116}, 1009 (1998)
  [arXiv:astro-ph/9805201].

\bibitem{Sasaki:1995aw}
  M.~Sasaki and E.~D.~Stewart,
  ``A General Analytic Formula For The Spectral Index Of The Density
  Perturbations Produced During Inflation,''
  Prog.\ Theor.\ Phys.\  {\bf 95}, 71 (1996)
  [arXiv:astro-ph/9507001].

\bibitem{Schwarz:1983qr}
  J.~H.~Schwarz,
  ``Covariant Field Equations Of Chiral N=2 D=10 Supergravity,''
  Nucl.\ Phys.\  B {\bf 226}, 269 (1983).

\bibitem{Slosar:2008hx}
  A.~Slosar, C.~Hirata, U.~Seljak, S.~Ho and N.~Padmanabhan,
  ``Constraints on local primordial non-Gaussianity from large scale
  structure,''
  arXiv:0805.3580 [astro-ph].

\bibitem{Smith:2006nka}
  T.~L.~Smith, E.~Pierpaoli and M.~Kamionkowski,
  ``A new cosmic microwave background constraint to primordial  gravitational
  waves,''
  Phys.\ Rev.\ Lett.\  {\bf 97}, 021301 (2006)
  [arXiv:astro-ph/0603144].

\bibitem{Starobinsky:1986fxa}
  A.~A.~Starobinsky,
  ``Multicomponent de Sitter (Inflationary) Stages and the Generation of
  Perturbations,''
  JETP Lett.\  {\bf 42}, 152 (1985)
  [Pisma Zh.\ Eksp.\ Teor.\ Fiz.\  {\bf 42}, 124 (1985)].

\bibitem{Steinhardt:2001vw}
  P.~J.~Steinhardt and N.~Turok,
  ``A cyclic model of the universe,''
  arXiv:hep-th/0111030.

\bibitem{Steinhardt:2001st}
  P.~J.~Steinhardt and N.~Turok,
  ``Cosmic evolution in a cyclic universe,''
  Phys.\ Rev.\  D {\bf 65}, 126003 (2002)
  [arXiv:hep-th/0111098].

\bibitem{Steinhardt:2002kw}
  P.~J.~Steinhardt and N.~Turok,
  ``The cyclic universe: An informal introduction,''
  Nucl.\ Phys.\ Proc.\ Suppl.\  {\bf 124}, 38 (2003)
  [arXiv:astro-ph/0204479].

\bibitem{Steinhardt:2002ih}
  P.~J.~Steinhardt and N.~Turok,
  ``A cyclic model of the universe,''
  Science {\bf 296}, 1436 (2002).

\bibitem{Steinhardt:2004gk}
  P.~J.~Steinhardt and N.~Turok,
  ``The cyclic model simplified,''
  New Astron.\ Rev.\  {\bf 49}, 43 (2005)
  [arXiv:astro-ph/0404480].

\bibitem{Steinhardt:2006bf}
  P.~J.~Steinhardt and N.~Turok,
  ``Why the cosmological constant is small and positive,''
  Science {\bf 312}, 1180 (2006)
  [arXiv:astro-ph/0605173].

\bibitem{Tolley:2002cv}
  A.~J.~Tolley and N.~Turok,
  ``Quantum fields in a big crunch / big bang spacetime,''
  Phys.\ Rev.\  D {\bf 66}, 106005 (2002)
  [arXiv:hep-th/0204091].

\bibitem{Tolley:2003nx}
  A.~J.~Tolley, N.~Turok and P.~J.~Steinhardt,
  ``Cosmological perturbations in a big crunch / big bang space-time,''
  Phys.\ Rev.\  D {\bf 69}, 106005 (2004)
  [arXiv:hep-th/0306109].

\bibitem{Tolley:2007nq}
  A.~J.~Tolley and D.~H.~Wesley,
  ``Scale-invariance in expanding and contracting universes from two-field
  models,''
  JCAP {\bf 0705}, 006 (2007)
  [arXiv:hep-th/0703101].

\bibitem{Tolman}
  R.~C.~Tolman,
  ``Relativity, Thermodynamics and Cosmology,''
{\it  Oxford, Uk: Univ. Pr. ( 1934)}

\bibitem{Tsujikawa:2001ad}
  S.~Tsujikawa,
  Phys.\ Lett.\  B {\bf 526}, 179 (2002)
  [arXiv:gr-qc/0110124].

\bibitem{Turok:2007ry}
  N.~Turok, B.~Craps and T.~Hertog,
  ``From Big Crunch to Big Bang with AdS/CFT,''
  arXiv:0711.1824 [hep-th].

\bibitem{Turok:2004gb}
  N.~Turok, M.~Perry and P.~J.~Steinhardt,
  ``M theory model of a big crunch / big bang transition,''
  Phys.\ Rev.\  D {\bf 70}, 106004 (2004)
  [Erratum-ibid.\  D {\bf 71}, 029901 (2005)]
  [arXiv:hep-th/0408083].

\bibitem{Turok:2004yx}
  N.~Turok and P.~J.~Steinhardt,
  ``Beyond inflation: A cyclic universe scenario,''
  Phys.\ Scripta {\bf T117}, 76 (2005)
  [arXiv:hep-th/0403020].

\bibitem{Verde:1999ij}
  L.~Verde, L.~M.~Wang, A.~Heavens and M.~Kamionkowski,
  ``Large-scale structure, the cosmic microwave background, and primordial
  non-gaussianity,''
  Mon.\ Not.\ Roy.\ Astron.\ Soc.\  {\bf 313}, L141 (2000)
  [arXiv:astro-ph/9906301].

\bibitem{Wesley:2005bd}
  D.~H.~Wesley, P.~J.~Steinhardt and N.~Turok,
  ``Controlling chaos through compactification in cosmological models with  a
  collapsing phase,''
  Phys.\ Rev.\  D {\bf 72}, 063513 (2005)
  [arXiv:hep-th/0502108].

\bibitem{Witten:1995ex}
  E.~Witten,
  ``String theory dynamics in various dimensions,''
  Nucl.\ Phys.\  B {\bf 443}, 85 (1995)
  [arXiv:hep-th/9503124].

\bibitem{Witten:1996mz}
  E.~Witten,
  ``Strong Coupling Expansion Of Calabi-Yau Compactification,''
  Nucl.\ Phys.\  B {\bf 471}, 135 (1996)
  [arXiv:hep-th/9602070].

\bibitem{Yadav:2007yy}
  A.~P.~S.~Yadav and B.~D.~Wandelt,
  ``Detection of primordial non-Gaussianity (fNL) in the WMAP 3-year data at
  above 99.5\% confidence,''
  arXiv:0712.1148 [astro-ph].





\end{thebibliography}

\end{document}